\documentclass{article}

\usepackage[nonatbib,final]{neurips_2020}

\usepackage[utf8]{inputenc} 
\usepackage[T1]{fontenc}    
\usepackage{hyperref}       
\usepackage{url}            
\usepackage{booktabs}       
\usepackage{amsfonts}       
\usepackage{nicefrac}       
\usepackage{microtype}      
\usepackage{bm}

\usepackage{xspace}
\usepackage{graphicx}
\usepackage[table,dvipsnames]{xcolor}
\usepackage{amsmath}
\usepackage[percent]{overpic}
\usepackage[font={small}]{caption}
\usepackage{gensymb}

\makeatletter
\DeclareRobustCommand\onedot{\futurelet\@let@token\@onedot}
\def\@onedot{\ifx\@let@token.\else.\null\fi\xspace}
\def\eg{\emph{e.g}\onedot} 
\def\ie{\emph{i.e}\onedot} 
 
\def\etc{\emph{etc}\onedot} \def\vs{\emph{vs}\onedot}
 
\def\etal{\emph{et al}\onedot}
\def\resp{\emph{resp}\onedot}
\makeatother

\newcommand{\bx}{{\mathbf{x}}}

\newcommand{\cX}{{\mathcal{X}}}
\newcommand{\bw}{{\mathbf{w}}}

\newcommand{\bh}{{\mathbf{h}}}
\newcommand{\bn}{{\mathbf{n}}}

\newcommand{\bg}{{\mathbf{g}}}
\newcommand{\bp}{{\mathbf{p}}}
\newcommand{\bc}{{\mathbf{c}}}
\newcommand{\bbr}{{\mathbf{r}}}
\newcommand{\cP}{{\mathcal{P}}}

\newcommand{\cR}{{\mathcal{R}}}
\newcommand{\northamerica}{UrbanCity}

\title{MuSCLE: Multi Sweep Compression of LiDAR using Deep Entropy Models}

\author{%
  Sourav Biswas$^{1,2}$
  \quad
  Jerry Liu$^{1}$
  \quad
  Kelvin Wong$^{1,3}$
  \quad
  Shenlong Wang$^{1,3}$
  \quad
  Raquel Urtasun$^{1,3}$ \\
  $^{1}$Uber Advanced Technologies Group
  \quad
  $^{2}$University of Waterloo
  \quad
  $^{3}$University of Toronto \\
  \texttt{\{souravb,jerryl,kelvin.wong,slwang,urtasun\}@uber.com}
}

\begin{document}

\maketitle


\begin{abstract}
We present a novel compression algorithm for reducing the storage of LiDAR sensor data streams. Our model exploits spatio-temporal relationships across multiple LiDAR sweeps to reduce the bitrate of both geometry and intensity values. Towards this goal, we propose a novel conditional entropy model that models the probabilities of the octree symbols by considering both coarse level geometry and previous sweeps' geometric and intensity information. We then use the learned probability to encode the full data stream into a compact one.
Our experiments demonstrate that our method significantly reduces the joint geometry and intensity bitrate over prior state-of-the-art LiDAR compression methods, with a reduction of 7--17\% and 6--19\% on the \northamerica~and SemanticKITTI datasets respectively.
\end{abstract}


\section{Introduction}

The past decade has witnessed numerous innovations in intelligent systems, thanks to an explosion of progress in sensing and AI algorithms. In particular, LiDAR sensors are extensively used in various applications such as indoor rovers, unmanned aerial vehicles, and self-driving cars to accurately capture the 3D geometry of the scene.
Yet the rapid adoption of LiDAR has brought about a key challenge---dealing with the mounting storage costs associated with the massive influx of LiDAR data.
For instance, a 64-line Velodyne LiDAR continuously scanning a given scene produces over \textit{3 billion points} in a single hour.
Hence, developing efficient and effective compression algorithms to store such 3D point cloud data streams is crucial to reduce the storage and communication bandwidth.

Unlike its well-studied image and video counterparts, point cloud stream compression is a challenging yet under-explored problem.
Many prior approaches have focused on encoding a point cloud stream as independent  \textit{sweeps}, where each sweep captures a rough 360-degree rotation of the sensor.
Early approaches exploit a variety of compact data structures to represent the point cloud in a memory-efficient manner, such as octrees \cite{mekuria2016}, KD-trees \cite{draco}, and spherical images \cite{tu2016}.
More recent works along this direction utilize powerful machine learning models to encode redundant geometric correlations within these data structures for better compression \cite{tu2019, tu2019b, huang2020}.
In general, most of these aforementioned approaches do not make effective use of temporal correlations within point clouds.
Moreover, these prior approaches have largely focused on compressing the geometric structure of the point cloud (the spatial coordinates); yet there has been little attention paid towards compression of other attributes, \eg LiDAR intensity, which are crucial for many downstream tasks. Compressing such attributes along with geometric structure can make a significant impact on reducing storage.

In this paper, we present a novel, learning-based compression algorithm that comprehensively reduces the storage of LiDAR sensor data streams.
Our method extends the recent success of octree-structured deep entropy models \cite{huang2020} for single LiDAR sweep compression to intensity-valued LiDAR streaming data. Specifically, we propose a novel \textit{deep conditional entropy model} that models the probabilities of the octree symbols and associated intensity values by exploiting spatio-temporal correlations within the data: taking both coarse level information at the current sweep, as well as relevant neighboring nodes information from the previous sweep.
Unlike prior approaches, our method models the joint entropy across an entire point cloud sequence, while unifying geometry and attribute compression into the same framework.

We validate the performance of our approach on two large datasets, namely \northamerica~\cite{liang_pnpnet} and SemanticKITTI~\cite{behley2019}.
The experiments demonstrate that our method significantly reduces the joint geometry and intensity bitrate over prior state-of-the-art LiDAR compression methods,
with a reduction of 7--17\% on \northamerica~and 6--19\% on SemanticKITTI.
We also conduct extensive experiments showcasing superior performance against prior works on numerous downstream perception tasks.


\section{Multi-Sweep LiDAR Compression}
\label{headings}

In this work, we propose a comprehensive framework for the \textit{lossy} compression of LiDAR point cloud streams, by exploiting the spatio-temporal redundancies through a \textit{learned entropy model}. We aim to maximize the reconstruction quality of these point clouds while reducing their joint bitrate.
Every point in a LiDAR point cloud contains both a spatial 3D location $(x,y,z)$, as well as an intensity value $r$, and we jointly compress both.

\begin{figure}[t]
    \centering
	\includegraphics[width=0.99\linewidth]{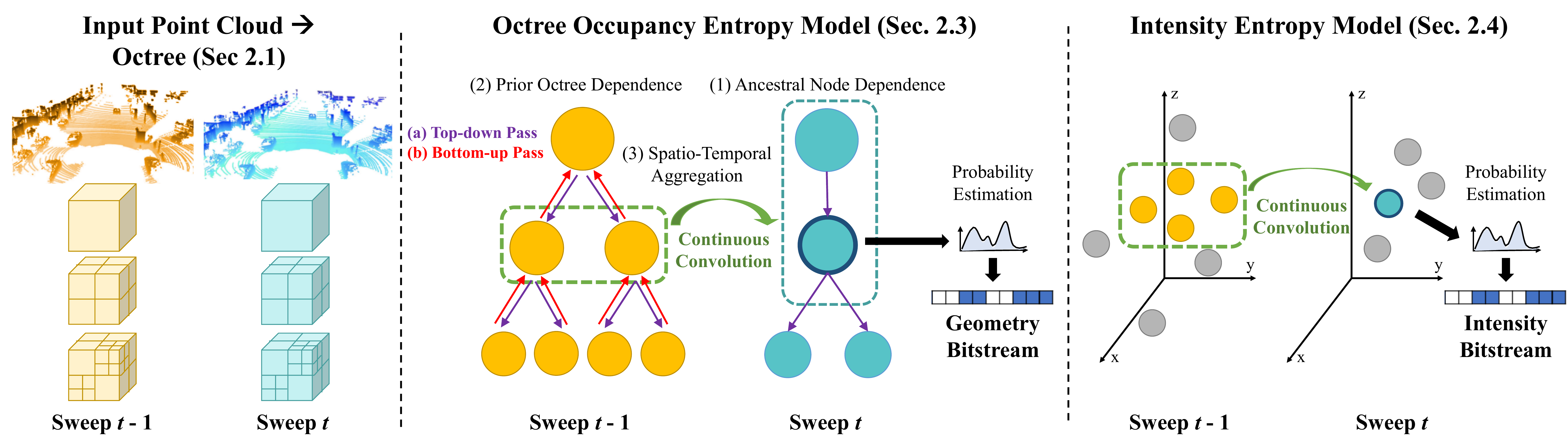}
	\caption{\small Comprehensive overview of our method. Our point cloud stream is serialized into an octree representation (Sec \ref{sec:method_octree}). We apply a spatio-temporal entropy model to the octree occupancy bytestream (Sec. \ref{sec:occ_entropy_model}), modeling ancestral dependence, prior octree dependence, and octree alignment. We also apply a deep entropy model to model the intensity stream (Sec. \ref{sec:att_entropy_model}). }
    \label{fig:main_layout}
\end{figure}

Our method is shown in Fig. \ref{fig:main_layout}.
We first quantize and encode all point spatial coordinates in the stream into an octree representation, where leaves represent the quantized points and intermediate nodes contain 8-bit symbols representing child occupancies (Sec. \ref{sec:method_octree}).
We then present a novel deep entropy model (Sec. \ref{sec:method_setup}): a probability model that utilizes \textit{spatio-temporal context} to predict occupancy symbols for each node (Sec. \ref{sec:occ_entropy_model}), as well as intensity values for each point for intensity compression (Sec. \ref{sec:att_entropy_model}). The outputs of these entropy models are finally fed into a lossless entropy coding algorithm, such as range coding, to produce the final bitstream (Sec. \ref{sec:entropy_coding}).

\subsection{Octree Representation} \label{sec:method_octree}

\paragraph{Octree Structure and Bit Representation:}
LiDAR point clouds are intrinsically challenging to process due to their sparsity and  inherently unstructured nature. A tool to counteract these challenges is to use a tree-based data structure, such as an octree or KD-tree, to efficiently partition the  space. Inspired by \cite{mekuria2016, huang2020}, we quantize and represent every point cloud in our stream as an octree with an associated depth value $D$, corresponding to the quantized precision of the point cloud.

Specifically, an octree can be constructed from a 3D point cloud by first partitioning the spatial region into 8 octants, and recursively partitioning each octant until each node contains at most one point, or until $D$ is reached. The resulting octree contains both intermediate nodes and leaf nodes. Each intermediate node can be represented by an 8-bit occupancy symbol $\bx$, representing the occupancies of its children; each node also has an implied spatial position.
Each leaf node contains one point of the point cloud, and stores the offset between the point and its corner position, as well as the point intensity.
We determine the intensity value of each point in the quantized point cloud by taking that of its nearest neighbor in the original point cloud.
The number of bits allocated to each leaf node is level-dependent; an octree with $D=k$ will store $k-i$ bits for a leaf node at level $i,  i \leq k$. Hence, the octree is memory-efficient---shared bits are encoded with intermediate nodes and residual bits with leaves.

\paragraph{Serialization:} We serialize the octree into two (uncompressed) bytestreams by traversing the octree in breadth-first order. The first bytestream
contains the intermediate node occupancy symbols in breadth-first order, and the second bytestream contains the leaf node offsets/intensities encountered during traversal. Our entropy model focuses primarily on the node occupancies/intensities---we demonstrate in our supplementary materials that leaf offsets do not contain meaningful patterns we can exploit. Hence for subsequent sections we denote $\cP^{(t)} = (\cX^{(t)}, \cR^{(t)})$, where $\cX^{(t)} = \{ \bx^{(t)}_1, ..., \bx^{(t)}_{m_t} \}$ is the set of occupancy symbols, and $\cR^{(t)} = \{  \bbr^{(t)}_1,...,\bbr^{(t)}_{n_t} \}$ is the set of intensities. The serialization is lossless; the only loss comes from $D$-dependent octree quantization. This gives a guarantee on reconstruction quality and allows compression efforts to solely focus on bitrate reduction.

\subsection{Octree-Based Conditional Entropy Module} \label{sec:method_setup}

The octree sequence is now fed into our \textit{entropy model}. Our entropy model is a probability model that approximates the unknown joint distribution of point clouds $p_{\mathrm{data}}$ with our own distribution $p( \cdot ; \bw)$. Since we convert our point clouds to octree representations, the probability model is equivalent to modeling $p(\cP^{(1)},...,\cP^{(n)}; \bw)$. According to the classic Shannon's source coding theorem \cite{shannon1948}, the expected bitrate for the point cloud stream is tightly approximated by the cross-entropy between the real point cloud stream distribution and our parametrized model: $\mathbb{E}_{ p_{\mathrm{data}} } [- \log p(\cP^{(1)},...,\cP^{(n)}; \bw)]$.

We then assume that the joint probability factorizes as follows:

\begin{align}
\log p(\cP^{(1)},...,\cP^{(n)}; \bw) & = \sum_t \log {p(\cP^{(t)} | \cP^{(t-1)}; \bw)} \\
& = \sum_t \{ \log p(\cX^{(t)} | \cP^{(t-1)}; \bw) + \log p(\cR^{(t)} | \cX^{(t)}, \cP^{(t-1)}; \bw) \}
\end{align}

We make a 1st-order Markov assumption: a given octree $\cP^{(t)}$ only depends on the sweep preceding it, $\cP^{(t-1)}$.
We then factor the octree into two entropy models: the node occupancy model $p(\cX^{(t)} | \cP^{(t-1)}; \bw)$, and the intensity model $p(\cR^{(t)} | \cX^{(t)}, \cP^{(t-1)}; \bw)$ conditioned on occupancies. The dependence only on past sweeps makes the model applicable to an online LiDAR stream setting.

\subsection{Occupancy Entropy Model} \label{sec:occ_entropy_model}
We obtain our node occupancy model by continuing to factorize the occupancy probabilities:

\begin{align}
&p(\cX^{(t)} | \cP^{(t-1)}; \bw) =  \prod_{i}{ p(\bx^{(t)}_i | \cX^{(t)}_{\textrm{ans}(i)}, \cP^{(t-1)}; \bw ) }
\end{align}

Here, $\cX^{(t)}_{\textrm{ans}(i)} = \{ \bx^{(t)}_{\textrm{pa}(i)}, \bx^{(t)}_{\textrm{pa}(\textrm{pa}(i))}, ..., \bx^{(t)}_{\textrm{pa}(...(\textrm{pa}(i)))} \}$ represents the set of ancestor nodes of $\bx^{(t)}_i$ and $\cP^{(t-1)}$ represents the point cloud from previous sweep.  As seen above, we simplify the autoregressive dependency on ancestors nodes on the octree for the given timestamp, as well as all the nodes at the previous timestamp.
We model $p(\cdot | \cX^{(t)}_{\textrm{ans}(i)}, \cP^{(t-1)}; \bw )$ with a deep neural network. The architecture has two backbones, namely the \emph{ancestral node dependence} module which encodes recurrent dependencies on the ancestor nodes $\cX^{(t)}_{\textrm{ans}(i)}$ from the current sweep's octree as well as a \emph{prior octree dependence} module which models information passed from the previous sweep. Fig.~\ref{fig:main_layout} depicts the architecture of such network.

\paragraph{Ancestral Node Dependence:}
Our ancestral node dependence module is a recurrent network defined over an ancestral, \textit{top-down} octree path.
Inspired by \cite{huang2020}, we feed a context feature $\bc_i$ for every node $\bx_i$ through a multi-layer perceptron (MLP) to extract an initial hidden embedding $\bh^{(t)}_{i, 0} = \sigma_0(\bc_i; \bw)$, where $\sigma_0(\cdot ; \bw)$ denotes a MLP with learnable parameter $\bw$.
Context features include the current octree level, octant spatial location, and parent occupancy; they are known beforehand per node $\bx_i$  and computed to facilitate representation learning.
We then perform $K_{\text{ans}}$ rounds of aggregation between every node's embedding and its parental embedding:
$\bh^{(t)}_{i, k} = \sigma_k([\bh^{(t)}_{i, k-1}, \bh^{(t)}_{\textrm{pa}(i), k-1}]; \bw)$.
As shorthand, we denote this entire tree-structured recurrent backbone branch as
$\bh^{(t)}_{i} = f_\textrm{ans}(\bx^{(t)}_i, \cX^{(t)}_{\textrm{ans}(i)})$.

\paragraph{Temporal Octree Dependence:}
We also incorporate the previous octree $\cP^{(t-1)}$ into the current entropy model at time $t$ through a \textit{temporal octree dependence} module.
We thus first align the previous octree into the sensor coordinate frame of the current octree.
Unlike the current octree where we only have access to parental information,  we can construct features that make use of \textit{all} information within the previous octree, containing both top-down ancestral information as well as \textit{bottom-up} child information.
We exploit this fact by designing a two-stream feature backbone to compute embeddings for every octree node at time $t-1$, inspired by tree-structured message passing algorithms \cite{scarselli_gnn, yedidia_gbp}.
The forward pass stream is the same as the ancestral dependence module above, generating top-down features from ancestors: $\bh^{(t-1)}_{j} = f_\textrm{ans}(\bx^{(t-1)}_j, \cX^{(t-1)}_{\textrm{ans}(j)})$.
After the top-down pass, we design a bottom-up aggregation pass, a recurrent network that produces aggregated features from descendants to the current node.
Unlike the ancestral module in which each node only has one parent, the number of children per node can vary, and we desire that the output is invariant to the ordering of the children. Hence, we resolve this by designing the following function inspired by deep sets \cite{zaheer_deepsets}: $\bg^{(t-1)}_{j} = f_{\textrm{agg}, 1}(\bh^{(t-1)}_{j} + \sum_{c \in \textrm{child}(j)} f_{\textrm{agg},2} (\bg^{(t-1)}_{c} ))$, which produces the final aggregated embedding feature containing both top-down and bottom-up context.

\paragraph{Spatio-Temporal Aggregation:}
The final step incorporates the set of aggregated features in the previous octree $\{ \bg_{j}^{(t-1)} \}$, with ancestral features in the current octree $\{ \bh_{i}^{(t)} \}$ to help with occupancy prediction in the current octree.
A key observation is that only a subset of spatially proximal nodes in the previous sweep can contribute to better prediction for a given node at time $t$; moreover, the relative location of each neighbor should define its relative importance. Inspired by this fact,  we employ \textit{continuous convolutions} \cite{wang2018} to process previous octree features at the current node.
A continuous conv. layer aggregates features from neighboring points to a given node in the following manner:
$
\bh_{i}= \sum_{j \in \mathcal{N}(i)} \sigma(\bp_j - \bp_i ) \bh_{j}
$
where $\mathcal{N}(i)$ is the $i$-th node's $k$-nearest neighboring nodes in 3D space from the $(t-1)$ sweep at the same level as $i$,  $\bp_i$ is the 3D position of each node, and $ \sigma $ denotes a learned MLP.  We use a separate MLP  with a continuous conv. layer per octree level to process the aggregated features in the previous octree  $\{ \bg_{j}^{(t-1)} \}_{j \in \mathcal{N}(i)}$ and produce an embedding feature $\bg_{i, \textrm{st}}^{(t)}$.

\paragraph{Entropy Header:}
Finally, the warped feature $\bg_{i, \textrm{st}}^{(t)}$ and ancestral features $\bh_{i}^{(t)}$ are aggregated through a final MLP to output a 256-dimensional softmax of probability values $p(\bx^{(t)}_i | \cX^{(t)}_{\textrm{ans}(i)}, \cP^{(t-1)}; \bw )$, corresponding to the predicted 8-bit occupancy for node $i$, time $t$.

\subsection{Intensity Entropy Model} \label{sec:att_entropy_model}

The goal of the intensity entropy model is to compress extraneous intensities tied to each spatial point coordinate. We assume these intensities are bounded and discrete, so compression is lossless; if they are continuous, there will be a loss incurred through discretization. The model factorizes as follows:

\begin{flalign}
p(\cR^{(t)} | \cX^{(t)}, \cP^{(t-1)}; \bw) =   \prod_{i}{ p(\bbr^{(t)}_i | \cX^{(t)}, \cP^{(t-1)}; \bw) }
\end{flalign}

The intent of conditioning on the occupancies $\cX^{(t)}$ is not to directly use their values \textit{per se}, but to emphasize that intensity decoding occurs \textit{after} the point spatial coordinates have already been reconstructed in $\mathbb{R}^3$.
Therefore, we can directly make use of the spatial position corresponding to each intensity $\cR^{(t)}_i$ in compression.
We aim to leverage temporal correlations between point intensities across consecutive timestamps to better model the entropy of $\bbr^{(t)}_i$. Similar to node occupancy prediction above, there is the challenge of how to incorporate previous intensity information when there are no direct correspondences between the two point clouds. We again employ continuous convolutions to resolve this challenge. Let $\cR_{\mathcal{N}(i)}$ be the set of nearest neighbor intensities $\{ \bbr^{(t-1)}_j\}_{j \in \mathcal{N}(i)}$, where nearest neighbor is defined by spatial proximity of previous point $j$ to the current point $i$.
We apply an MLP with a continuous conv. layer that takes the past intensities $\bbr^{(t-1)}_j$ as input and outputs an embedding feature for each node $i$.
This feature is then fed through a linear layer and softmax to output intensity probability values. In our setting we assume our intensity value is an 8-bit integer, so the resulting probability vector is 256-dimensional $p(\bbr^{(t)}_i  | \cX^{(t)}, \cP^{(t-1)}; \bw)$.


\begin{figure}[!t]
\centering
\includegraphics[height=0.25\textwidth]{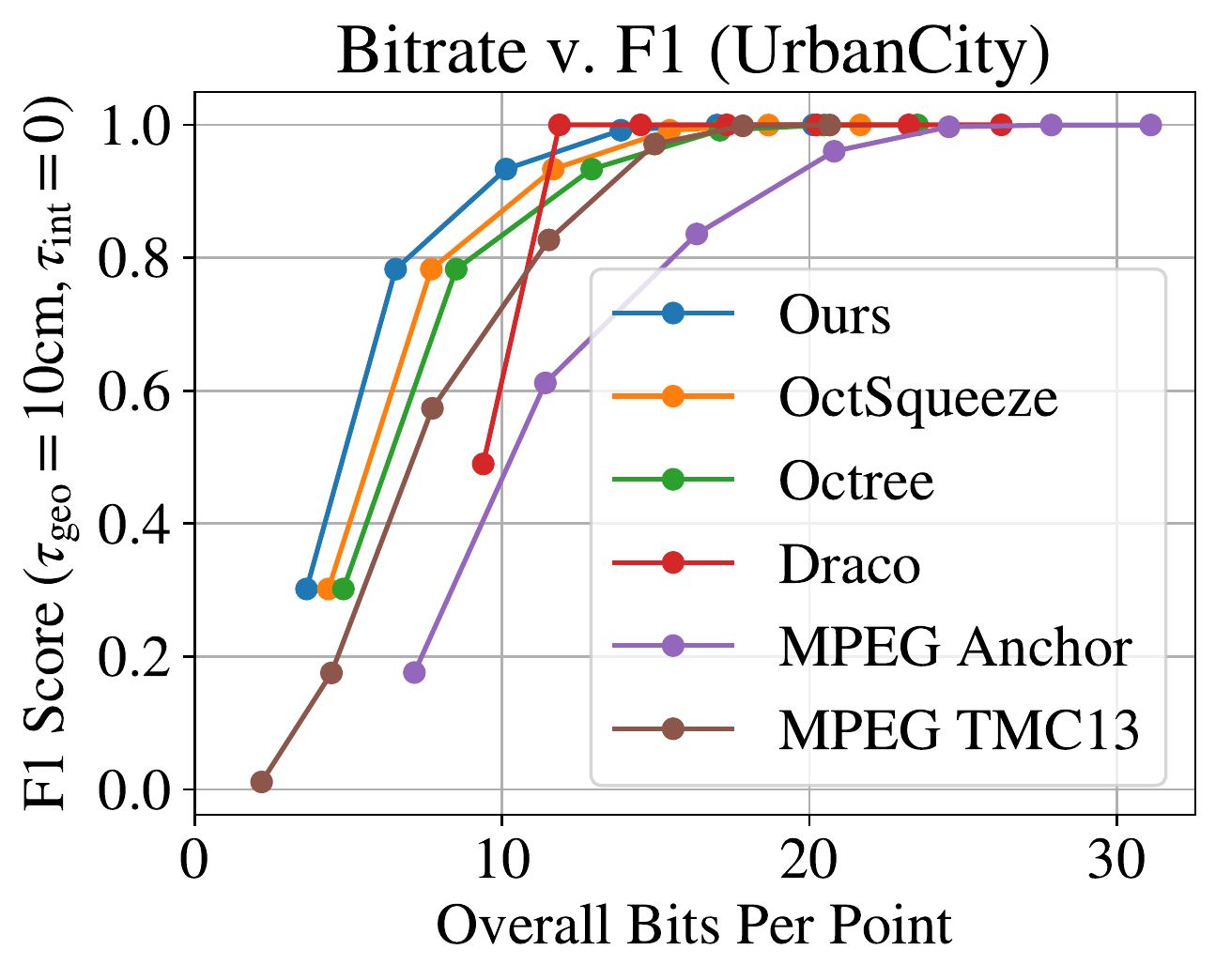}
\includegraphics[height=0.25\textwidth]{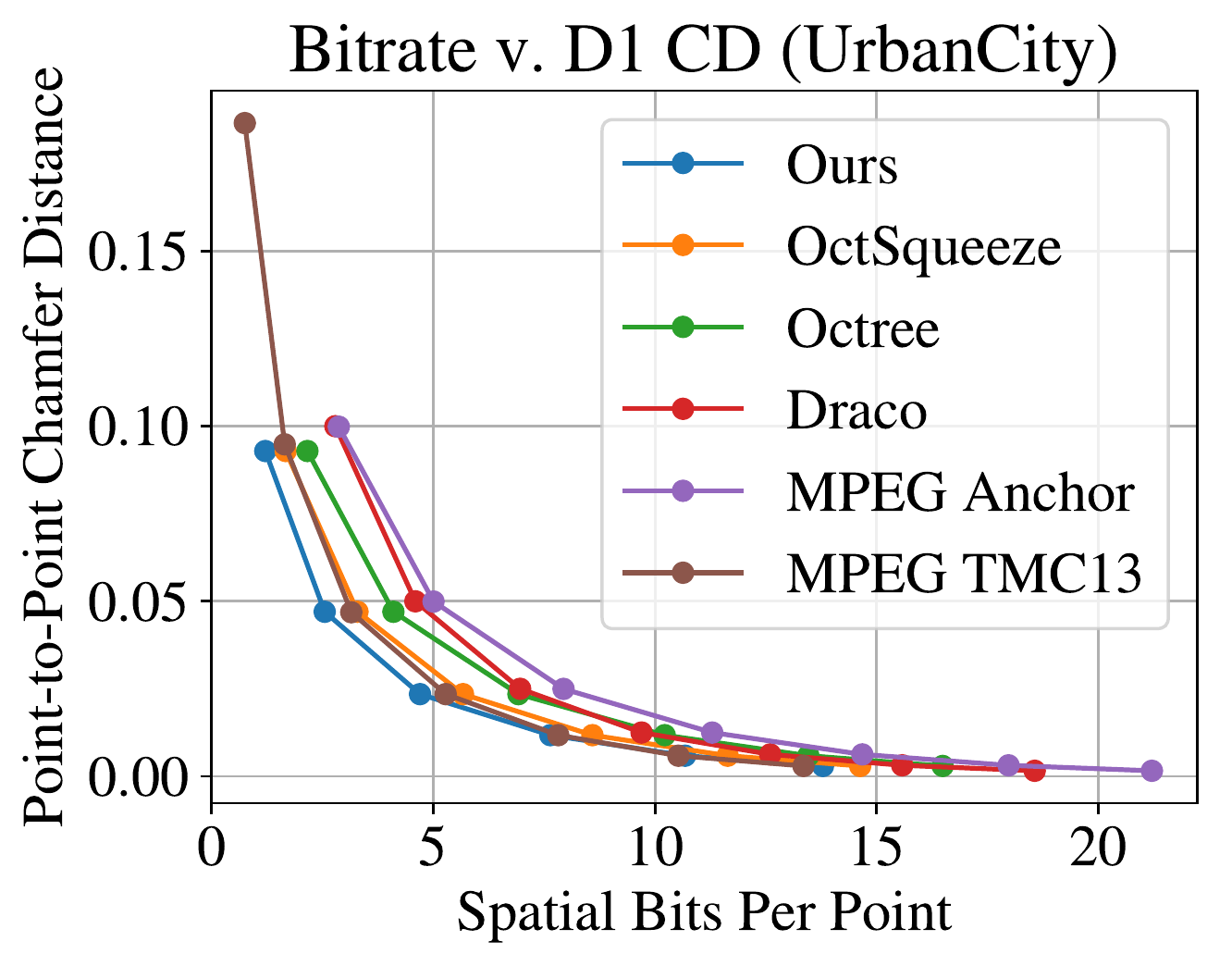}
\includegraphics[height=0.25\textwidth]{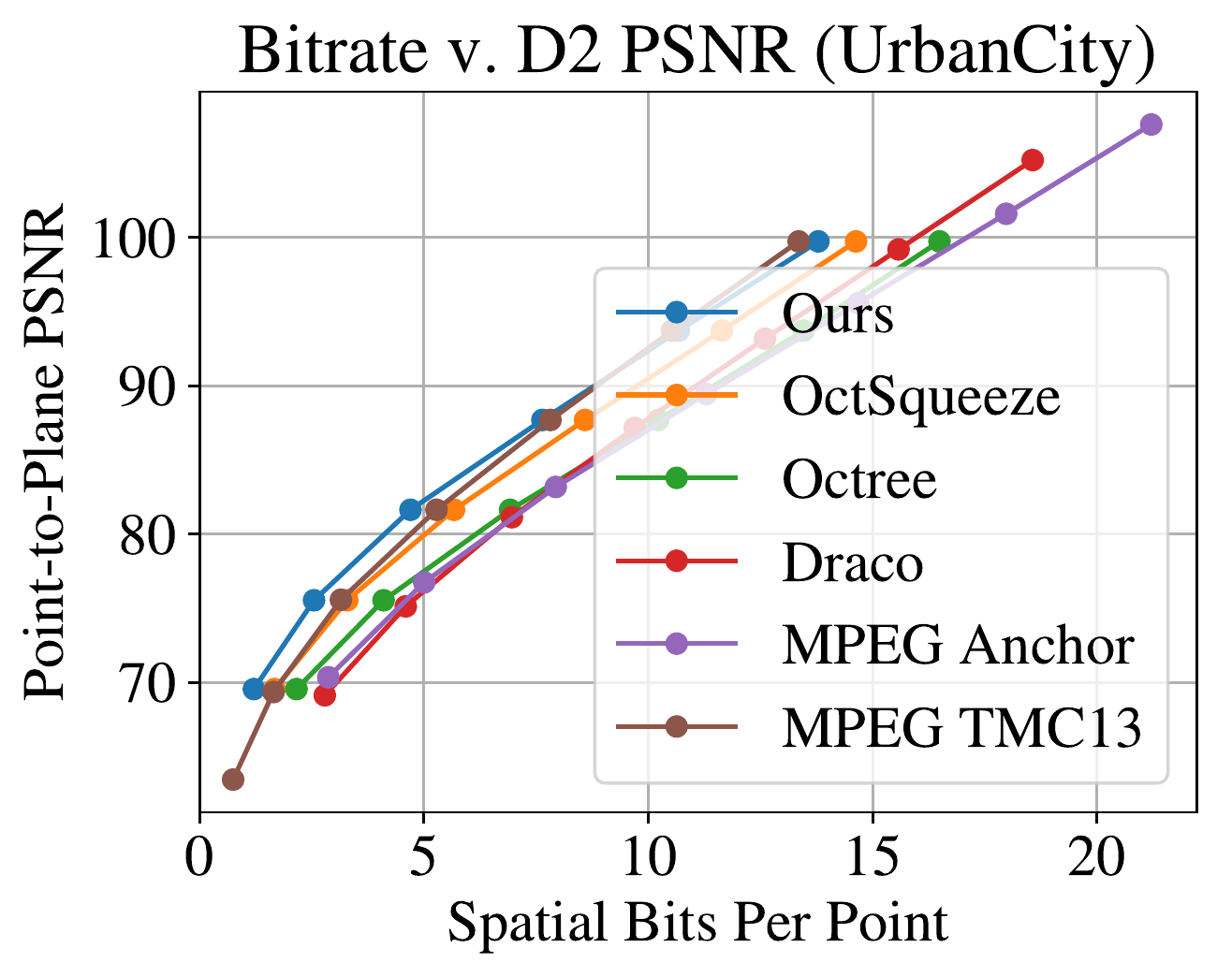}

\includegraphics[height=0.25\textwidth]{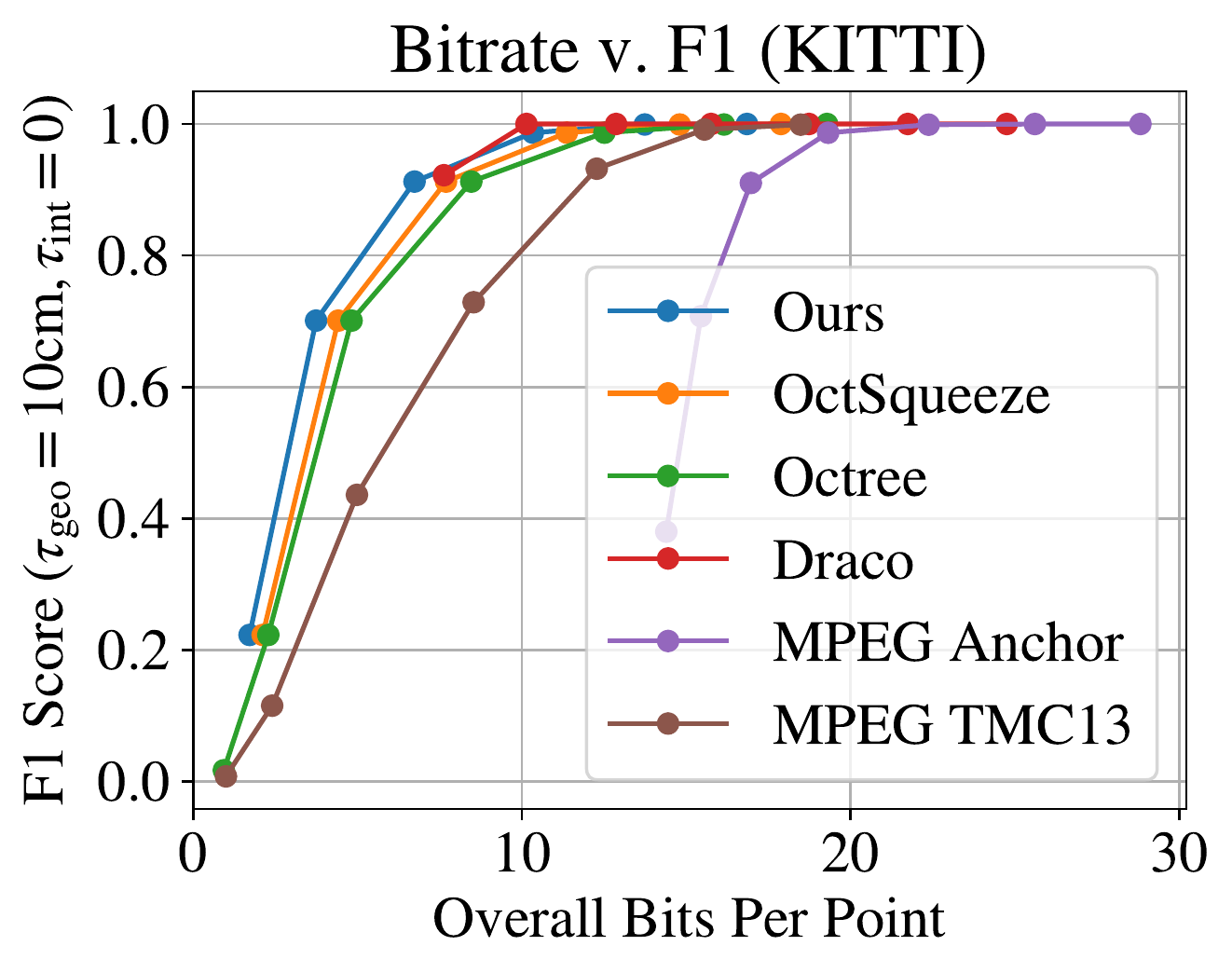}
\includegraphics[height=0.25\textwidth]{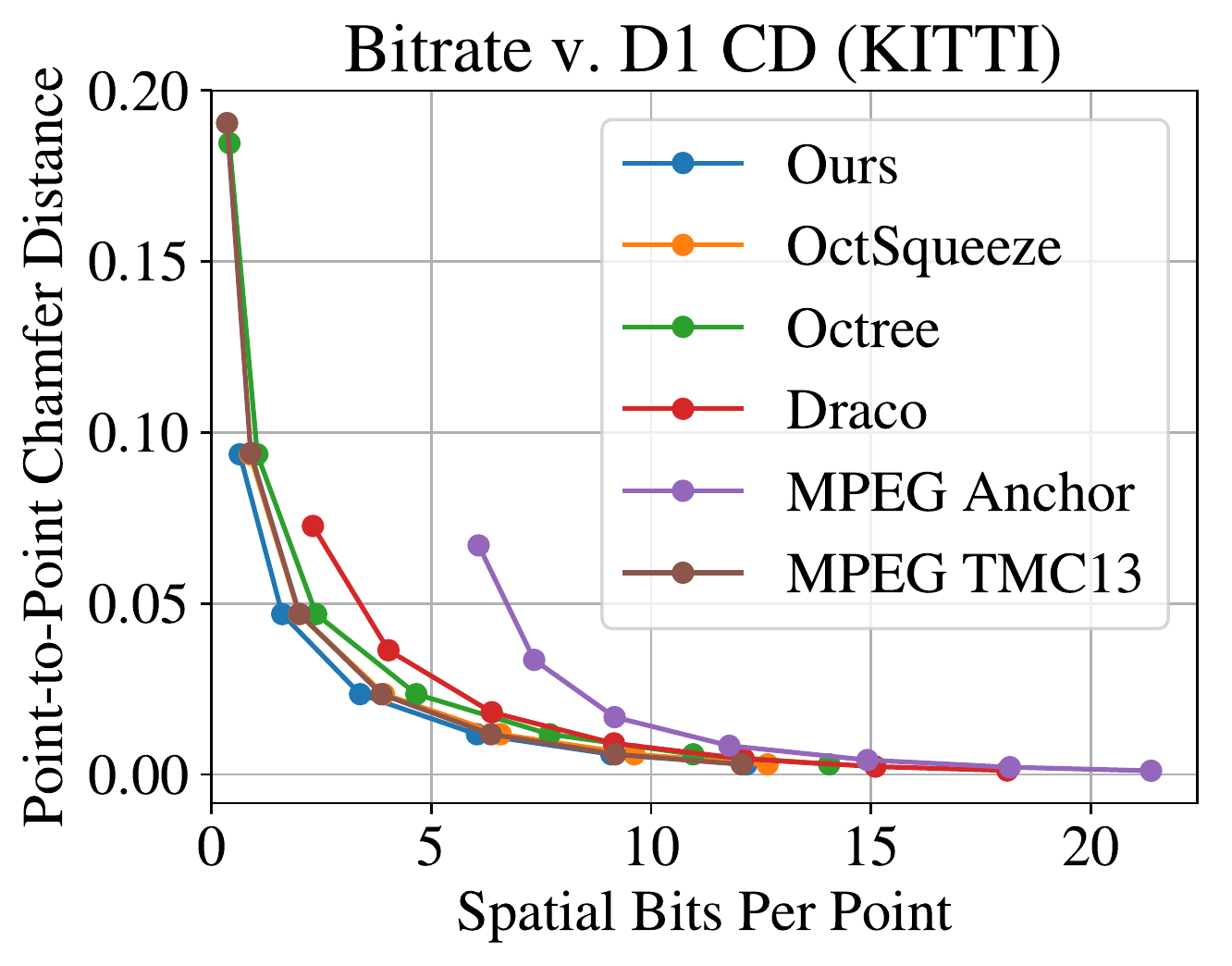}
\includegraphics[height=0.25\textwidth]{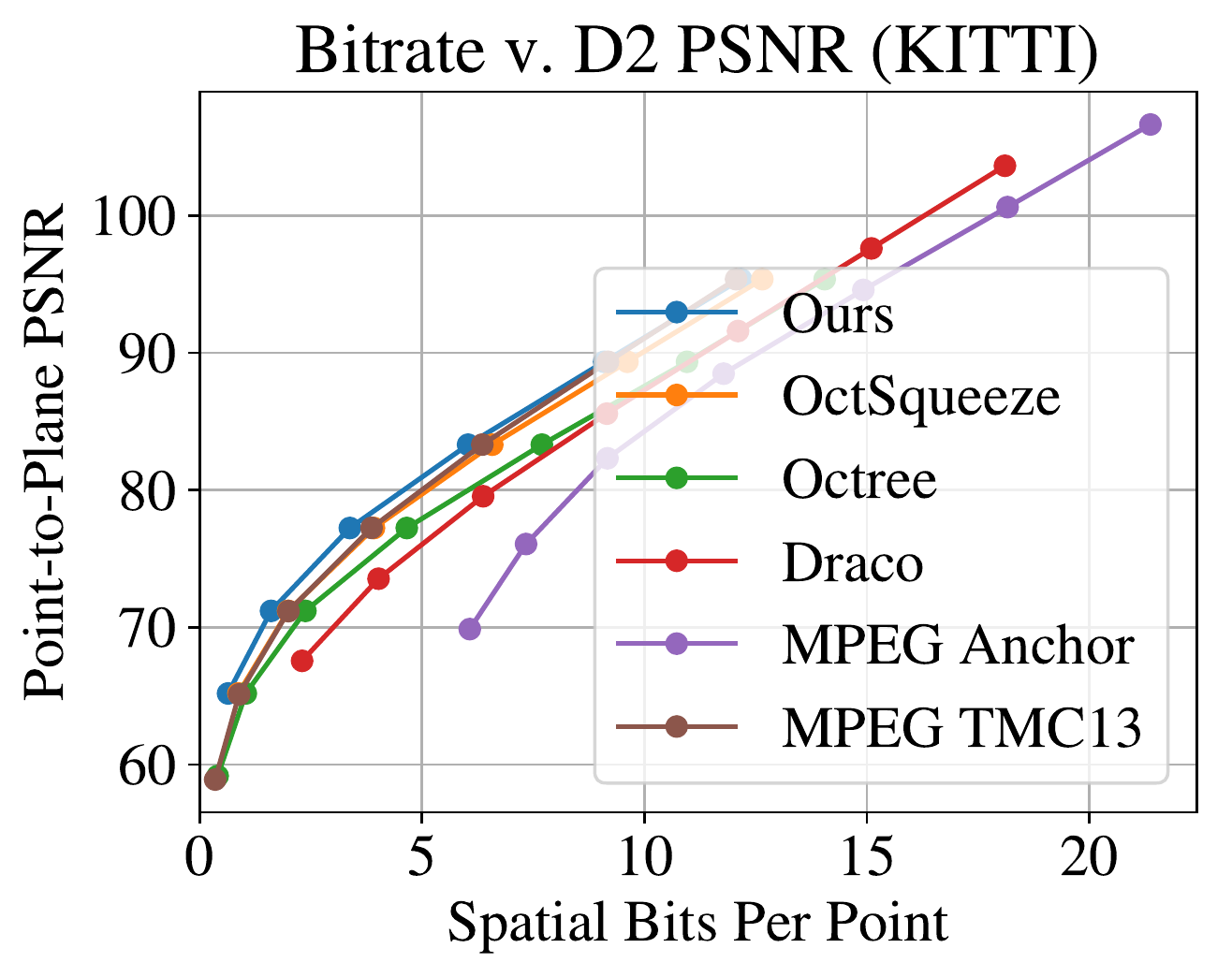}
\caption{
    \small
     Bitrate \vs reconstruction quality curves on \northamerica~(top) and KITTI (bottom).
    From left-to-right:
    $ \mathrm{F}_1 $ with $ \tau_\mathrm{geo} = 10\mathrm{cm} $ and $ \tau_\mathrm{int} = 0 $ ($ \uparrow $),
    point-to-point chamfer distance ($ \downarrow $),
    point-to-plane PSNR ($ \uparrow $).
}
\label{figure:quantitative-results}
\vspace{-10pt}
\end{figure}

\subsection{Entropy Coding Process} \label{sec:entropy_coding}
\paragraph{Encoding:}  We integrate our entropy model with an entropy coding algorithm (range coding \cite{martin_rangecoding}) to produced the final compressed bitstream. During the encoding pass, for every octree in the stream, the entropy model is applied across the octree occupancy bytestream, as well as across the intensities per point, to predict the respective probability distributions. We note that encoding only requires one batch GPU pass per sweep for the occupancy and intensity models. The resulting distributions are then passed to range coding which compresses the occupancies and intensities into two bitstreams.

\paragraph{Decoding:} The same entropy models are used during decoding. First, the occupancy entropy model is run, given the already decoded past octree, to produce distributions that recover the occupancy serialization and spatial coordinates of the current point cloud. Then, the intensity entropy model is run, given the already decoded intensities in the past point cloud, to produce distributions that recover the current point intensities.
Note that our model is well-setup for parallel computation during decoding, for both the occupancies and intensities. As mentioned in Sec. \ref{sec:occ_entropy_model}, the dependence on ancestral nodes instead of all past nodes allows us to only run at most $O(D)$ GPU passes for the occupancy model per sweep. Moreover, the assumed independence between intensities in the current sweep, given the past, allows us to only run 1 GPU pass per sweep for the intensity entropy model.

\subsection{Learning}

Both our occupancy and intensity entropy models are trained end-to-end with cross-entropy loss, for every node $\bx^{(t)}_i \in \cX^{(t)}_i$ and intensity $\bbr^{(t)}_i \in \cR^{(t)}_i$, for every point cloud in a stream:

\begin{flalign}
\ell = \mathbb{E}_{\bm{P} \sim p_{\mathrm{data}} } \left[ - \sum_{t} {\sum_{i}{  \log p(\bx^{(t)}_{i, gt} | \cX^{(t)}_{\textrm{ans}(i)}, \cP^{(t-1)}; \bw) } } - \sum_{t} {\sum_{i}{  \log p(\bbr^{(t)}_{i, gt} | \cX^{(t)}, \cP^{(t-1)}; \bw ) } } \right]
\end{flalign}

Here, $\bx^{(t)}_{i, gt}$ and $\bbr^{(t)}_{i, gt}$ denote the ground-truth values of the node occupancies/intensities, respectively. As mentioned above, minimizing cross-entropy loss is equivalent to our goal of reducing expected bitrate of the point cloud stream.

\subsection{Discussion and Related Works}
\label{section:method/related-works}

Our approach belongs to a family of point cloud compression algorithms based on
tree data structures~\cite{devillers2000,draco,schnaebel2006,huang2008,
zhang2014,thanou2015,deQueiroz2016,mekuria2016,deQueiroz2016,garcia2017,
shao2017,garcia2018,garcia2020}.
Tree-based algorithms are advantageous since they use spatial-partioning data structures
that can efficiently represent sparse and non-uniformly dense 3D point clouds.
Two notable examples are Google's Draco~\cite{draco}
and the MPEG anchor~\cite{schwarz2019}, which use a KD-tree codec~\cite{devillers2000}
and an octree codec~\cite{mekuria2016} respectively.
To exploit temporal redundancies, the MPEG anchor encodes successive point
clouds as block-based rigid transformations of previous point clouds; this,
however, narrows its usefulness to scenes with limited motion.
Moreover, these prior works use simple entropy models that do not fully
exploit redundant information hidden in LiDAR point clouds; \eg, repetitive
local structures, objects with strong shape priors, \etc.
In contrast, we use a learned entropy model to directly capture these dependencies.

Our approach is also related to work in deep point cloud
compression~\cite{tu2019,tu2019b,quach2019,wang2019,yan2019,huang2019,huang2020}.
In particular, both our approach and the prior state-of-the-art~\cite{huang2020}
use deep entropy models that operate on octree structures directly.
However, they do not model temporal redundancies between successive point clouds
and compress LiDAR geometry only.
In this work, we propose a unified framework that aggregates spatio-temporal
context to jointly compress both LiDAR geometry and intensity.

Finally, our work is inspired by recent successes in deep image compression \cite{balle_varhyperprior, minnen_jointpriors, toderici_fullimgcomp, mentzer_condprobimg, theis_imgcomp_ae, mentzer_lossless, townsend_bitsback} and video compression \cite{wu_vidinterpolation, lu_dvc, rippel_learnedvidcomp, habibian_rdauto, djelouah_interframe, yang_hier_rec, lin_mlvc}, many of which use deep entropy models.


\section{Experimental Evaluation}
\label{section:experiments}


\begin{figure}[!t]
\centering

\begin{overpic}[clip,trim=0cm 5cm 0cm 5cm,width=0.24\textwidth]{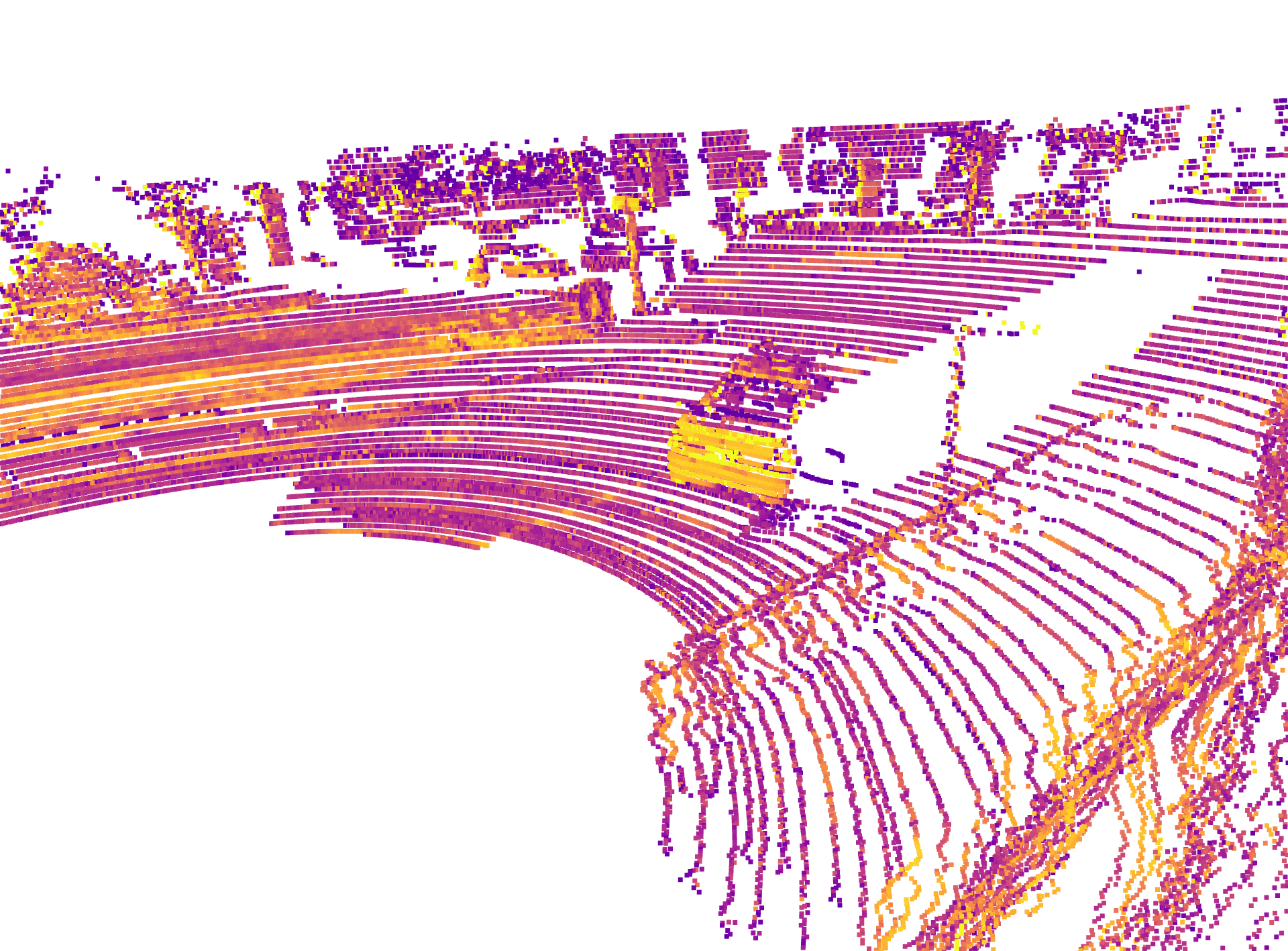}
    \put (0,51) {\colorbox{gray!30}{\scriptsize Oracle (\northamerica): Bitrate 104}}
\end{overpic}
\begin{overpic}[clip,trim=0cm 5cm 0cm 5cm,width=0.24\textwidth]{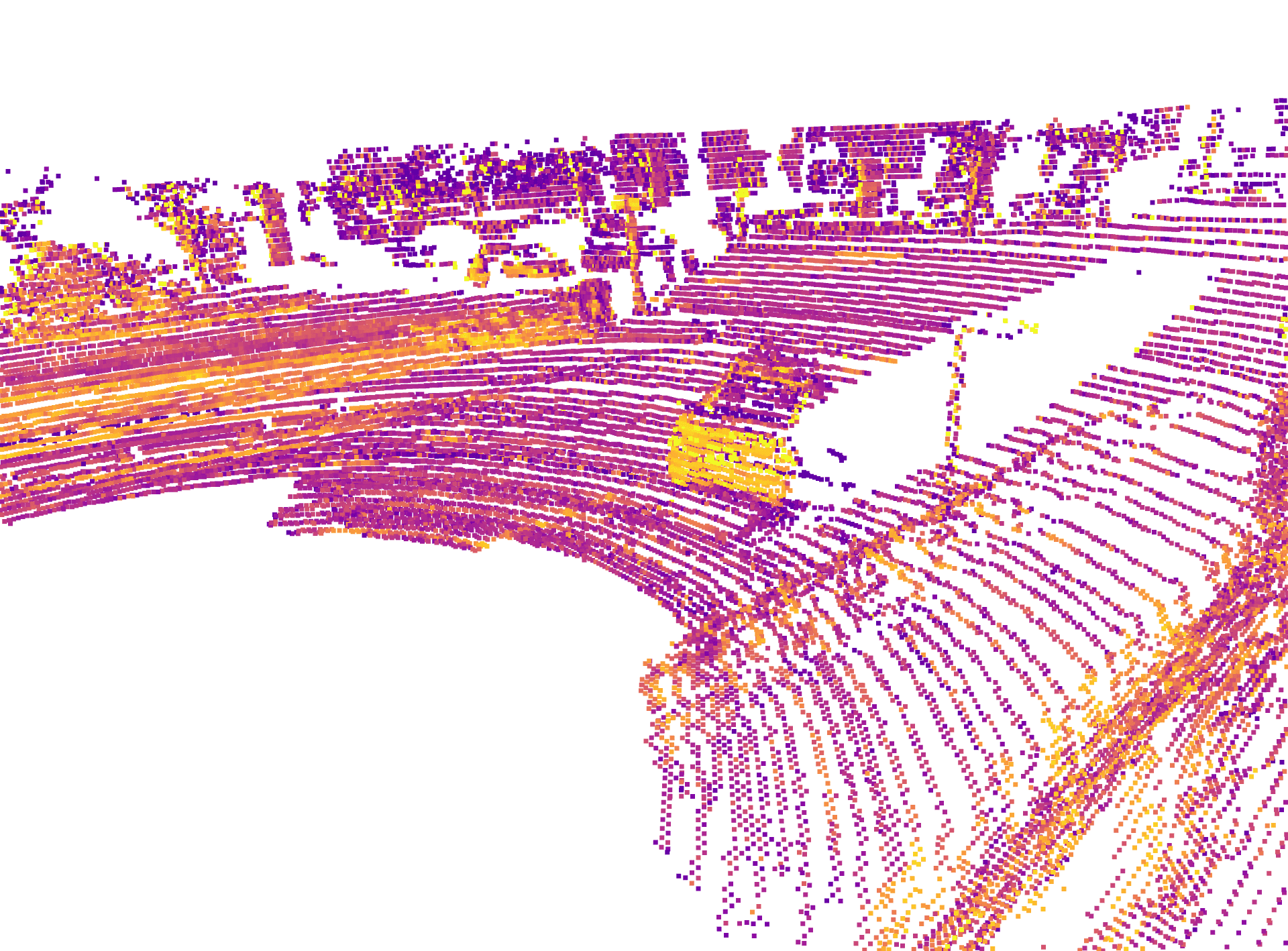}
    \put (0,51) {\colorbox{gray!30}{\scriptsize Ours: F1 92.4, Bitrate 9.3}}
\end{overpic}
\begin{overpic}[clip,trim=0cm 5cm 0cm 5cm,width=0.24\textwidth]{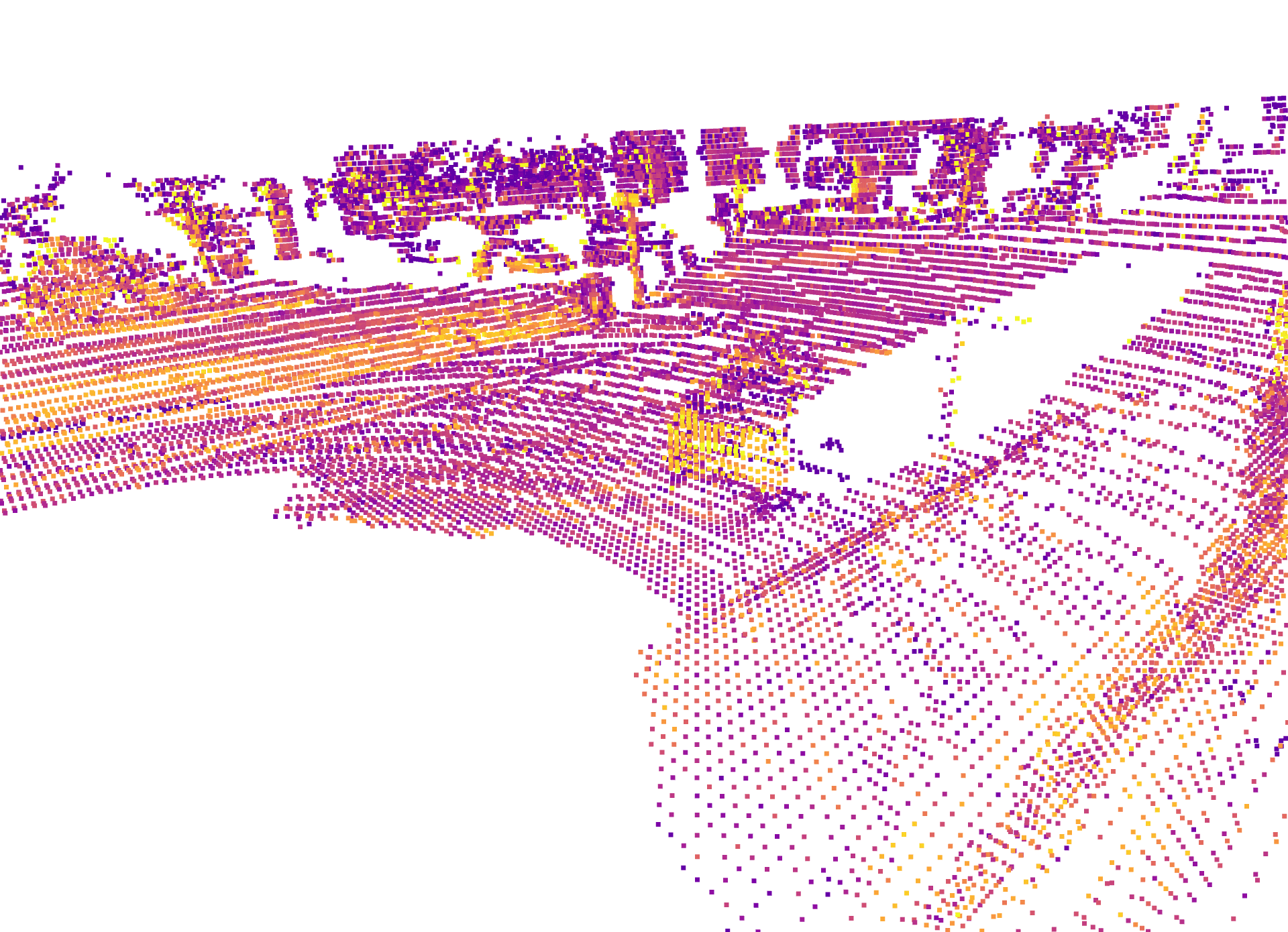}
    \put (0,51) {\colorbox{gray!30}{\scriptsize Draco: F1 80.8, Bitrate 9.4}}
\end{overpic}
\begin{overpic}[clip,trim=0cm 5cm 0cm 5cm,width=0.24\textwidth]{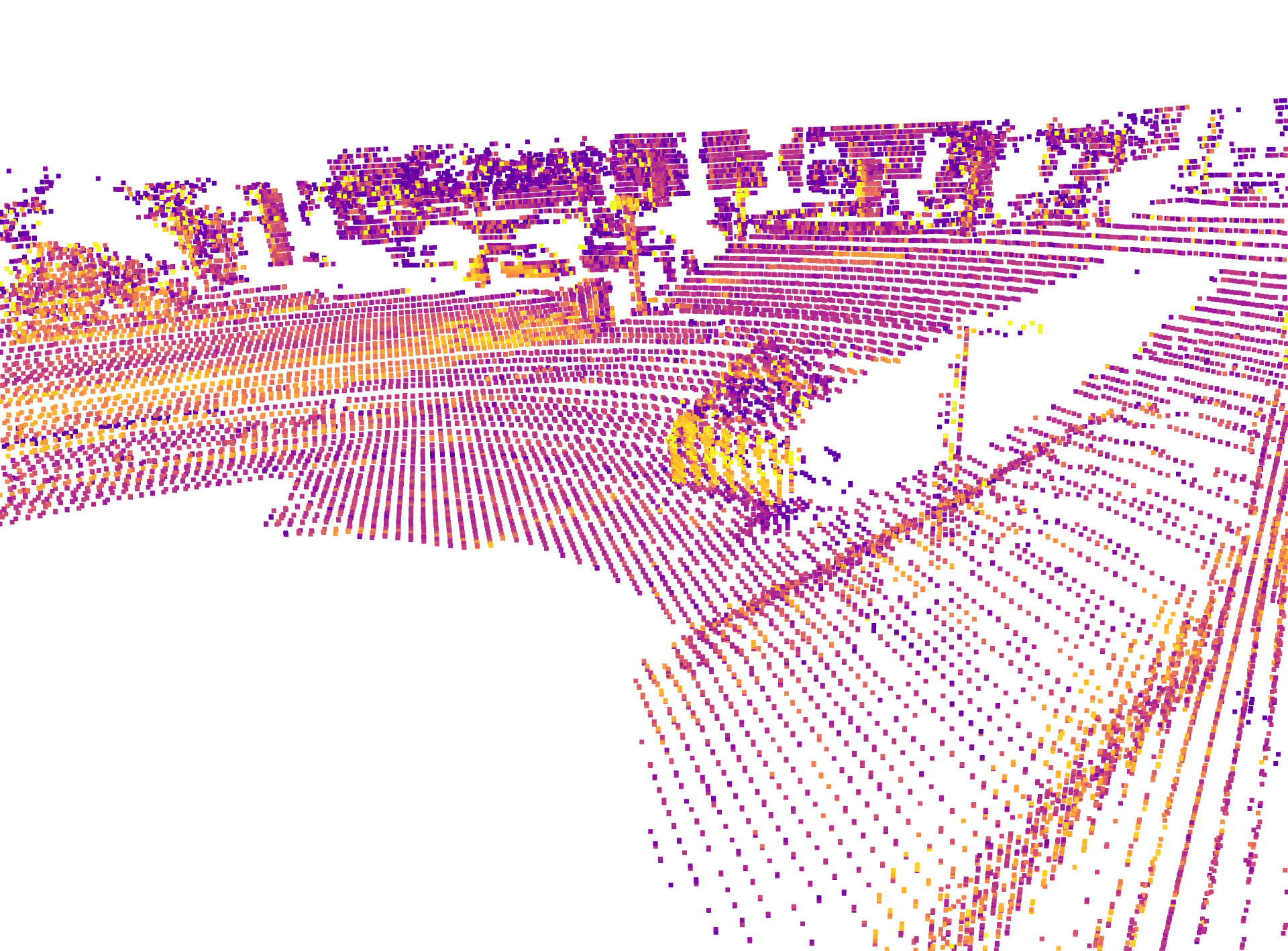}
    \put (0,51) {\colorbox{gray!30}{\scriptsize MPEG: F1 59.1, Bitrate 11.4}}
\end{overpic}

\begin{overpic}[clip,trim=0cm 10cm 0cm 0cm,width=0.24\textwidth]{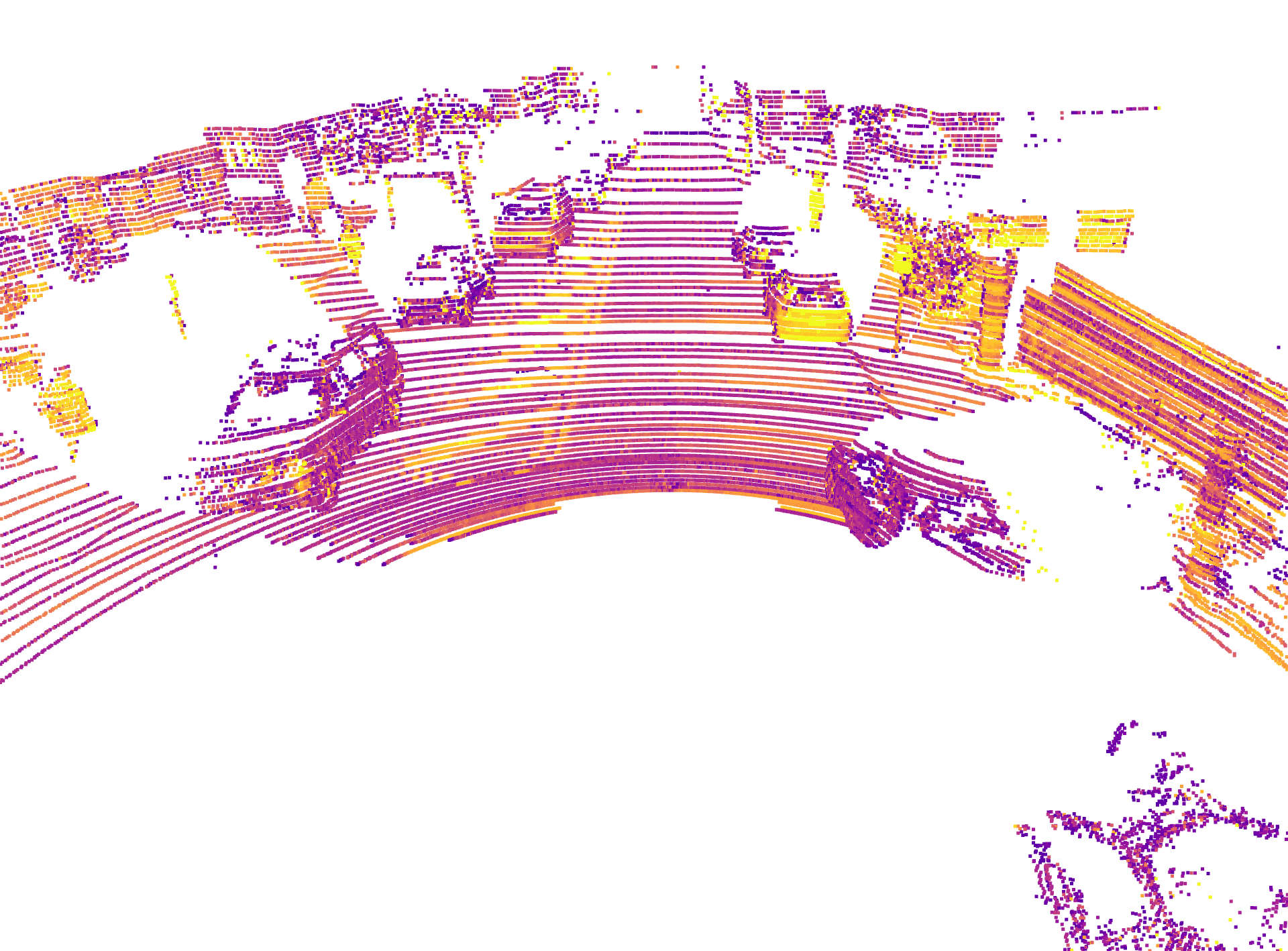}
    \put (0,51) {\colorbox{gray!30}{\scriptsize Oracle (\northamerica): Bitrate 104}}
\end{overpic}
\begin{overpic}[clip,trim=0cm 10cm 0cm 0cm,width=0.24\textwidth]{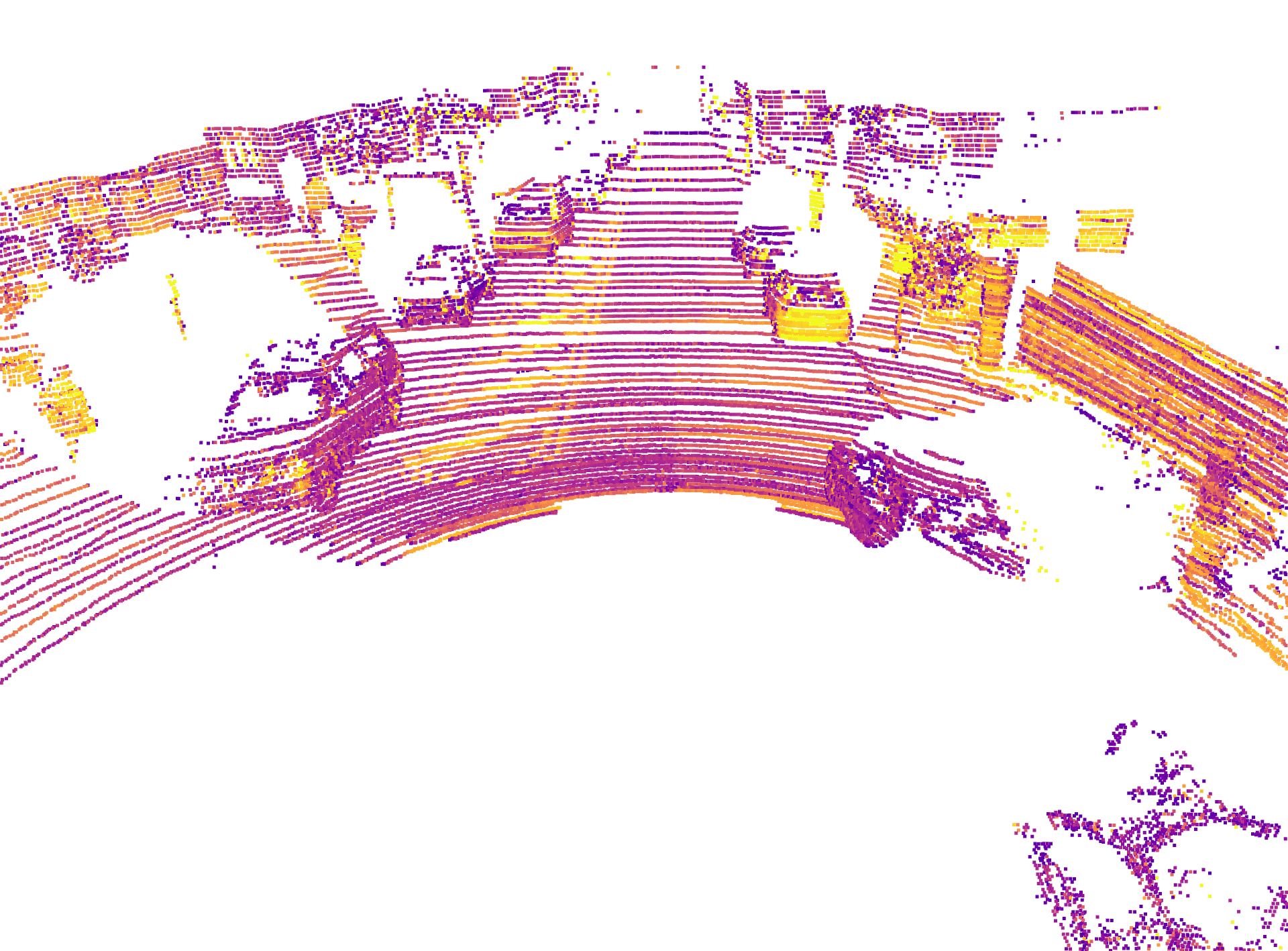}
    \put (0,51) {\colorbox{gray!30}{\scriptsize Ours: F1 99.2, Bitrate 13.7}}
\end{overpic}
\begin{overpic}[clip,trim=0cm 10cm 0cm 0cm,width=0.24\textwidth]{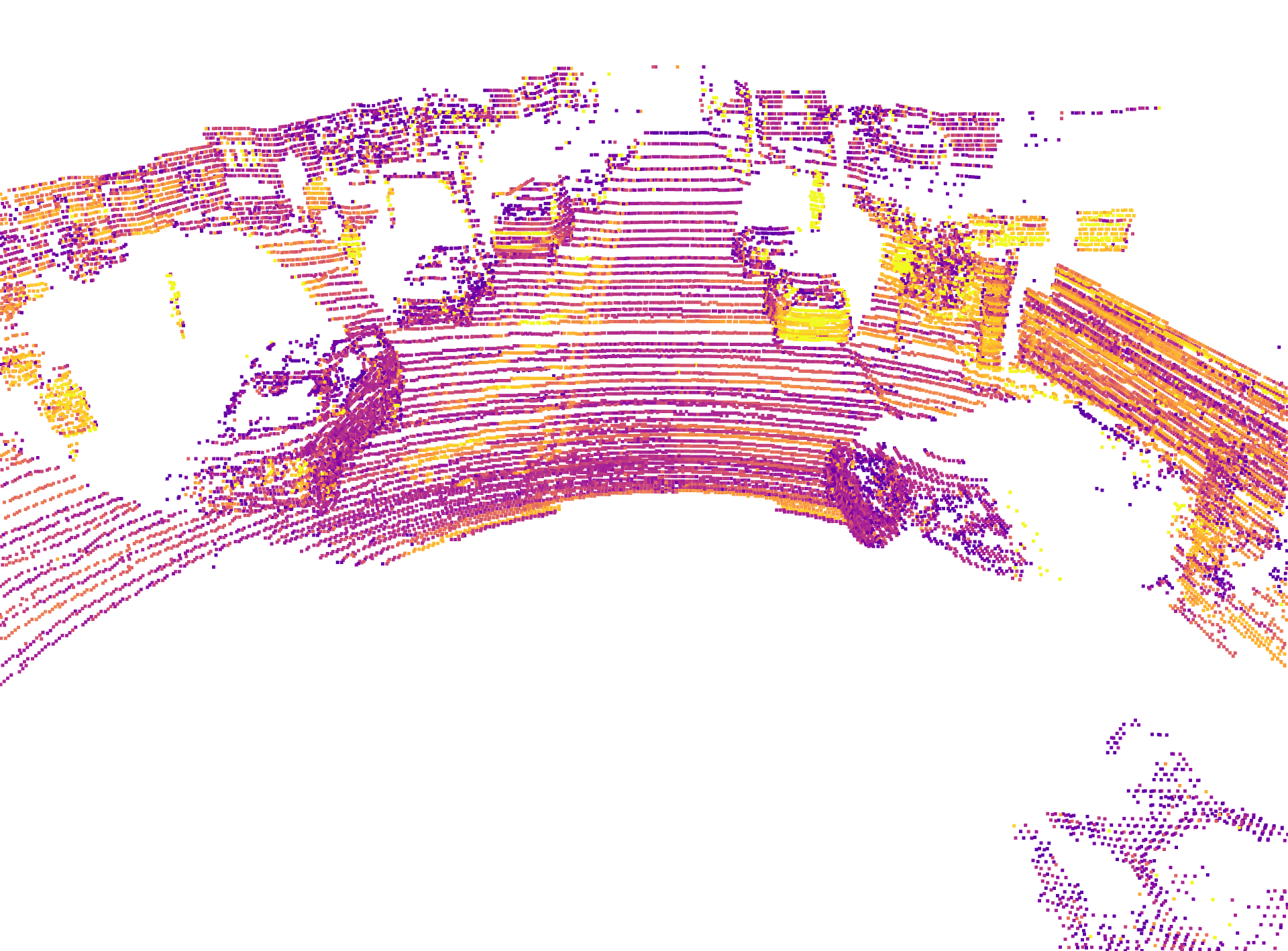}
    \put (0,51) {\colorbox{gray!30}{\scriptsize Draco: F1 92.3, Bitrate 13.8}}
\end{overpic}
\begin{overpic}[clip,trim=0cm 10cm 0cm 0cm,width=0.24\textwidth]{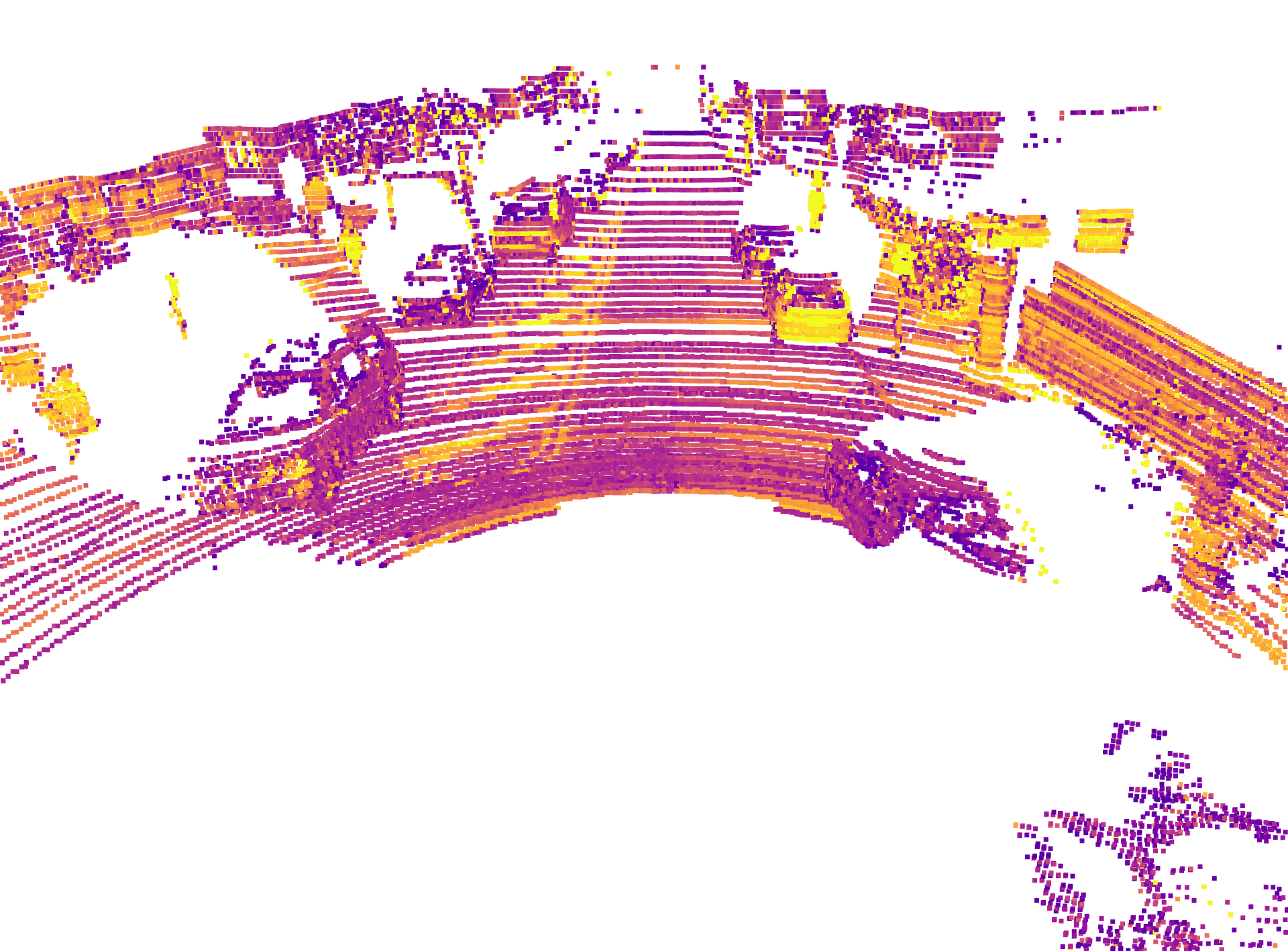}
    \put (0,51) {\colorbox{gray!30}{\scriptsize MPEG: F1 81.5, Bitrate 16.2}}
\end{overpic}

\begin{overpic}[clip,trim=0cm 10cm 0cm 0cm,width=0.24\textwidth]{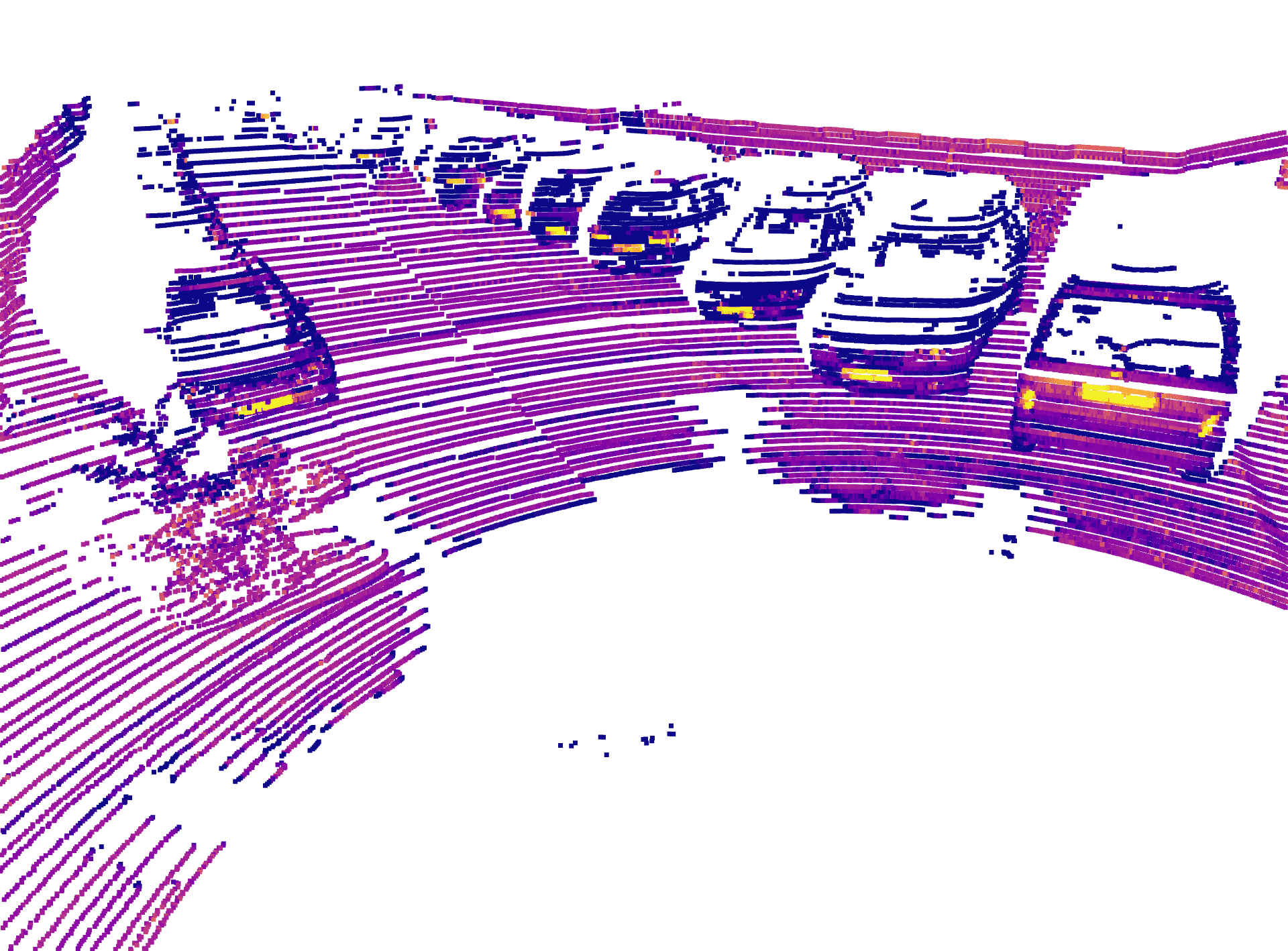}
    \put (0,50) {\colorbox{gray!30}{\scriptsize Oracle (KITTI): Bitrate 104}}
\end{overpic}
\begin{overpic}[clip,trim=0cm 10cm 0cm 0cm,width=0.24\textwidth]{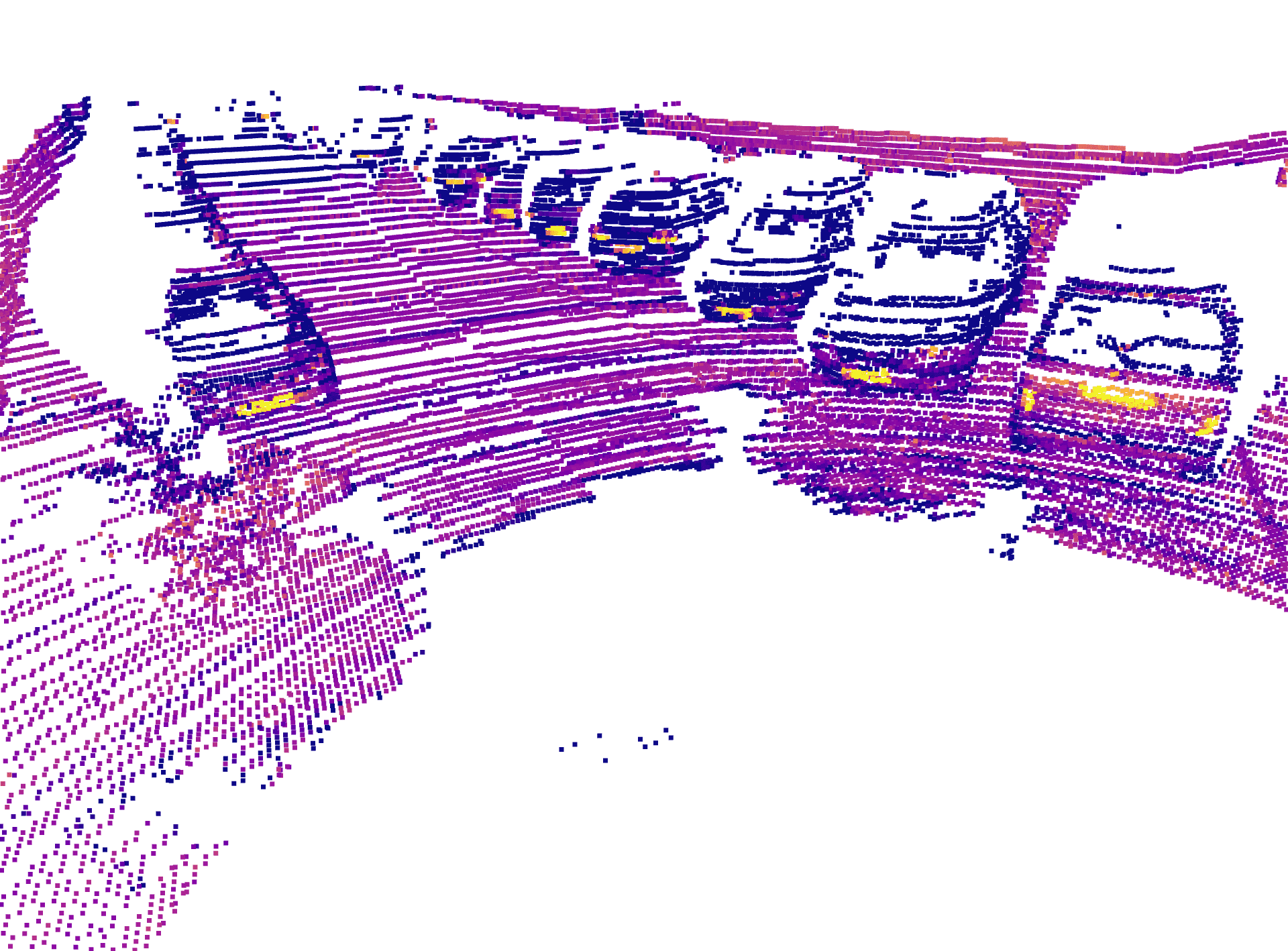}
    \put (0,50) {\colorbox{gray!30}{\scriptsize Ours: F1 90.2, Bitrate 5.6}}
\end{overpic}
\begin{overpic}[clip,trim=0cm 10cm 0cm 0cm,width=0.24\textwidth]{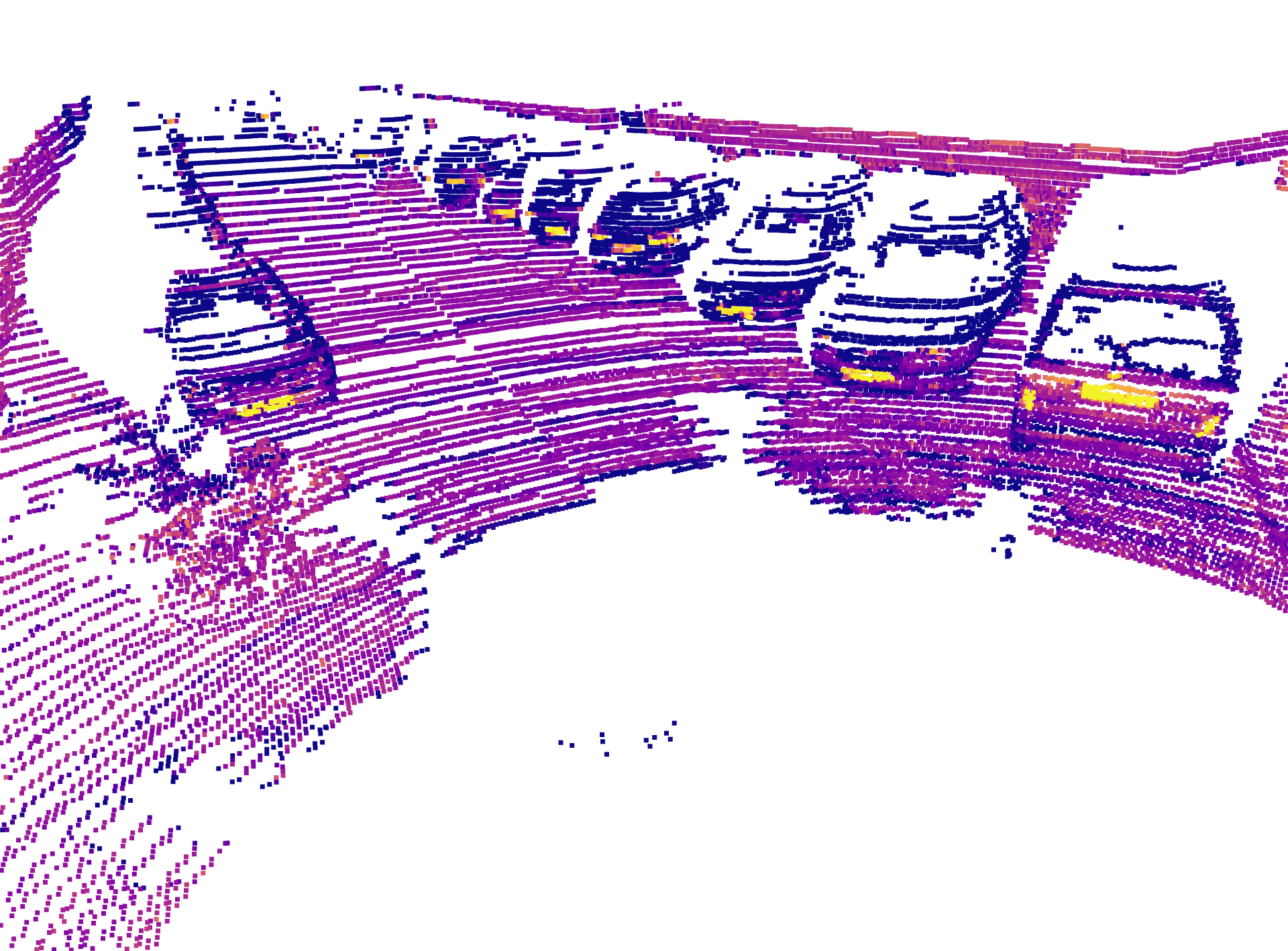}
    \put (0,50) {\colorbox{gray!30}{\scriptsize Draco: F1 87.1, Bitrate 5.7}}
\end{overpic}
\begin{overpic}[clip,trim=0cm 10cm 0cm 0cm,width=0.24\textwidth]{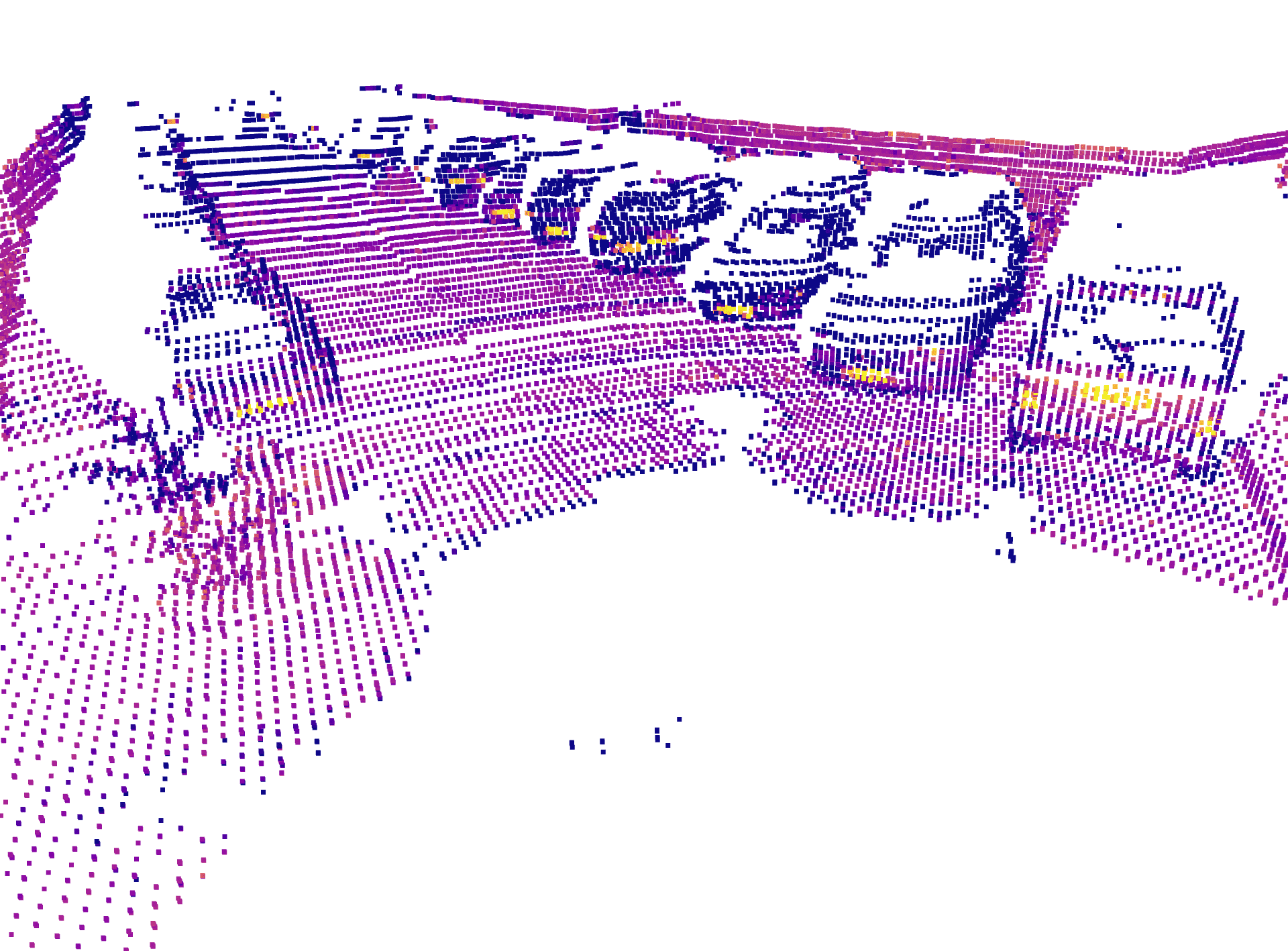}
    \put (0,50) {\colorbox{gray!30}{\scriptsize MPEG: F1 61.0, Bitrate 9.5}}
\end{overpic}

\begin{overpic}[clip,trim=0cm 10cm 0cm 0cm,width=0.24\textwidth]{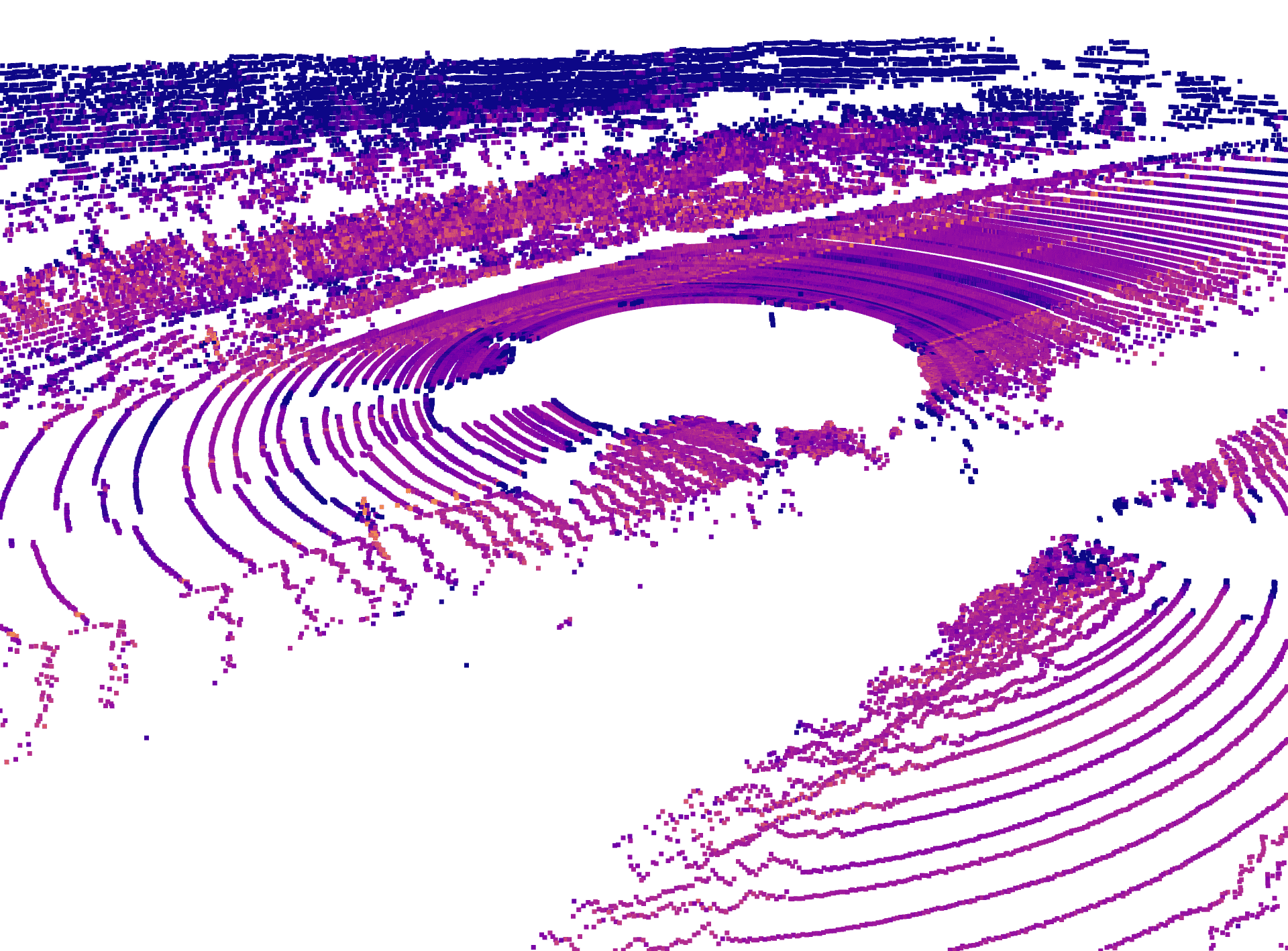}
    \put (0,51) {\colorbox{gray!30}{\scriptsize Oracle (KITTI): Bitrate 104}}
\end{overpic}
\begin{overpic}[clip,trim=0cm 10cm 0cm 0cm,width=0.24\textwidth]{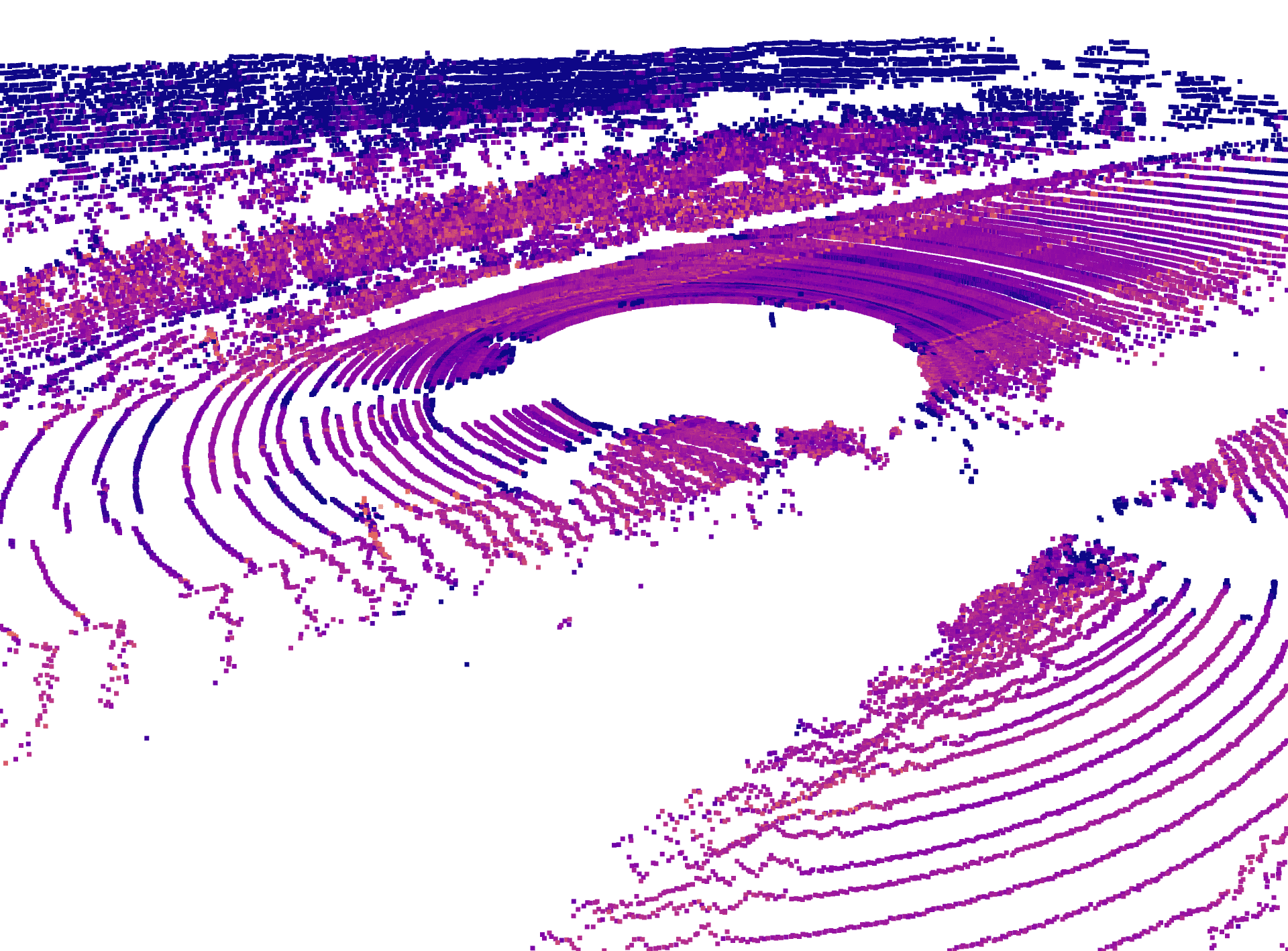}
    \put (0,51) {\colorbox{gray!30}{\scriptsize Ours: F1 98.6, Bitrate 10.1}}
\end{overpic}
\begin{overpic}[clip,trim=0cm 10cm 0cm 0cm,width=0.24\textwidth]{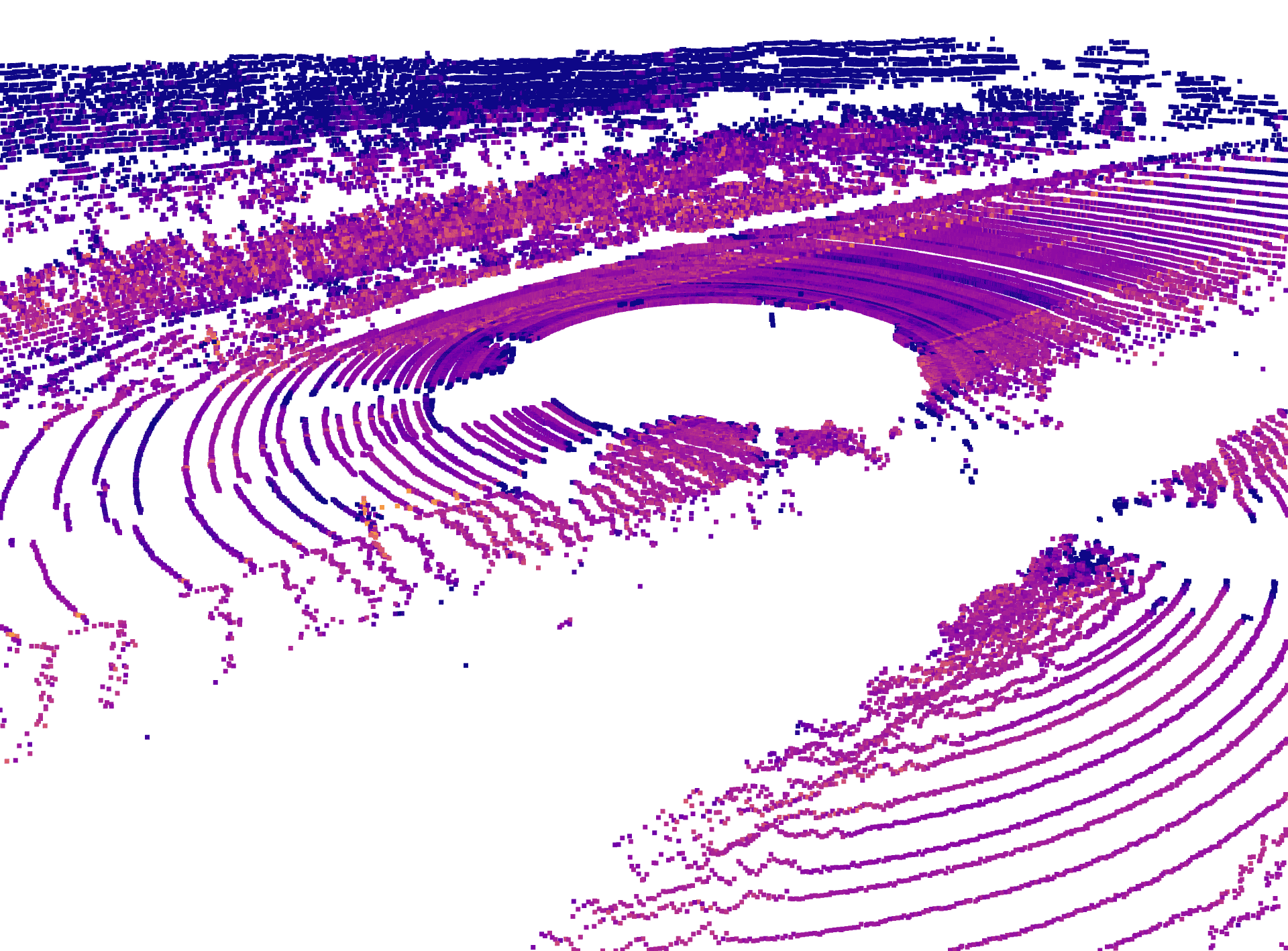}
    \put (0,51) {\colorbox{gray!30}{\scriptsize Draco: F1 96.9, Bitrate 10.1}}
\end{overpic}
\begin{overpic}[clip,trim=0cm 10cm 0cm 0cm,width=0.24\textwidth]{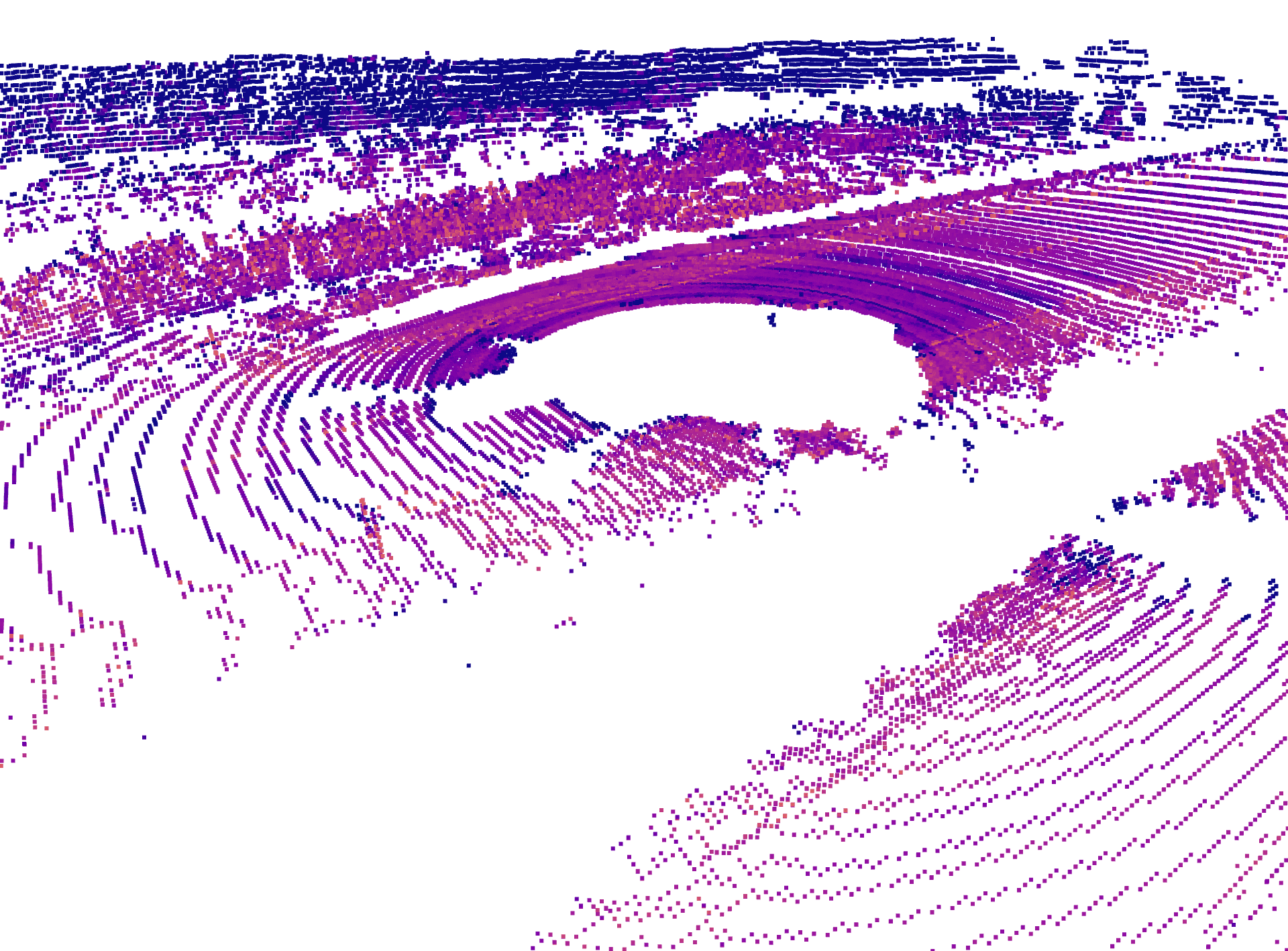}
    \put (0,51) {\colorbox{gray!30}{\scriptsize MPEG: F1 79.9, Bitrate 12.9}}
\end{overpic}

\caption{
\small Qualitative results on \northamerica~and KITTI.
Points are colored by intensity.
}
\label{figure:qualitative-results}
\vspace{-10pt}
\end{figure}

We evaluate our LiDAR compression method
on two large-scale datasets.
Holding reconstruction quality equal, our framework for
joint geometry and intensity compression achieves a 7--17\% and 6--19\% bitrate
reduction over OctSqueeze \cite{huang2020}, the prior state-of-the-art in deep point cloud compression, on \northamerica~and SemanticKITTI.
Holding bitrate equal, our method's reconstructions also have a smaller realism gap on downstream tasks.

\subsection{Experimental Details}
We validate the performance of our approach on two datasets:
\northamerica~\cite{liang_pnpnet} and SemanticKITTI~\cite{behley2019}.

\paragraph{\northamerica:} {\northamerica} is a large-scale dataset collected by a fleet
of self-driving vehicles in several cities across North America~\cite{liang_pnpnet}.
Every sequence consists of 250 Velodyne HDL-64E LiDAR sweeps sampled at 10Hz, each
containing a 3D point cloud (as 32-bit floats) and their intensity values
(as 8-bit unsigned integers).
The average size of each sweep is 80,156 points.
We train our entropy models on 5000 sequences and evaluate on a test set of 500.
Every sweep in \northamerica~is annotated with per-point semantic
labels for the vehicle, pedestrian, motorbike, road, and background classes,
as well as bird's eye view bounding boxes for the first three classes.
We use these labels to perform downstream perception experiments on the same
train/test split.

\paragraph{SemanticKITTI:}
We also conduct compression and downstream perception experiments on
SemanticKITTI~\cite{behley2019}, which enhances the KITTI \cite{Geiger2013IJRR}
dataset with dense semantic labels for each LiDAR sweep. It consists of 22 driving sequences containing
a total of 43,552 Velodyne HDL-64E LiDAR sweeps sampled at 10Hz.
The average size of each sweep is 120,402 points.
In our experiments, we use the official train/test splits: sequences 00 to 10
(except for 08) for training and sequences 11 to 21 to evaluate reconstruction quality.
Since semantic labels for the test split are unavailable, we evaluate
downstream tasks on the validation sequence 08.


\begin{table}[!t]
\centering
\parbox{.48\linewidth}{
	\centering
\resizebox{0.8 \linewidth}{!}{%
\begin{tabular}{cccc|c|c|c}
\toprule
           &            &            &            & \multicolumn{3}{c}{Spatial Bitrate}             \\
O          & T          & B          & CC         & D = 12         & D = 14        & D = 16         \\
\midrule
\checkmark &            &            &            & 2.91           & 8.12          & 14.16          \\
\checkmark & \checkmark &            &            & 2.87           & 8.04          & 14.08          \\
\checkmark & \checkmark & \checkmark &            & 2.72           & 7.90          & 13.95          \\
\checkmark & \checkmark & \checkmark & \checkmark & \textbf{2.55}  & \textbf{7.64} & \textbf{13.79} \\
\bottomrule
\end{tabular}%
}
\caption{
\small Abalation study of occupancy entropy model on \northamerica.
\textbf{O}, \textbf{T}, and \textbf{B} stand for using past
occupancy bytes, top-down aggregated features, and
bottom-up aggregated features.
\textbf{CC} indicates using continuous conv.
\textbf{D} stands for the octree's max depth.
}
\label{table:occupancy-model-ablation-results}
}
\hfill
\parbox{.48\linewidth}{
\centering
\resizebox{0.8 \linewidth}{!}{%
\begin{tabular}{c|c|c|c|c}
\toprule
                          &            & \multicolumn{3}{c}{Intensity Bitrate} \\
Encoder                   & P          & D = 12         & D = 14        & D = 16        \\
\midrule
\texttt{zlib}~\cite{zlib} &            & 2.42           & 4.79          & 5.23          \\
MLP                       & \checkmark & 2.31           & 4.62          & 5.01          \\
CC                        & \checkmark & \textbf{2.13}  & \textbf{4.30} & \textbf{4.68} \\
\bottomrule
\end{tabular}%
}
\caption{
\small Ablation study of intensity entropy model on SemanticKITTI.
\texttt{zlib} is an off-the-shelf library~\cite{zlib};
\textbf{MLP} is our model without continuous conv.; and
\textbf{CC} is our final model.
\textbf{P} stands for using past intensity information.
\textbf{D} stands for the octree's max depth.
}
\label{table:intensity-model-ablation-results}
}
\end{table}

\paragraph{Baselines:}
We compare against a number of state-of-the-art LiDAR compression algorithms:
Huang \etal's deep octree-based method (\textbf{OctSqueeze})~\cite{huang2020},
Google's KD-tree based method (\textbf{Draco})~\cite{draco},
Mekuria \etal's octree-based MPEG anchor (\textbf{MPEG Anchor})~\cite{mekuria2016}\footnote{We use the authors' implementation: \url{https://github.com/cwi-dis/cwi-pcl-codec}.},
and \textbf{MPEG TMC13}\footnote{MPEG TMC13 reference implementation: \url{https://github.com/MPEGGroup/mpeg-pcc-tmc13}}.
From discussions with the authors, ``MPEG Anchor'' in~\cite{huang2020} is a
custom implementation that uses an empirical histogram distribution to compress
octree occupancy symbols; we report this baseline as \textbf{Octree}.
As OctSqueeze and Octree do not compress LiDAR intensities, we augment
them with an off-the-shelf lossless compression algorithm~\cite{zlib}.
In particular, we first assign an intensity to each encoded point
based on the intensity of its nearest neighbour in the original point cloud.
Then, we compress the resulting bytestream.
For MPEG Anchor, we use the built-in PCL color coder in the authors'
implementation, which encodes the average intensity at each leaf node in the
octree with range coding.
Similarly, for Draco and MPEG TMC13, we use their built-in attributes coders.
We also compare against a video compression based algorithm using LiDAR's range
image representation (\textbf{MPEG Range}).
As this baseline was uncompetitive, we report its results in the supplementary.

\paragraph{Implementation Details:}
In our experiments, we construct octrees over a 400m $\times$ 400m $\times$ 400m
region of interest centered on the ego-vehicle.
By varying the octree's maximum depth from 11 to 16, we can control our method's
bitrate-distortion tradeoff, with spatial quantization errors ranging from 9.75cm
(at depth 11) to 0.3cm (at depth 16).
We train and evaluate individual entropy models at each depth from 11 to 16,
which we found gave the best results.
Our models use $ K_\text{ans} = 4 $ rounds of aggregation and $ k = 5 $
nearest neighbors for continuous convolution.
Our method is implemented in PyTorch~\cite{pytorch2019} and we use
Horovod~\cite{sergeev2018} to distribute training over 16 GPUs.
We train our models over 150,000 steps using the Adam optimizer~\cite{kingma2015}
with a learning rate of $1\mathrm{e}{-4}$ and a batch size of 16.

\paragraph{Metrics:}
We report reconstruction metrics in terms of $ \mathrm{F}_1 $ score,
point-to-point (D1) Chamfer distance \cite{huang2020},
and point-to-plane (D2) PSNR \cite{tian2017}.
Point-to-point and point-to-plane errors are standard MPEG metrics~\cite{schwarz2019}.
But whereas they measure reconstruction quality in terms of geometry only,
$ \mathrm{F}_1 $ measures this in terms of both geometry and intensity.
Reconstruction metrics are averaged across sweeps and
bitrate is the average number of bits used to store a LiDAR point.
Following standard practice, we do not count the
one-time transmission of network weights since it is negligible compared to
the size of long LiDAR streams; \eg 1 hour.
See our supplementary materials for details.

\subsection{Results}
\label{section:experiments/results}

\paragraph{Quantitative Results:}
In Fig.~\ref{figure:quantitative-results}, we report bitrate \vs reconstruction
quality curves for all competing methods on \northamerica~and SemanticKITTI.
The leftmost figures show the trade-off between overall bitrate \vs $ \mathrm{F}_1 $.
Here, we see that our method outperforms the prior state-of-the-art and, holding
reconstruction quality equal, achieves a 7--17\% (\resp, 6--19\%) relative reduction in bitrate
versus OctSqueeze on \northamerica~(\resp, SemanticKITTI).
Our model also outperforms MPEG TMC13---the MPEG point cloud compression standard---especially at lower bitrates.
The right two figures show the trade-off between spatial bitrate \vs Chamfer
distance and PSNR respectively.
Although our method shares a common octree data structure with OctSqueeze (\resp, Octree),
and thus have the same reconstruction quality, we achieve a 5--30\% (\resp, 15--45\%)
reduction in spatial bitrates on \northamerica~by additionally exploiting temporal information;
similar results also hold in SemanticKITTI.
These results validate our unified framework for geometry
and intensity compression using spatial-temporal information.

\paragraph{Qualitative Results:}
In Fig.~\ref{figure:qualitative-results}, we show reconstructions
from our method, Draco, and MPEG Anchor on \northamerica~and SemanticKITTI.
At similar bitrates, our method yields higher quality reconstructions than
the competing methods in terms of both geometry and intensity.
For example, from the first and third rows of Fig.~\ref{figure:qualitative-results},
we can see that our method produces faithful reconstructions even at high
compression rates.
In contrast, Draco and MPEG Anchor produce apparent artifacts.


\begin{figure}[!t]
\centering
\includegraphics[height=0.235\textwidth]{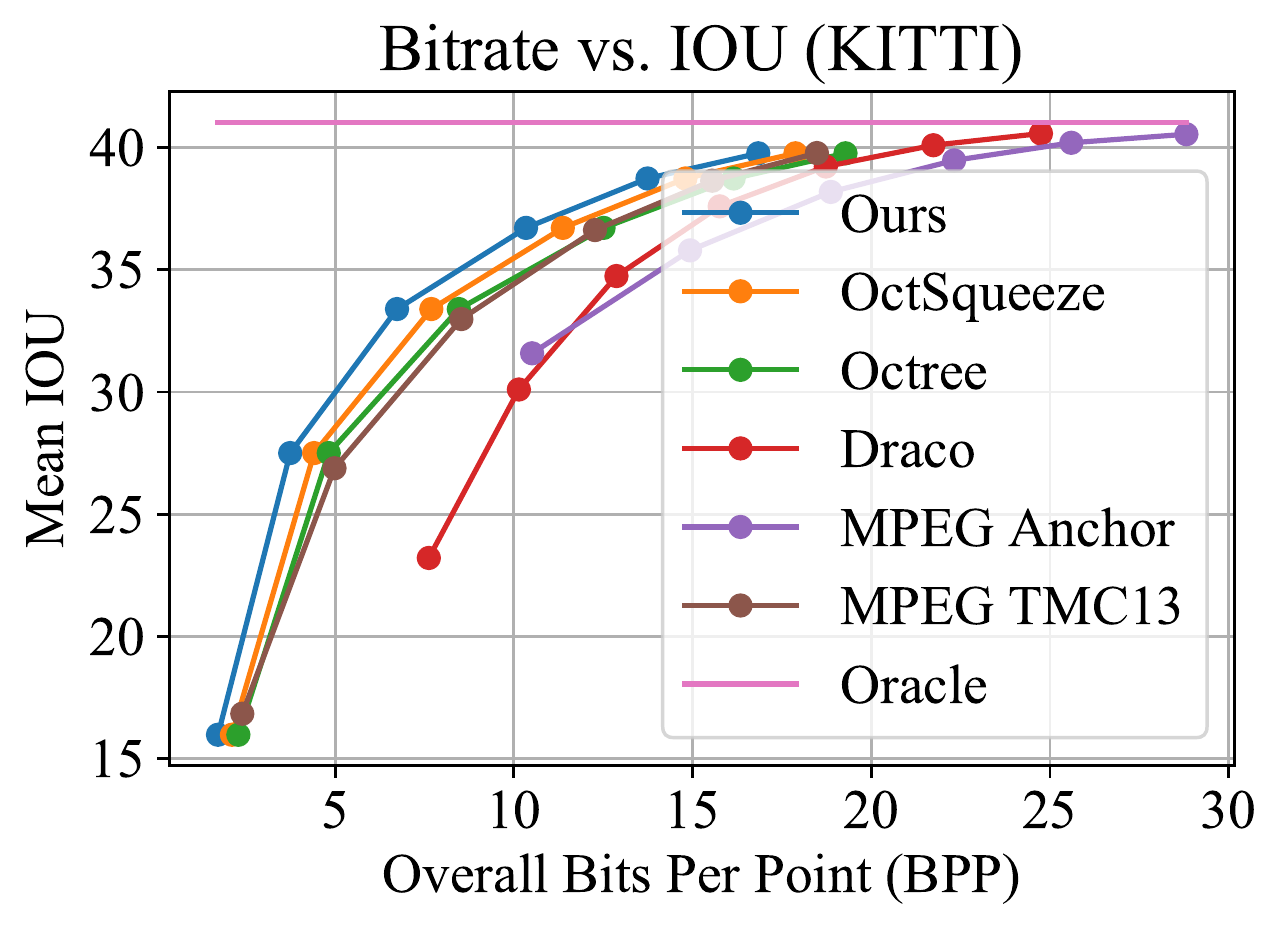}
\includegraphics[height=0.235\textwidth]{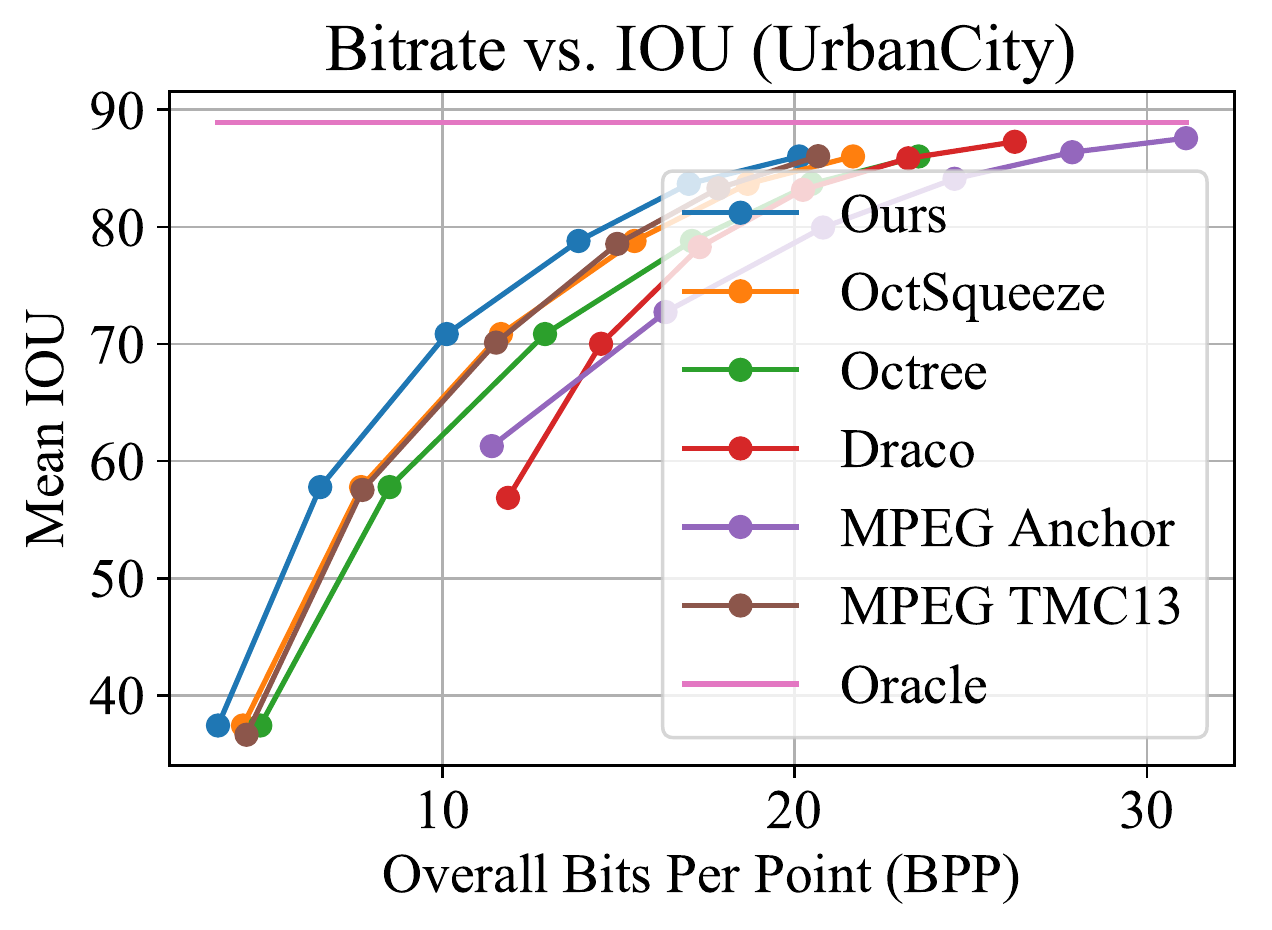} \\
\includegraphics[height=0.235\textwidth]{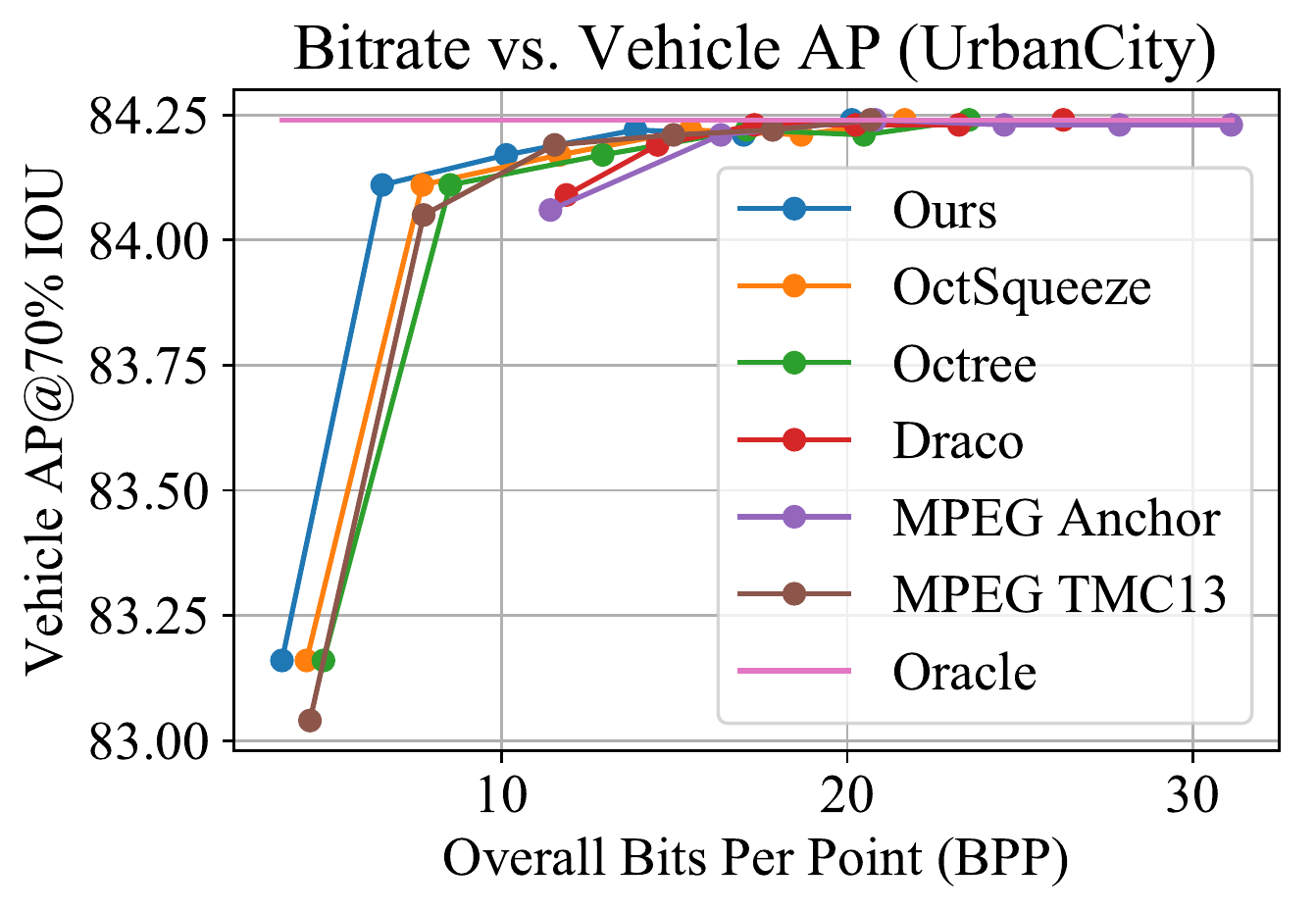}
\includegraphics[height=0.235\textwidth]{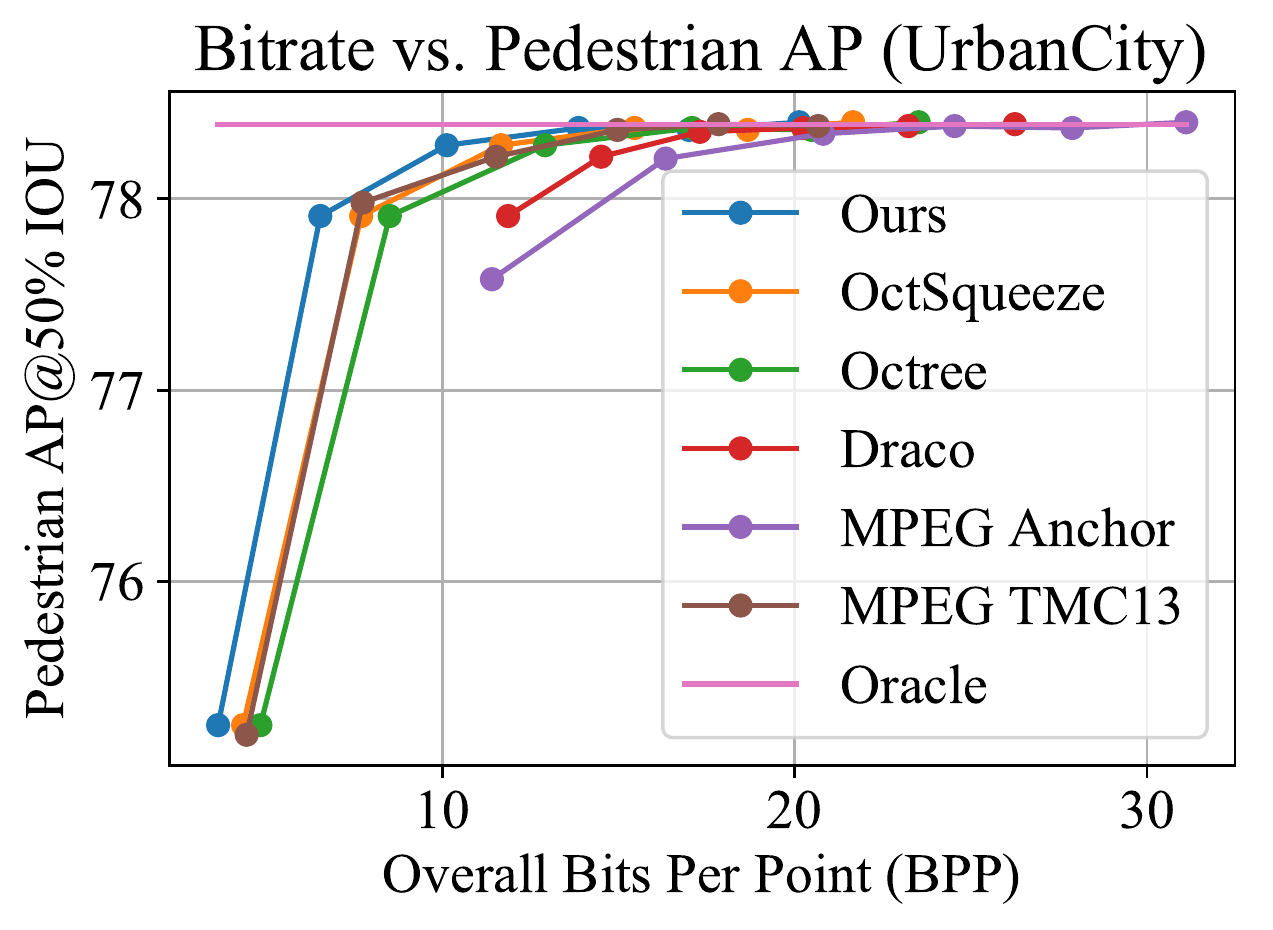}
\includegraphics[height=0.235\textwidth]{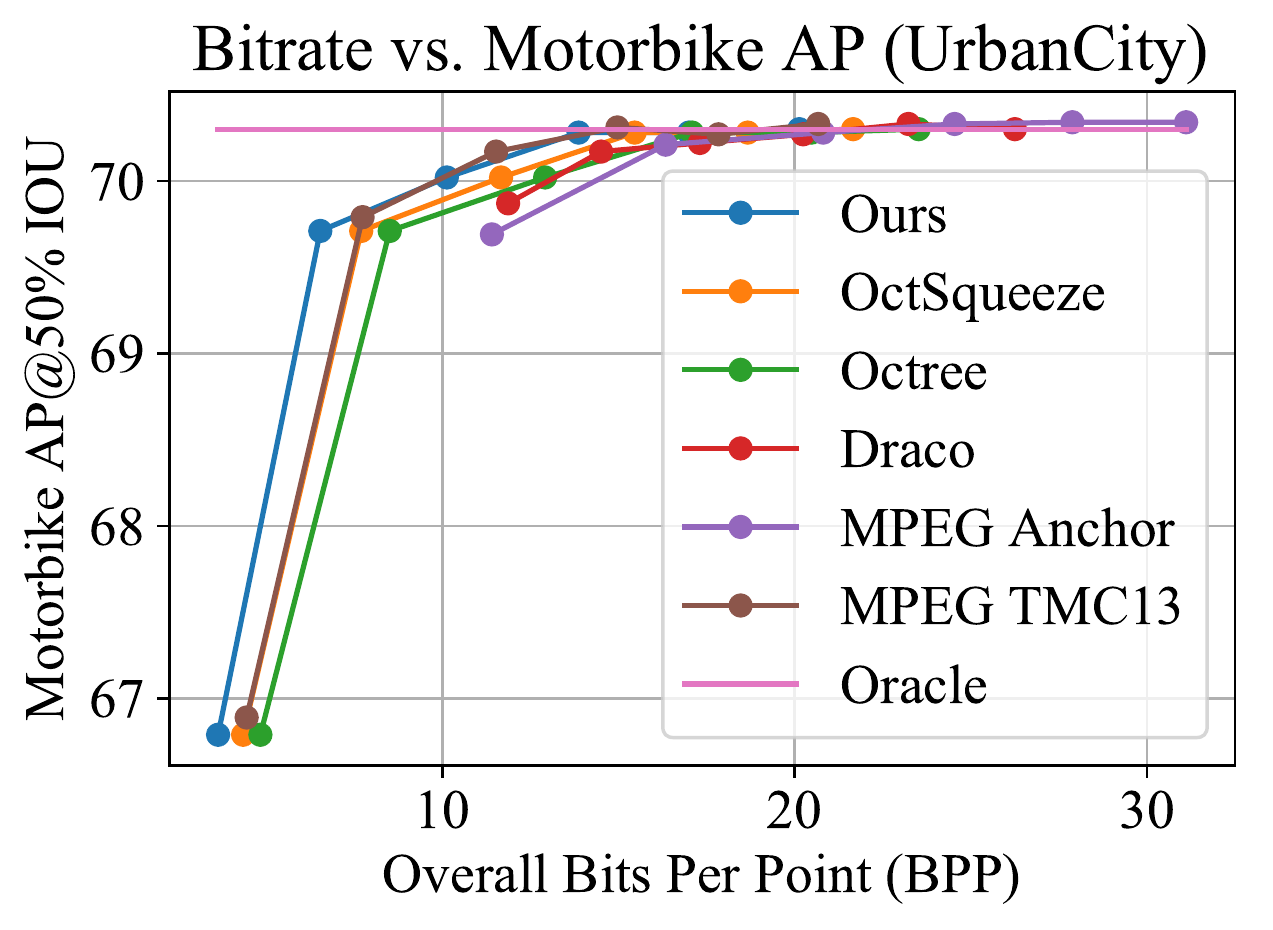}
\caption{
\small
Bitrate \vs downstream task performance.
Top: mean IoU for semantic segmentation on SemanticKITTI (left) and \northamerica~(right).
Bottom: AP for vehicle, pedestrian, and motorbike detection on \northamerica.
}
\label{figure:downstream-quantitative-results}
\end{figure}


\begin{figure}[!t]
\centering
\begin{overpic}[clip,trim=0cm 10cm 0cm 0cm,width=0.24\textwidth]{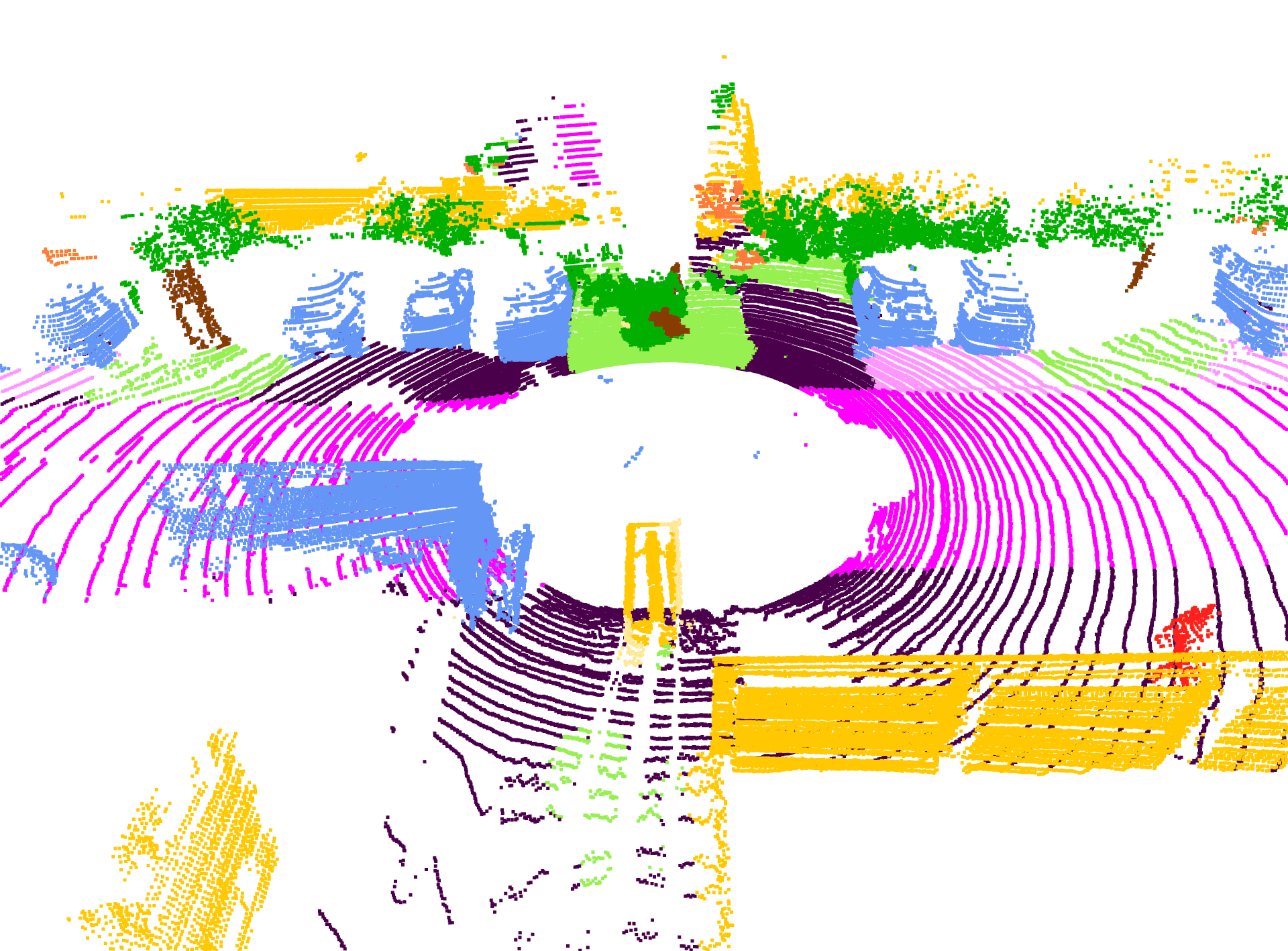}
    \put (0,51) {\colorbox{gray!30}{\scriptsize Oracle IOU: 37.5, Bitrate: 104.0}}
\end{overpic}
\begin{overpic}[clip,trim=0cm 10cm 0cm 0cm,width=0.24\textwidth]{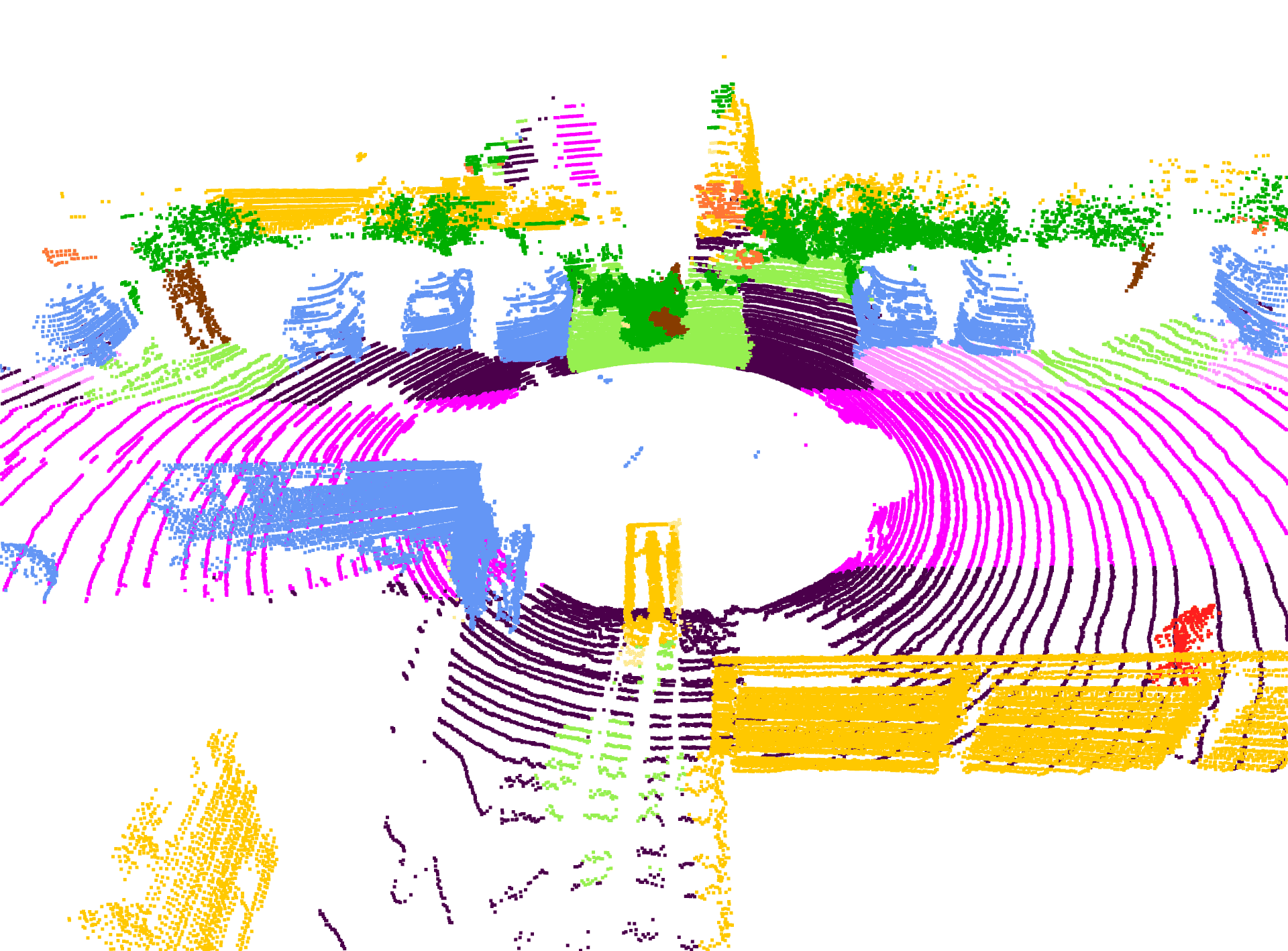}
    \put (0,51) {\colorbox{gray!30}{\scriptsize Ours IOU: 36.8, Bitrate: 13.1}}
\end{overpic}
\begin{overpic}[clip,trim=0cm 5cm 0cm 5cm,width=0.24\textwidth]{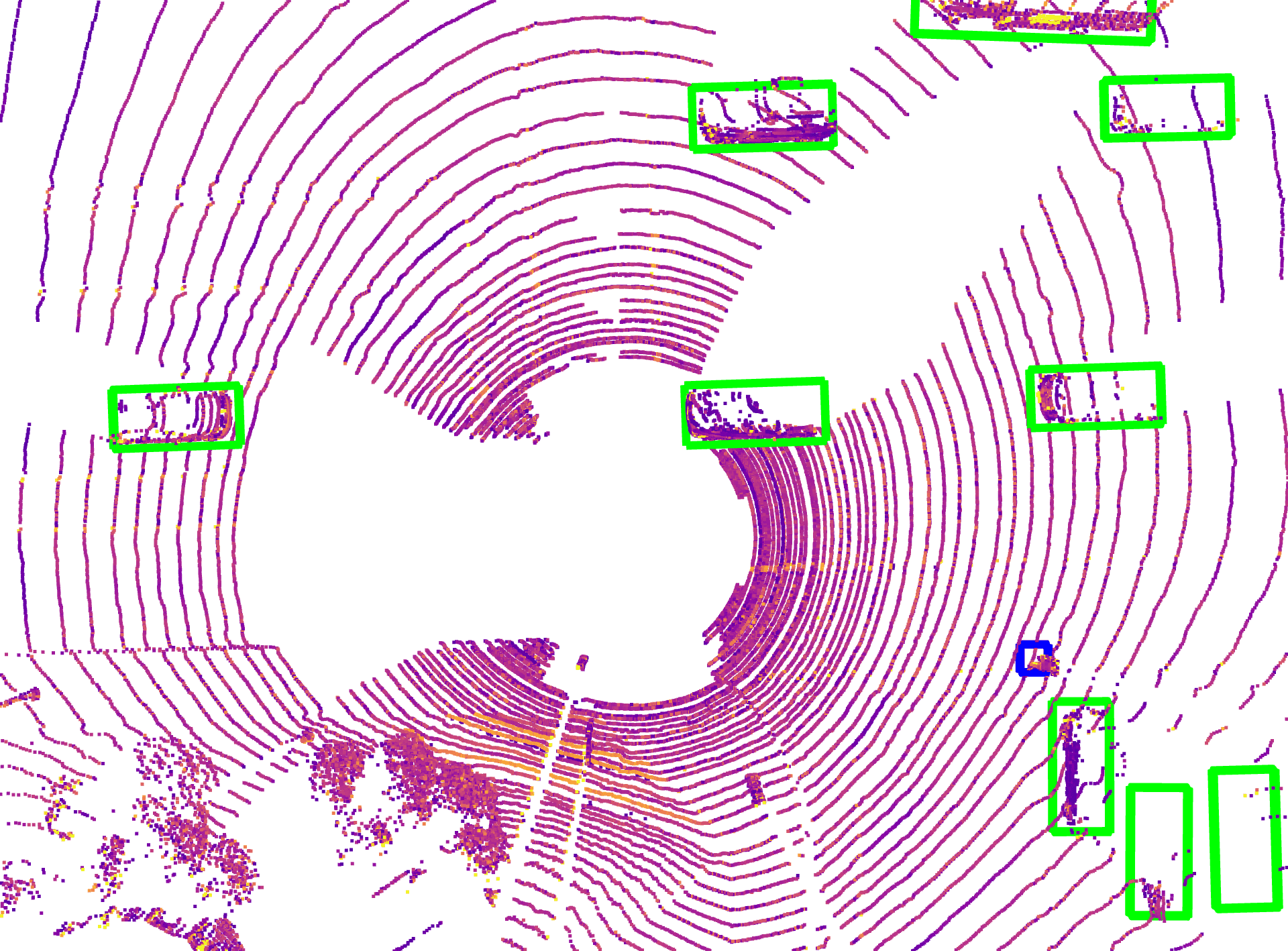}
    \put (0,51) {\colorbox{gray!30}{\scriptsize Oracle AP: 88.5, Bitrate: 104.0}}
\end{overpic}
\begin{overpic}[clip,trim=0cm 5cm 0cm 5cm,width=0.24\textwidth]{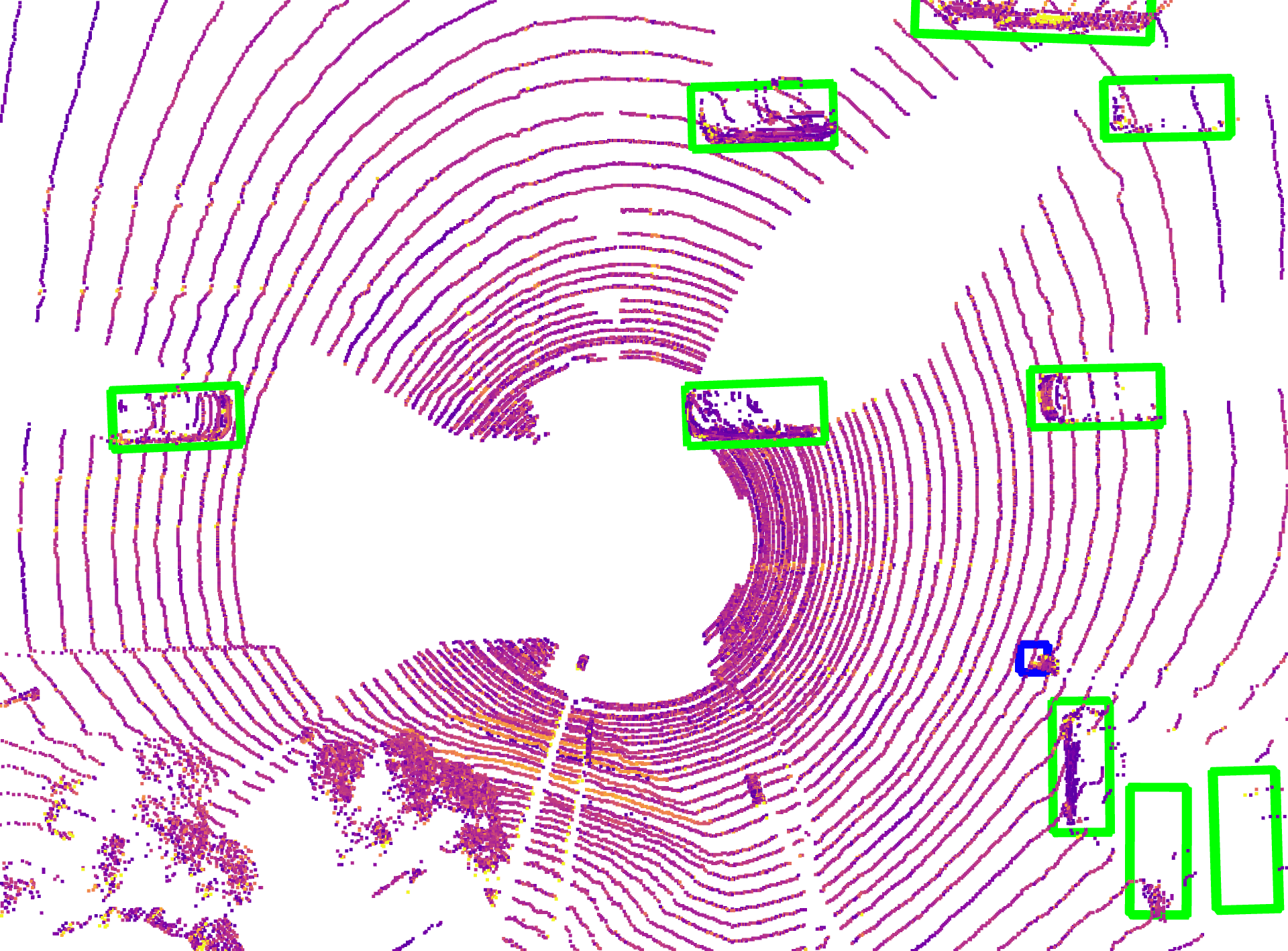}
    \put (0,51) {\colorbox{gray!30}{\scriptsize Ours AP: 88.5, Bitrate: 9.9}}
\end{overpic}

\caption{
\small
Left: Semantic segmentation on SemanticKITTI.
Right: Object detection on \northamerica.
Even at very low bitrates, our reconstructions have a minimal realism gap for downstream tasks.
}
\label{figure:downstream-qualitative-results}
\vspace{-10pt}
\end{figure}

\paragraph{Ablation Studies:}
We perform two ablation studies on our occupancy and intensity entropy models.
In Tab.~\ref{table:occupancy-model-ablation-results}, we ablate how to
incorporate past information to lower the entropy of our occupancy model.
We start with using the past octree's occupancy bytes (\textbf{O}) and then
progressively add top-down and bottom-up aggregated features (\textbf{T} and \textbf{B} respectively),
and finally continuous convolutions (\textbf{CC}).
We see that, holding reconstruction quality equal, each aspect of our
model consistently reduces bitrates.
In Tab.~\ref{table:intensity-model-ablation-results}, we compare three compression
methods for the intensity model: the \texttt{zlib} library, a multi-layer perceptron
entropy model (\textbf{MLP}), and our final model (\textbf{CC}).
Note that both MLP and CC conditions on context from neighboring points in
the past sweep; \texttt{zlib} does not.
However, whereas MLP uses context from one neighbor only,
CC aggregates context from multiple neighbors via continuous convolutions.
Our results show that learning to incorporate past context
reduces intensity bitrates by 4--5\%,
and that this improvement is strengthened to 11--12\%
by using continuous convolutions to align information across space and time.
Please see our supplementary for details.

\paragraph{Impact on Downstream Tasks:}
To study the impact of compression on downstream perception tasks,
we first train segmentation and detection models on uncompressed
LiDAR for SemanticKITTI and \northamerica.
Note that these models use both LiDAR geometry and intensity as input
(see supplementary for details).
Next, we evaluate the models on LiDAR reconstructions obtained from
various compression schemes and report their performance as a function of overall bitrate.
For segmentation on SemanticKITTI and \northamerica, we report mean
IOU using voxelized ground truth labels at a 10cm resolution.
For detection on \northamerica, we report AP at 50\%
IOU for pedestrians and motorbikes and 70\% IOU for vehicles.
In Fig.~\ref{figure:downstream-quantitative-results} and~\ref{figure:downstream-qualitative-results},
we see that our method's reconstructions have the smallest realism gap for downstream tasks across all bitrates.
This result is especially pronounced for segmentation models,
which are more sensitive to fine-grained geometric and intensity details.


\section{Conclusion}

We have presented a novel LiDAR point cloud compression algorithm using a deep entropy model which exploits spatio-temporal redundancies between successive LiDAR point clouds.
We showed that we can compress point clouds at identical reconstruction quality to the state-of-the-art while lowering bitrate significantly, as well as compress LiDAR intensity values effectively which was not as extensively explored by prior works. Furthermore, we showed our compression can be applied to downstream self-driving perception tasks without hindering performance.
Looking forward, we plan to extend our method to jointly compress data streams from entire sensor suites.


\section*{Broader Impact}

On an immediate level, our contributions are directly applicable as a data compression algorithm in a novel problem setting: the greater we can maximize the performance of such an algorithm, the more we can reduce the storage cost and space required by point clouds. We hope that this in turn unlocks a milestone towards fulfilling our ultimate vision: scaling up the research and deployment of intelligent robots, such as self-driving vehicles, that will revolutionize the safety, efficiency, and convenience of our transportation infrastructure. By capturing the 3D geometry of the scene, LiDAR sensors have proven to be crucial in effective and safe prediction/planning of these robots. Currently, LiDAR sensors are not only expensive due to the upfront cost, but also due to the recurring costs of the massive quantities of data they generate. Good point cloud and LiDAR compression algorithms will thus help to democratize the usage of LiDAR by making it more feasible for people to own and operate.
Perhaps just as importantly, our responsibility as researchers in a novel problem area led us to carefully consider the downstream impacts of such a compression algorithm---if the primary usage of LiDAR currently is on perception tasks, such as detection and segmentation, then we need to demonstrate how compression bitrate affects perception performance, helping the community determine the acceptable bitrate at which compression can be used for safe vision and robotics applications.
We hope that our work inspires the community to further advance sensor compression in addition to the traditional image and video settings.

\bibliographystyle{unsrt}
\bibliography{refs}

\newpage
\appendix

\section{Additional Experiments}
\label{section:additional-experiments}


\subsection{Compression of Leaf Offsets}
\label{section:leaf}

We mention in Sec. 2.1 of the main paper that we do not attempt to compress the leaf offsets from the octree. The reason is that we experimented with a few compression baselines and were not able to obtain a bitrate improvement over the uncompressed leaf offsets. 
We experiment with the zlib \cite{zlib}, LZMA \cite{lzma}, and bzip2 \cite{bz2} compression algorithms on the leaf offset stream from UrbanCity. The results are shown in Tab. \ref{table:leaf-offset-table}; we  surprisingly found that in all cases the compressed string was longer than the uncompressed one.

\begin{table}[h]
	\centering
	\parbox{\linewidth}{
		\centering
		\resizebox{0.7 \linewidth}{!}{%
			\begin{tabular}{c | c | c | c | c}
				\toprule
				& Uncompressed      &     zlib \cite{zlib}      &      LZMA \cite{lzma}      &  bzip2 \cite{bz2}            \\ \hline
				Avg. Bytes / Sweep & 102429.31        	&    102468.93        		& 	  102493.84   				&  103242.28      \\ \hline
			\end{tabular}%
		}
		\caption{
		Comparison of compression algorithms on leaf offsets from UrbanCity, in terms of average bytes per sweep. }
		\label{table:leaf-offset-table}
	}
\end{table}

There can be room for future work in entropy modeling the leaf offsets, but our current hypothesis is that since the intermediate octree nodes already encode the shared bits between points, the leaf offsets represent residual bits that can be considered ``higher-frequency'' artifacts (similar to residual frames in video compression), and are therefore harder to compress.


\subsection{Using a Range Image Representation}
\label{section:range}

We mention in Sec. 3.1 of the main paper that we designed a range image-based compression baseline. Towards this goal, we first converted point cloud streams in UrbanCity and KITTI into range image representations, which store LiDAR packet data into a 2D matrix. We consider two possible range image representations. The first contains dimensions $H_{\text{lid}} \times W_{\text{azm}}$, where the height dimension represents the separate \textit{laser ID's} of the LiDAR sensor, and the width dimension represents the discretized azimuth bins between -180\degree and 180\degree. Each pixel value represents the distance returned by the laser ID at the specific azimuth angle. Such a representation requires sufficient auxiliary calibration and vehicle information in order to reconstruct the points in Euclidean space---for instance, a separate transform matrix per laser and velocity information to compensate for rolling shutter effects. We use this representation for UrbanCity because we have access to most required information; unfortunately, not every log contains detailed calibration or precise velocity information, requiring us to use approximations.

The second representation simply projects the spatial coordinates of the point cloud sweep into the coordinate frame of the sensor, and does not require a map between laser ID and Euclidean space. Such an image contains dimensions $H_{\text{pitch}} \times W_{\text{azm}}$, where the height dimension now represents discretized pitch angles; each pixel value now represents the distance of a given point from the sensor frame at a given pitch and azimuth bin. We use this representation for our KITTI point clouds, since the dataset does not provide detailed laser calibration information.

We explore both geometry-only and geometry + intensity representations. Spatial positions are encoded in the 8-bit R,G channels of the png image (16 bits total). If intensity is encoded, it is encoded in the B channel. We run H.264 on the png image sequence as our compression algorithm. We evaluate on the same reconstruction metrics: point-to-point Chamfer distance and point-to-plane PSNR (geometry), and $\text{F}_1$ score (geometry + intensity).

\begin{figure}[!t]
\centering
\includegraphics[height=0.25\textwidth]{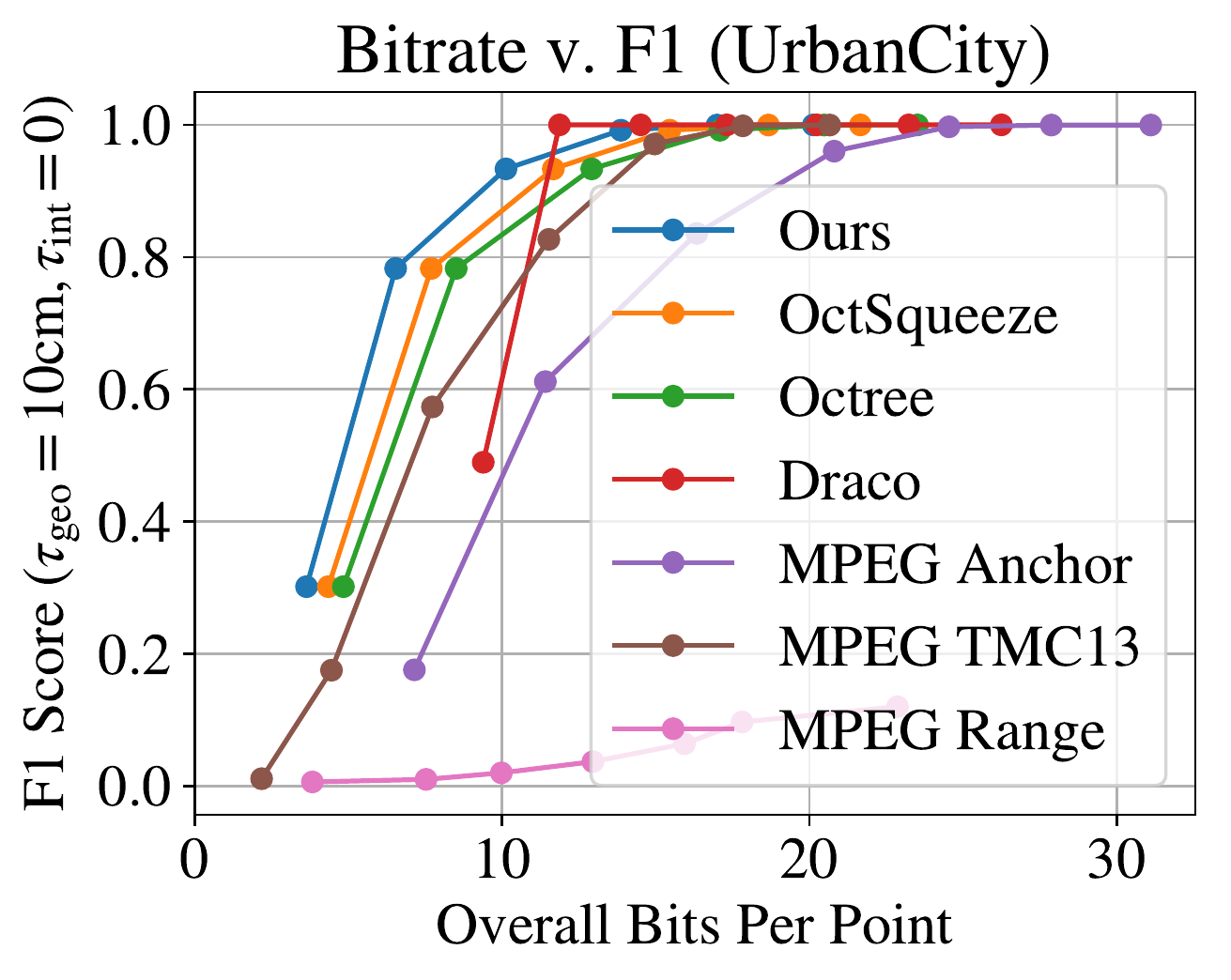}
\includegraphics[height=0.25\textwidth]{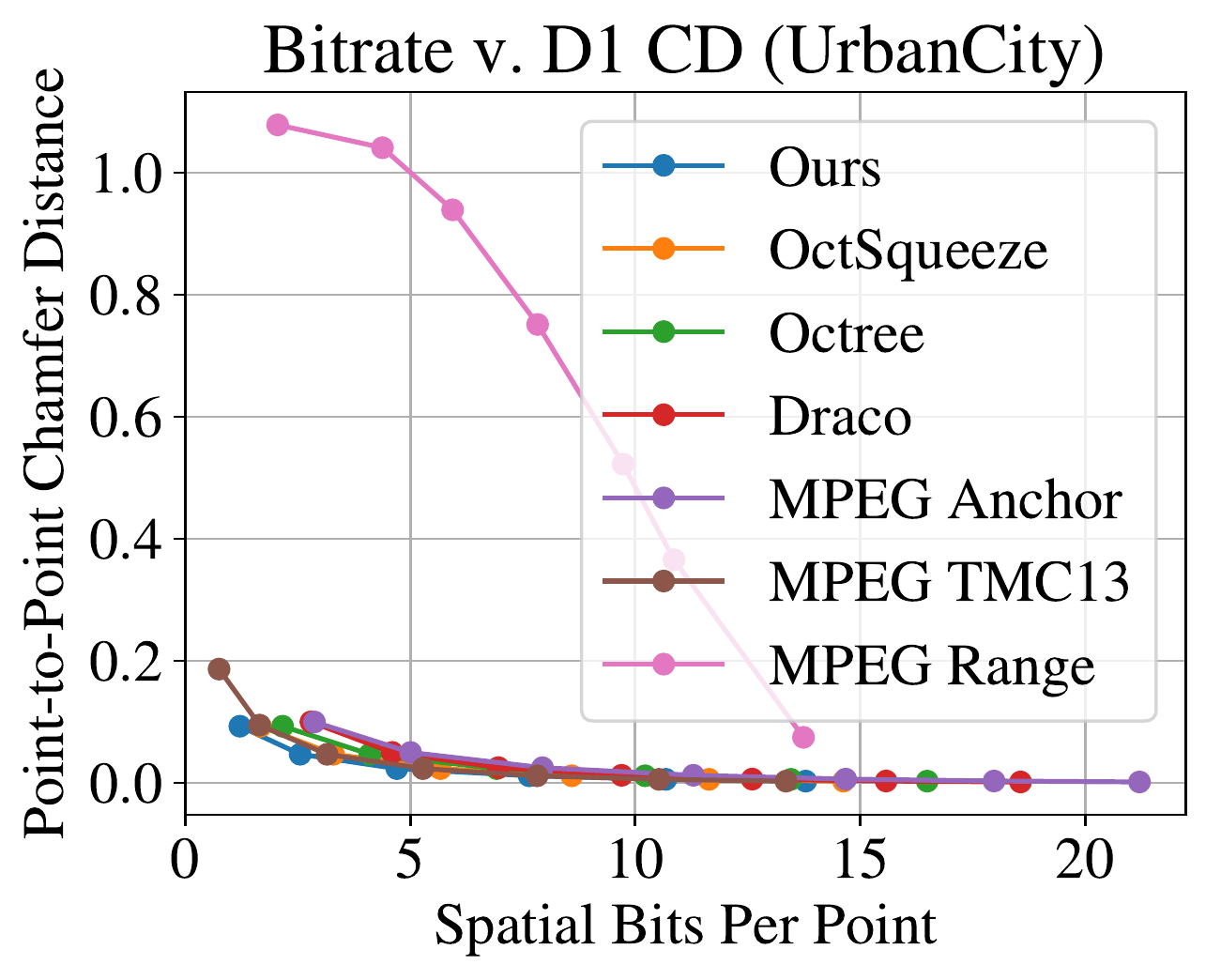}
\includegraphics[height=0.25\textwidth]{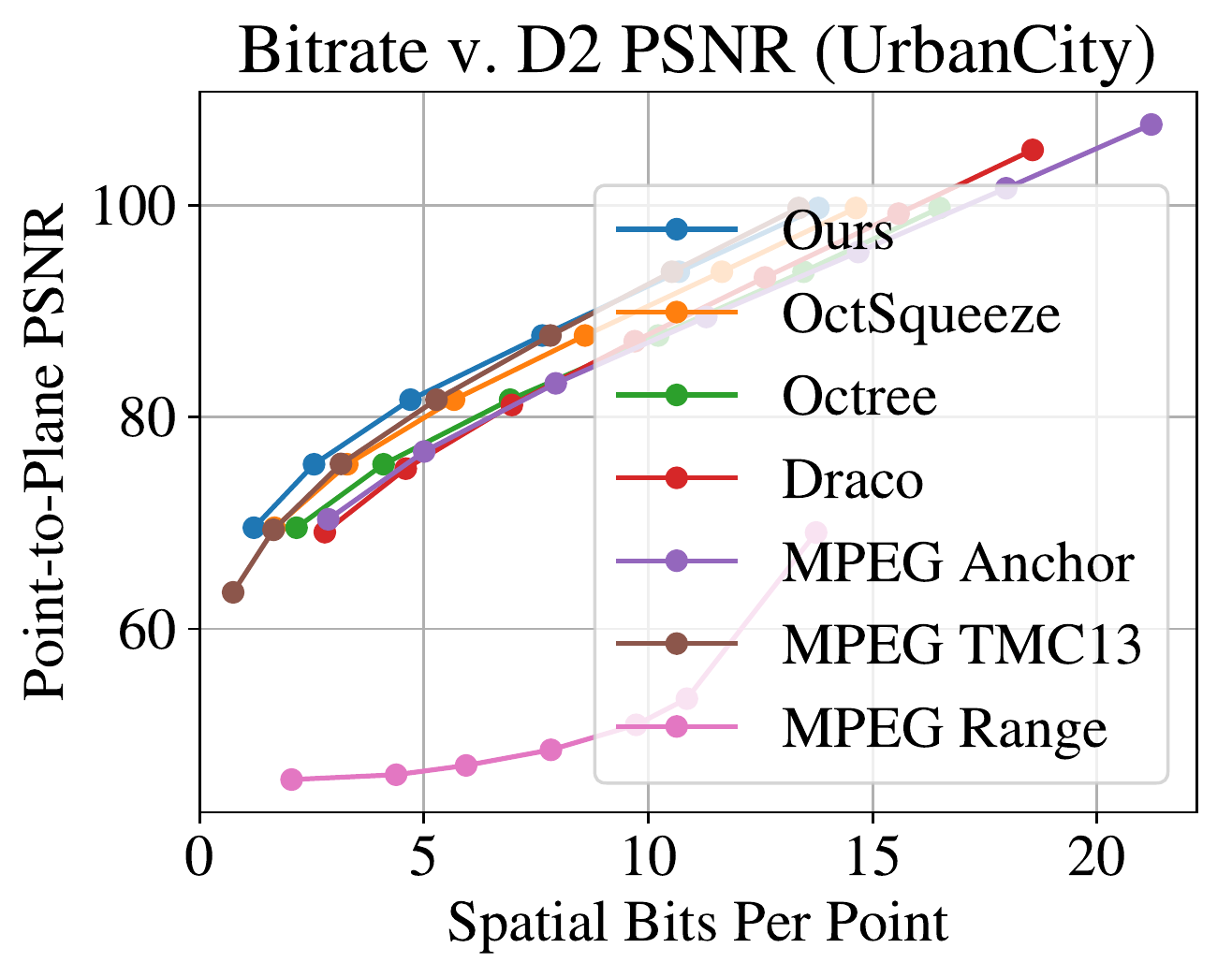}

\includegraphics[height=0.25\textwidth]{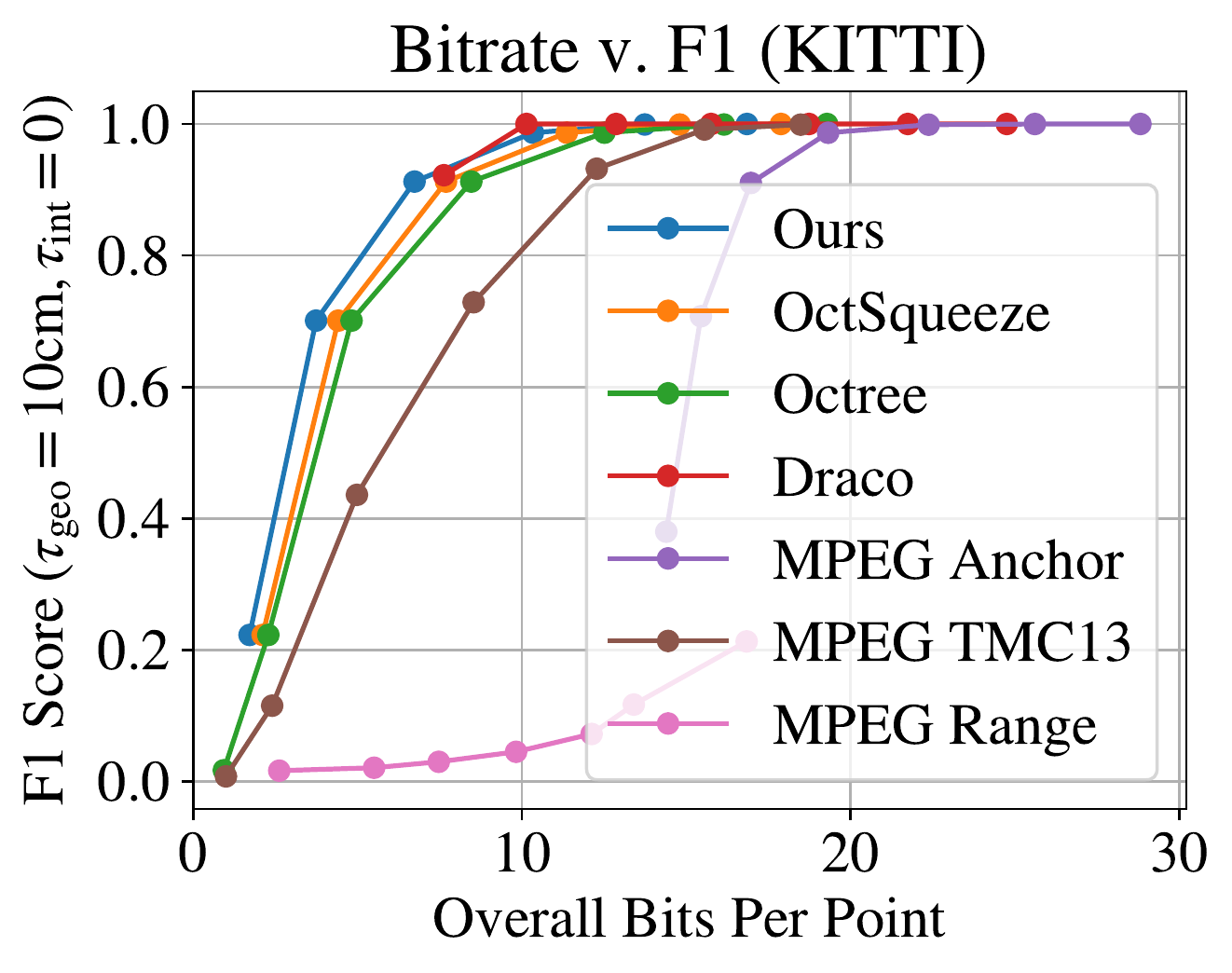}
\includegraphics[height=0.25\textwidth]{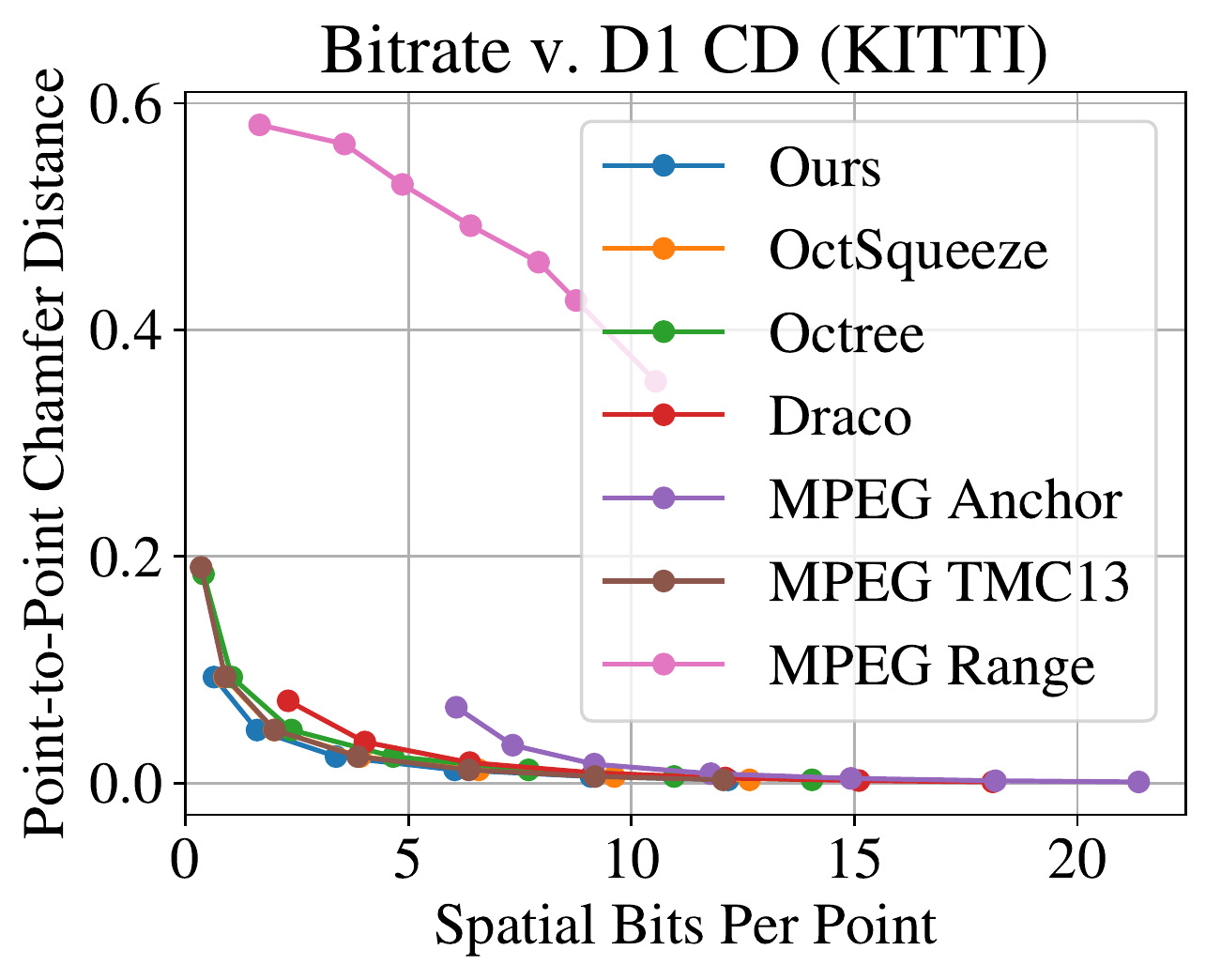}
\includegraphics[height=0.25\textwidth]{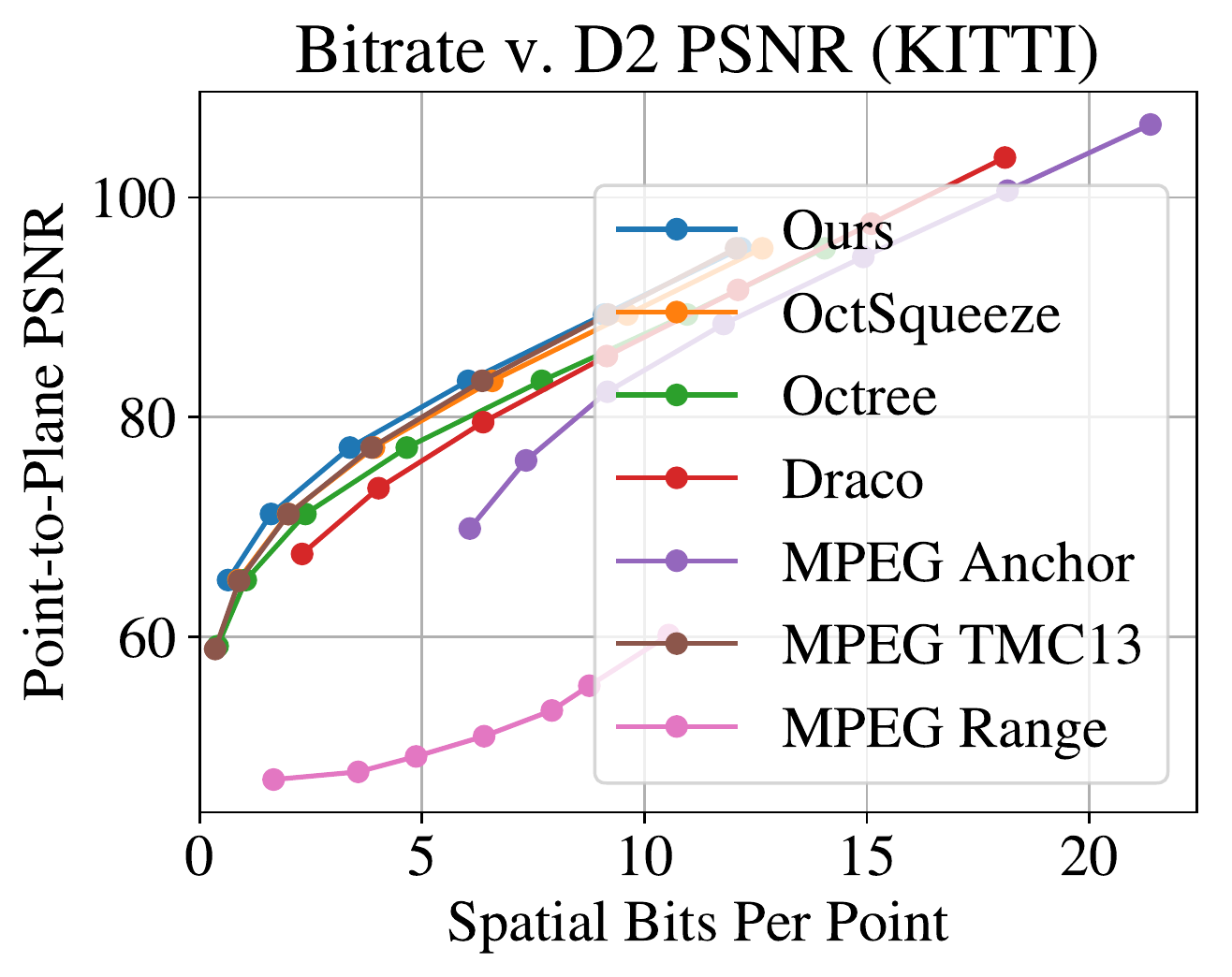}
\caption{
\small
Bitrate \vs reconstruction quality curves on \northamerica~(top) and KITTI (bottom).
From left-to-right:
$ \mathrm{F}_1 $ with $ \tau_\mathrm{geo} = 10\mathrm{cm} $ and $ \tau_\mathrm{int} = 0 $ ($ \uparrow $),
point-to-point chamfer distance ($ \downarrow $),
point-to-plane PSNR ($ \uparrow $).
}
\label{figure:additional-quantitative-results}
\vspace{-10pt}
\end{figure}

We show here in Fig.~\ref{figure:additional-quantitative-results}, that the results were uncompetitive---the range image representation underperforms other baselines and our approach on every evaluation metric. We observe that even the ``lossless'' representation (the right-most point on the curves) does not yield perfect reconstruction metrics. This can be surprising for the laser ID representation in UrbanCity. But we hypothesize that the errors come from approximations of the true calibration values (which are not obtainable for every log), as well as the velocity used in rolling shutter compensation---we found that small perturbations in these calibration values yield a large variance in reconstruction quality and metrics.

\section{Additional Architecture Details}
\label{section:architecture-details}

In this section we provide additional architecture details of our octree occupancy and intensity entropy models (Secs. 2.3 and  2.4 in main paper). We also provide architecture details of the models used in the ablation studies of the occupancy and intensity model (Tab. 1, Tab. 2 in main paper).

\subsection{Occupancy Entropy Model}
\label{section:occupancy-details}
\paragraph{Ancestral Node Dependence:} The context feature $\mathbf{c}_i$ consists of the octree level of the current node (1-- 16), spatial location of the node's octant $(x, y, z)$, octant index of the node relative to its parent (0--8), and parent occupancy byte (0--255), as well as occupancy byte in the corresponding node in the previous octree (0--255 if exists, 0 otherwise). The initial feature extractor is a 4-layer MLP with fully-connected (fc) layers and intermediate ReLU activations. The hidden layer dimension is 128. Then, every aggregation round consists of a 2-layer fc/ReLU MLP with a 256-dimensional input (concatenating with the ancestor feature), and a hidden dimension of 128. We set the number of aggregation rounds, $K_{\text{ans}}$, to 4.

\paragraph{Temporal Octree Dependence:} The top-down pass to generate $\mathbf{h}_j^{(t-1)}$ has essentially the same architecture as the ancestral node dependence module above. The one difference is that each context feature additionally includes the ``ground-truth'' occupancy byte of each node, since each node in sweep $t-1$ has already been decoded. Moreover, each hidden dimension is 64 instead of 128.

Next, recall that the bottom-up aggregation pass has the following formulation: $$\mathbf{g}^{(t-1)}_{j} = f_{\textrm{agg}, 1}(\mathbf{h}^{(t-1)}_{j} + \sum_{c \in \textrm{child}(j)} f_{\textrm{agg},2} (\mathbf{g}^{(t-1)}_{c} ))$$ Here, $f_{\textrm{agg},2}$ is a 2-layer fc/ReLU MLP taking a 64-dim input and outputting a 32-dim intermediate embedding. $f_{\textrm{agg}, 1}$ is a 2-layer fc/ReLU MLP taking a (32 + 64)-dim embedding (child embedding + top-down embedding), and outputting a 64-dim embedding for the current node $j$. The bottom-up pass is run starting from the lowest level $D$ (where there are no children) back up to level 0.

\paragraph{Spatio-Temporal Aggregation and Entropy Header:} Recall that the continuous convolution layer has the formulation $$\mathbf{h}_{i}= \sum_{j \in \mathcal{N}(i)} \sigma(\mathbf{p}_j - \mathbf{p}_i ) \mathbf{h}_{j}$$ where $\mathcal{N}(i)$ is the $i$-th node's $k$-nearest neighbors in sweep $t-1$, at the same octree level as node $i$, and $p_i$ is the 3D position of each node. Here, $\sigma$ is a learned \textit{kernel function}, and it is parameterized by an MLP, inspired by \cite{wang2018}. The MLP contains  3 fc/ReLU layers (no ReLU in last layer), with output dimensions 16, 32, and 64 respectively. The continuous conv layer produces the warped feature $\mathbf{g}^{(t)}_{i, \text{st}}$. The warped feature $\mathbf{g}^{(t)}_{i, \text{st}}$ and ancestral feature $\mathbf{h}_i^{(t)}$ are aggregated through a final, 4-layer fc/ReLU MLP with hidden dim 128. The prediction header outputs a softmaxed, 256-dim vector of occupancy predictions.

\subsection{Intensity Entropy Model} \label{sec:int_entropy}
The input to the intensity entropy MLP consists of the $k$-nearest neighbor intensities in sweep $t-1$: $\{ \mathbf{r}^{(t-1)}_j \}_{j \in \mathcal{N}(i)}$. We set $k=5$. In addition to the raw intensity value, we include the following features per $\mathbf{r}^{(t-1)}_j$: spatial $(x,y,z)$ position $\in \mathbb{R}^3$, delta vector to current point $\in \mathbb{R}^3$, and 1-D distance value. Hence each point contains an 8-dimensional feature.

Each feature per $\mathbf{r}^{(t-1)}_j$ is then independently given to a 4-layer MLP, consisting of fc layers and ReLU activations. The dimension of each hidden layer is 128. Then, the $k$ output features are input to a continuous convolution layer to produce a single 128-dimensional embedding. The kernel function $\sigma$ of the continuous conv. is parameterized with the same MLP as the one used in spatio-temporal aggregation in the occupancy model. The final predictor is a fc layer and softmax with a 256-dim. output.

\subsection{Ablation Study Architectures}
\label{section:ablation-details}
We first describe the architectures of the occupancy ablation in Tab. 1 of the main paper.
\begin{itemize}
	\item $\mathbf{O}$ uses the past occupancy byte to model temporal dependence. The byte is taken from the corresponding node in the previous octree if it exists; if it does not, the feature is zeroed out. This past occupancy byte is then appended to the context feature $\mathbf{c}_i$ (along with parent occupancy byte, octree level, etc.) and fed to the ancestral dependence module. There is no temporal octree dependence module or spatio-temporal aggregation; the final prediction header is directly attached to the ancestral feature.
	\item $\mathbf{O},\mathbf{T}$ includes the temporal octree dependence module, but removes the bottom-up pass. Hence the final feature produced from this module is $\mathbf{h}_j^{(t-1)}$(as opposed to $\mathbf{g}_j^{(t-1)}$). There does not exist a spatio-temporal aggregation module using continuous convolutions to produce an embedding for every node $i$. Instead, we use a simpler ``exact matching'' heuristic similar to including the occupancy bytes---$\mathbf{h}_j^{(t-1)}$ will only be included as a feature for node $i$ in sweep $t$, if node $j$ corresponds to the same octant in sweep $(t-1)$ as node $i$ in sweep $t$. If there is no exact correspondence, the feature is zeroed out.
	\item $\mathbf{O},\mathbf{T},\mathbf{B}$ includes the full temporal octree dependence module, including the bottom-up pass to produce $\mathbf{g}_j^{(t-1)}$. As with the above, we do not include our spatio-temporal aggregation module but rather use the exact matching heuristic to include $\mathbf{g}_j^{(t-1)}$ in the corresponding nodes $i$ in sweep $t$ only if the correspondence exists.
	\item $\mathbf{O},\mathbf{T},\mathbf{B}, \mathbf{CC}$ includes our full model, including using spatio-temporal aggregation with continuous convolutions to produce an embedding feature for every node $i$.
\end{itemize}

We now describe the architectures of the intensity ablation in Tab. 2 of the main paper.

\begin{itemize}
	\item $\mathbf{MLP}$ only utilizes context from one neighbor in sweep $t-1$. First, the nearest neighbor to node $i$ is obtained in sweep $t-1$. We take the neighbor's corresponding intensity, the delta vector to the current position $\in \mathbb{R}^3$, and 1-D distance value as inputs, and feed it through a 4-layer fc/ReLU MLP and a final softmax predictor head to output 256-dim probabilities.
	\item $\mathbf{CC}$ contains the full intensity model with continuous convolutions. For architecture details see \ref{sec:int_entropy}.
\end{itemize}


\section{Additional Experiment Details}
\label{section:experiment-details}


\begin{figure}[!t]
\centering
\includegraphics[height=0.25\textwidth]{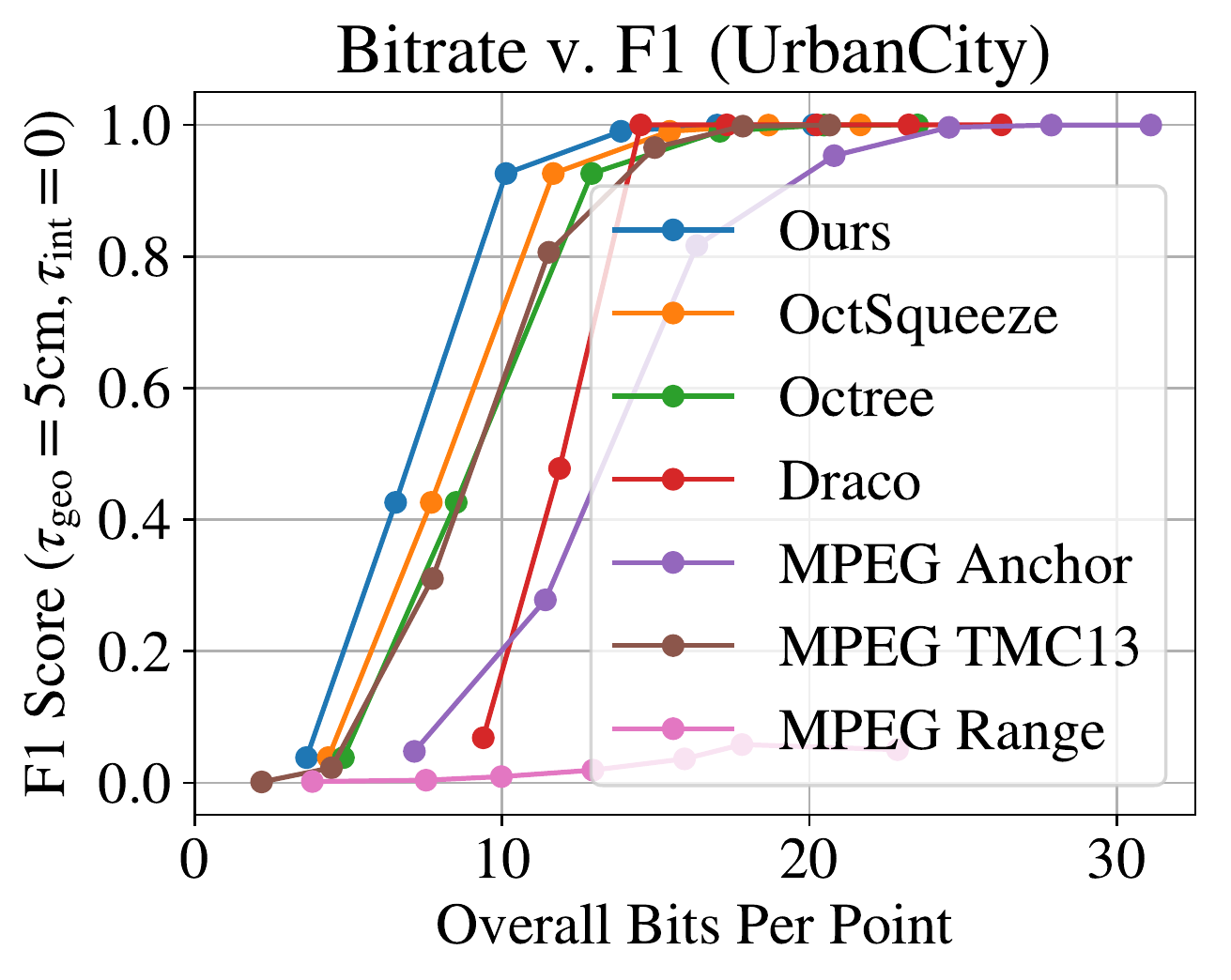}
\includegraphics[height=0.25\textwidth]{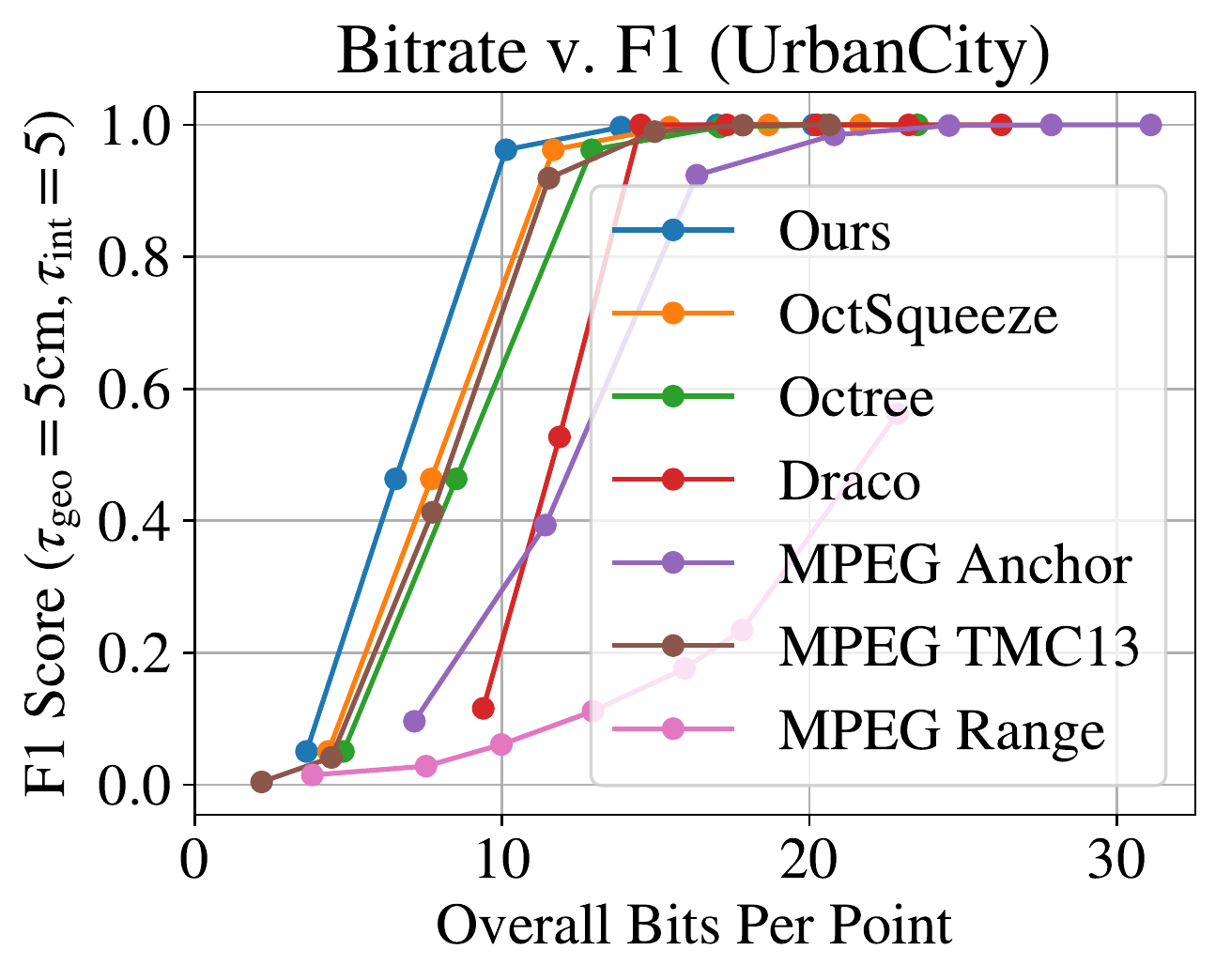}
\includegraphics[height=0.25\textwidth]{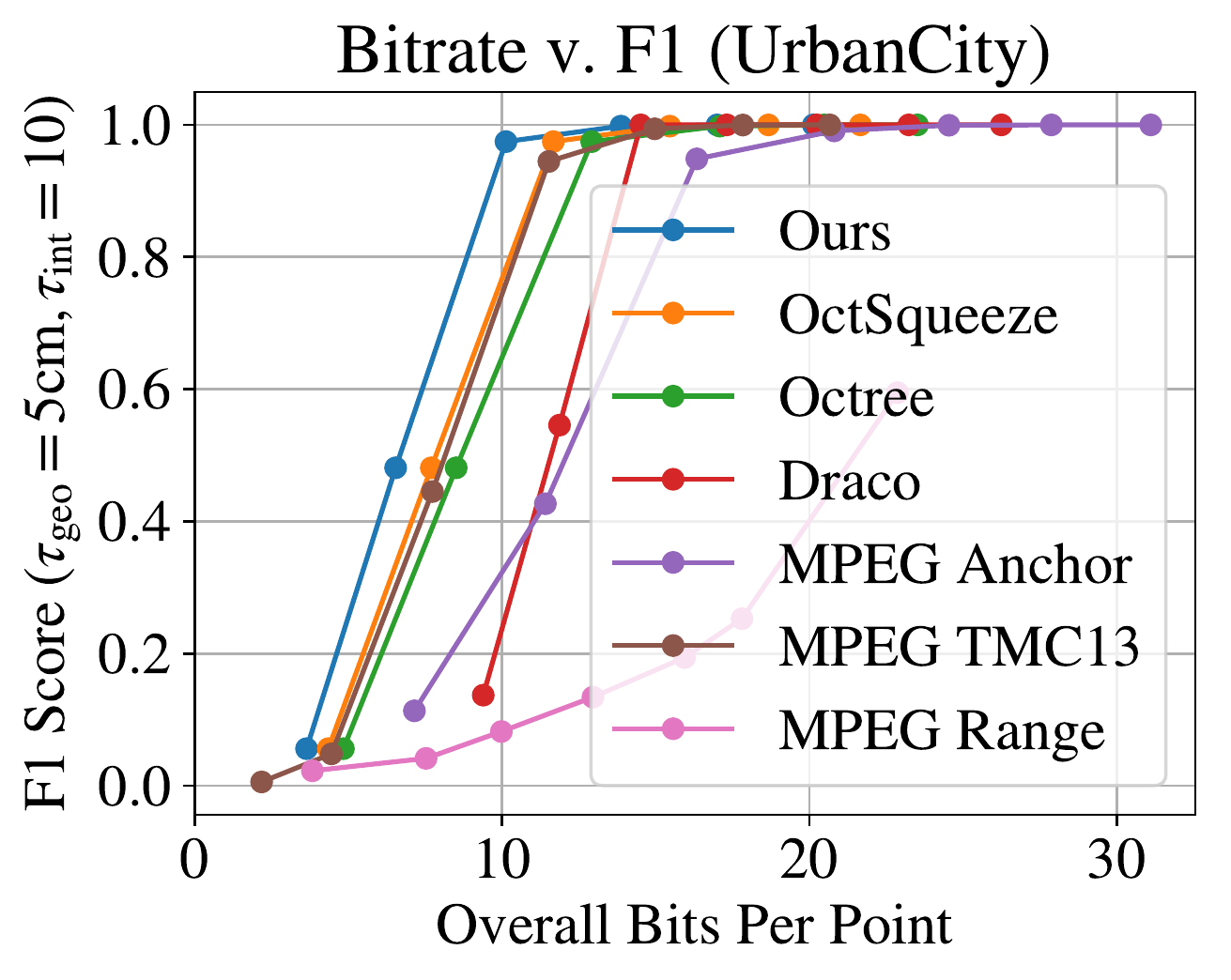} \\
\includegraphics[height=0.25\textwidth]{./figures/quantitative_results/na_supp_f1_010_000.pdf}
\includegraphics[height=0.25\textwidth]{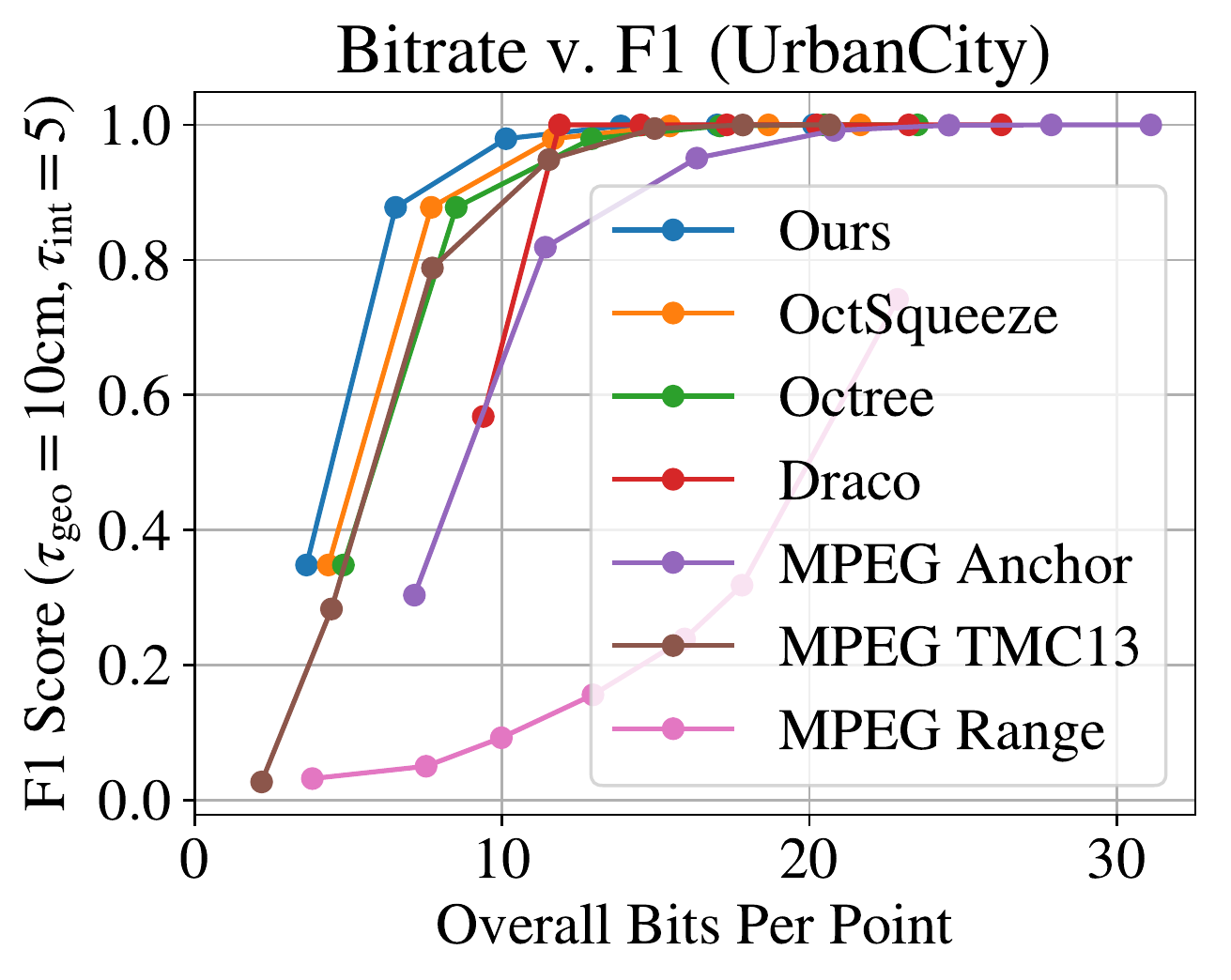}
\includegraphics[height=0.25\textwidth]{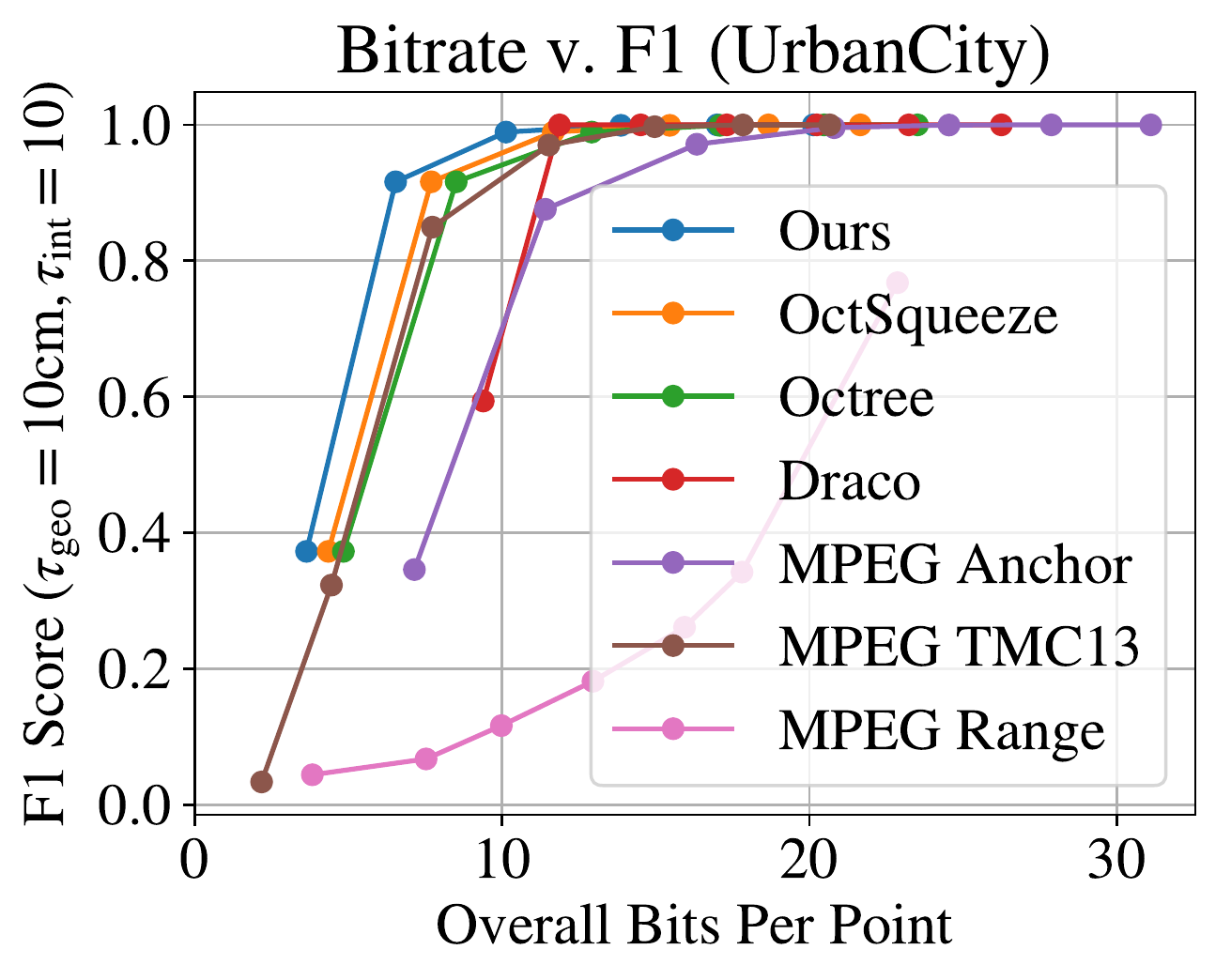} \\
\includegraphics[height=0.25\textwidth]{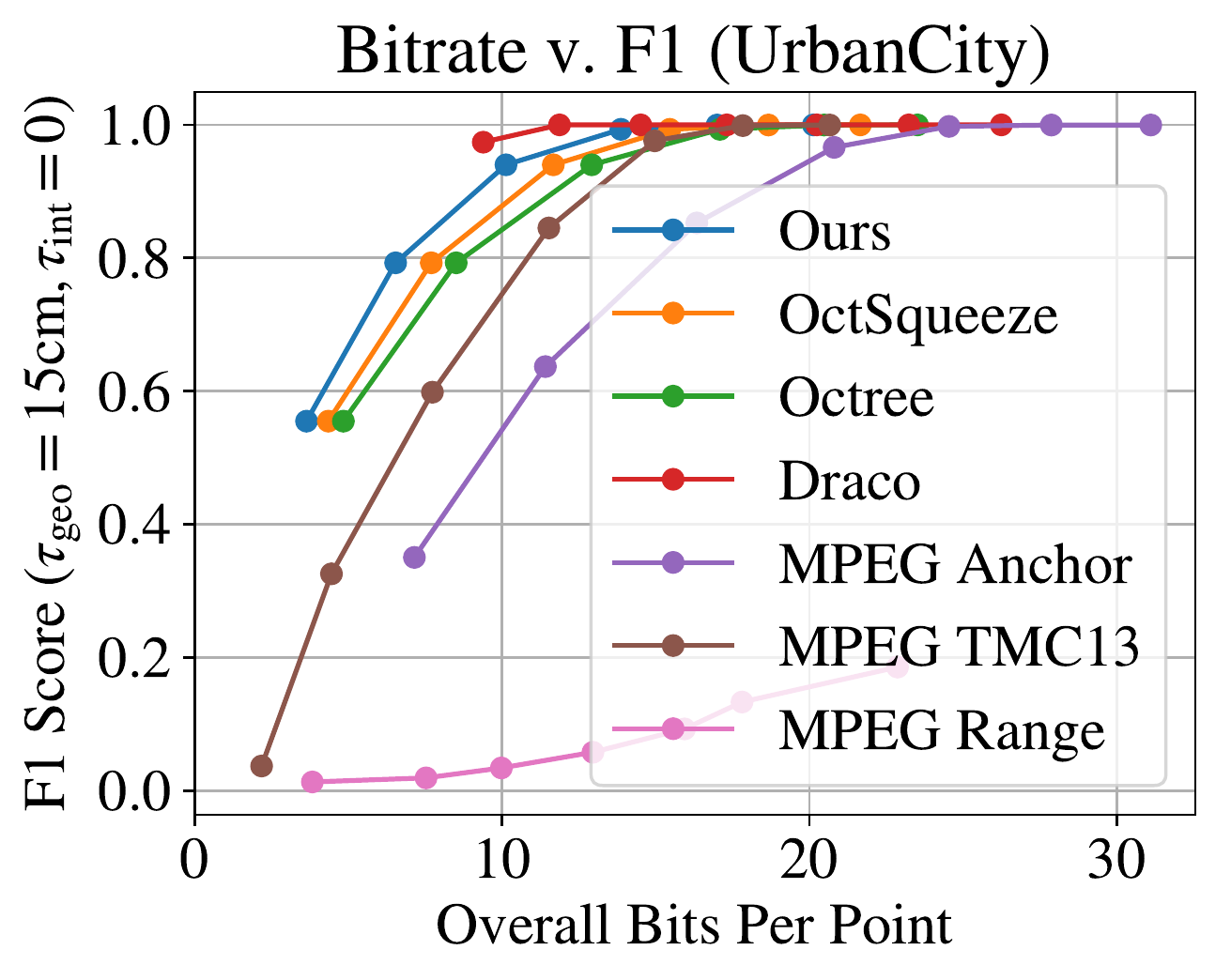}
\includegraphics[height=0.25\textwidth]{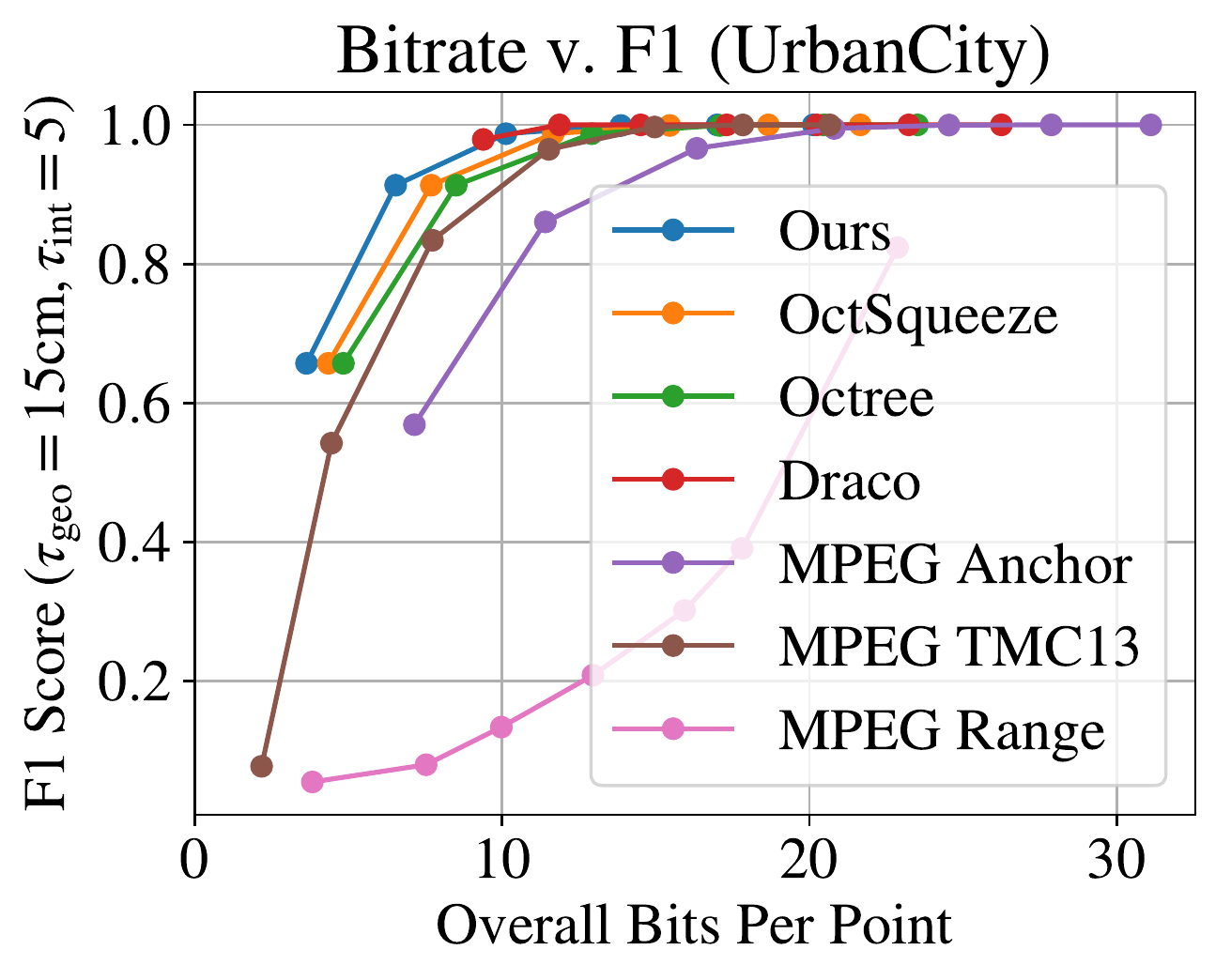}
\includegraphics[height=0.25\textwidth]{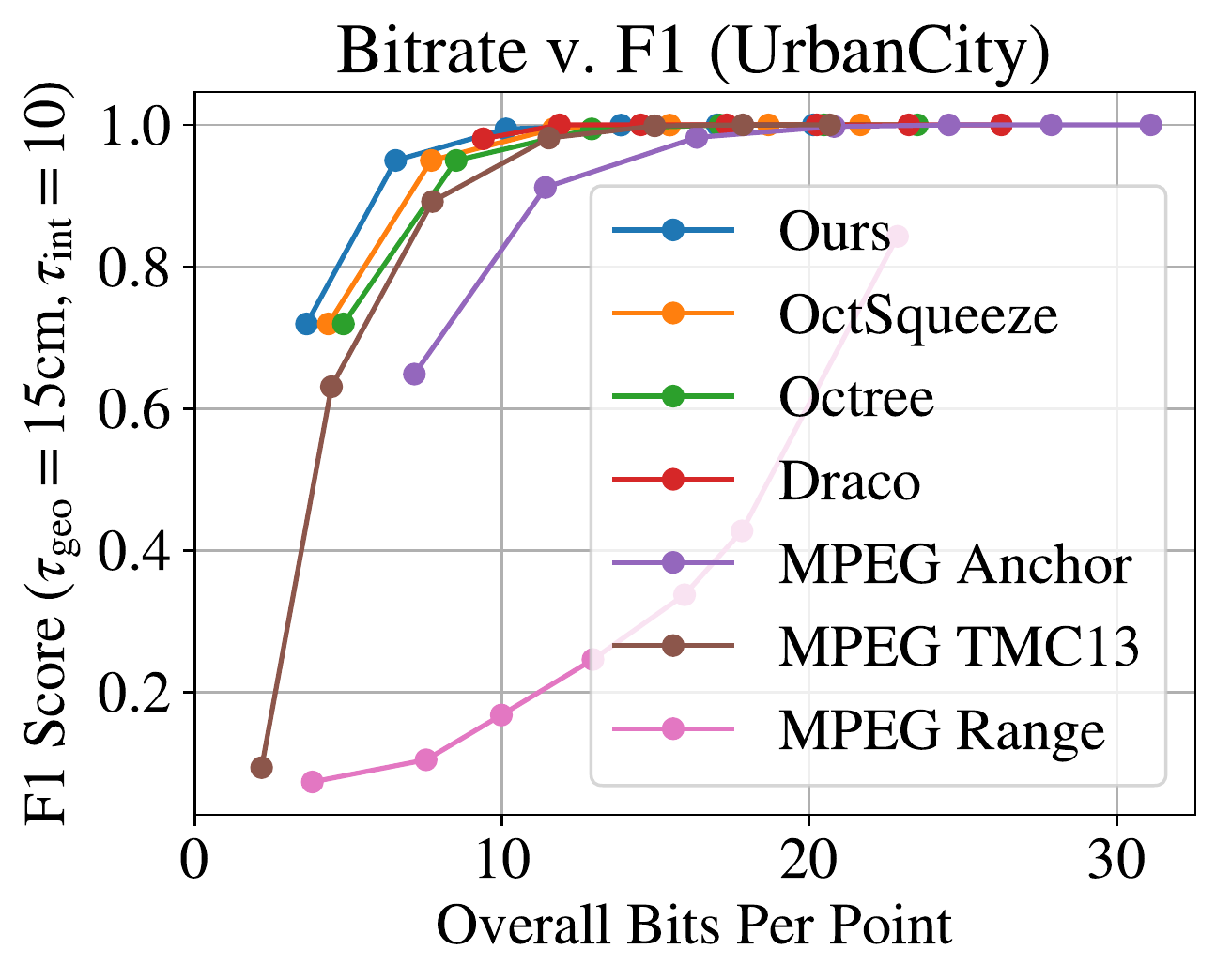} \\

\includegraphics[height=0.25\textwidth]{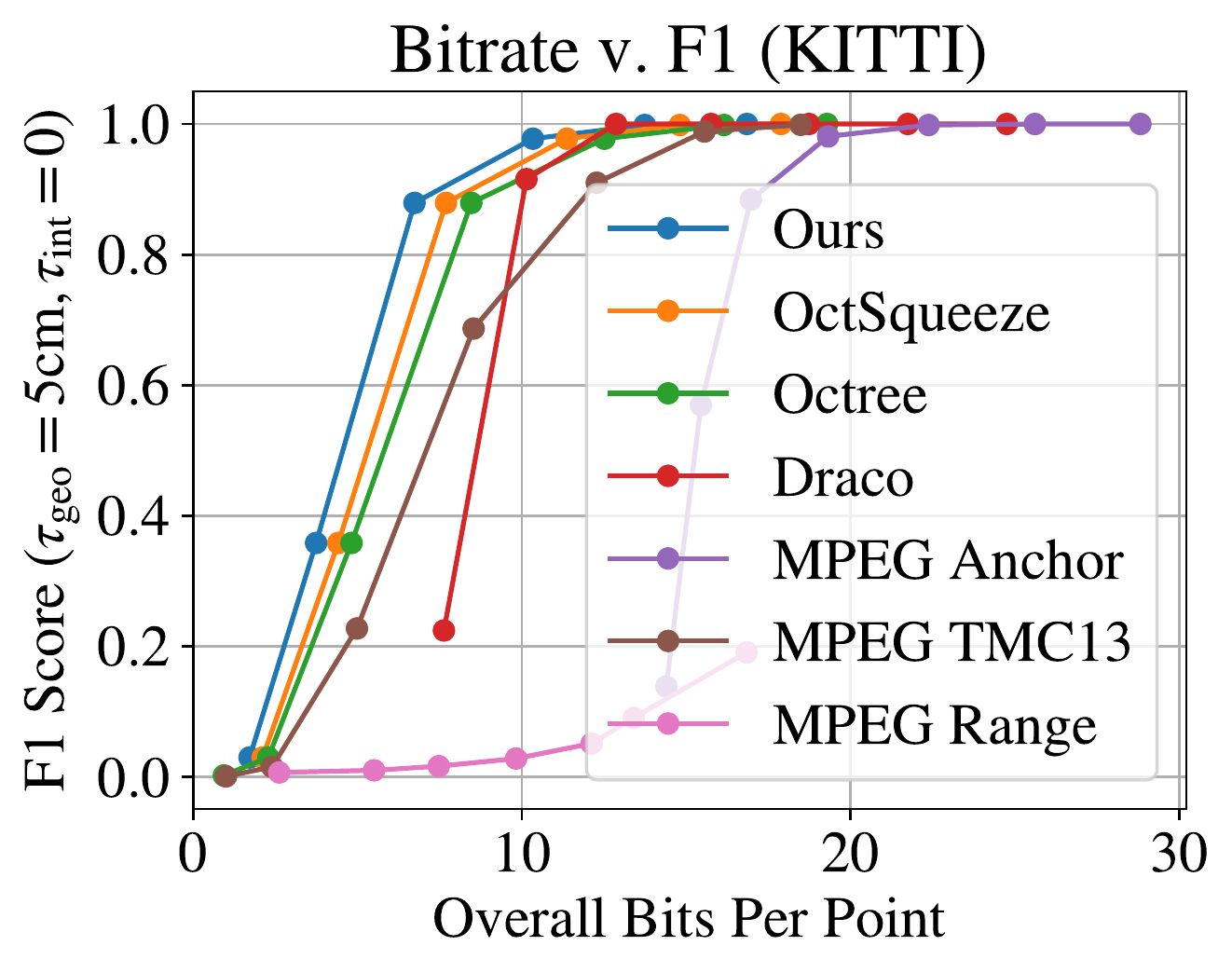}
\includegraphics[height=0.25\textwidth]{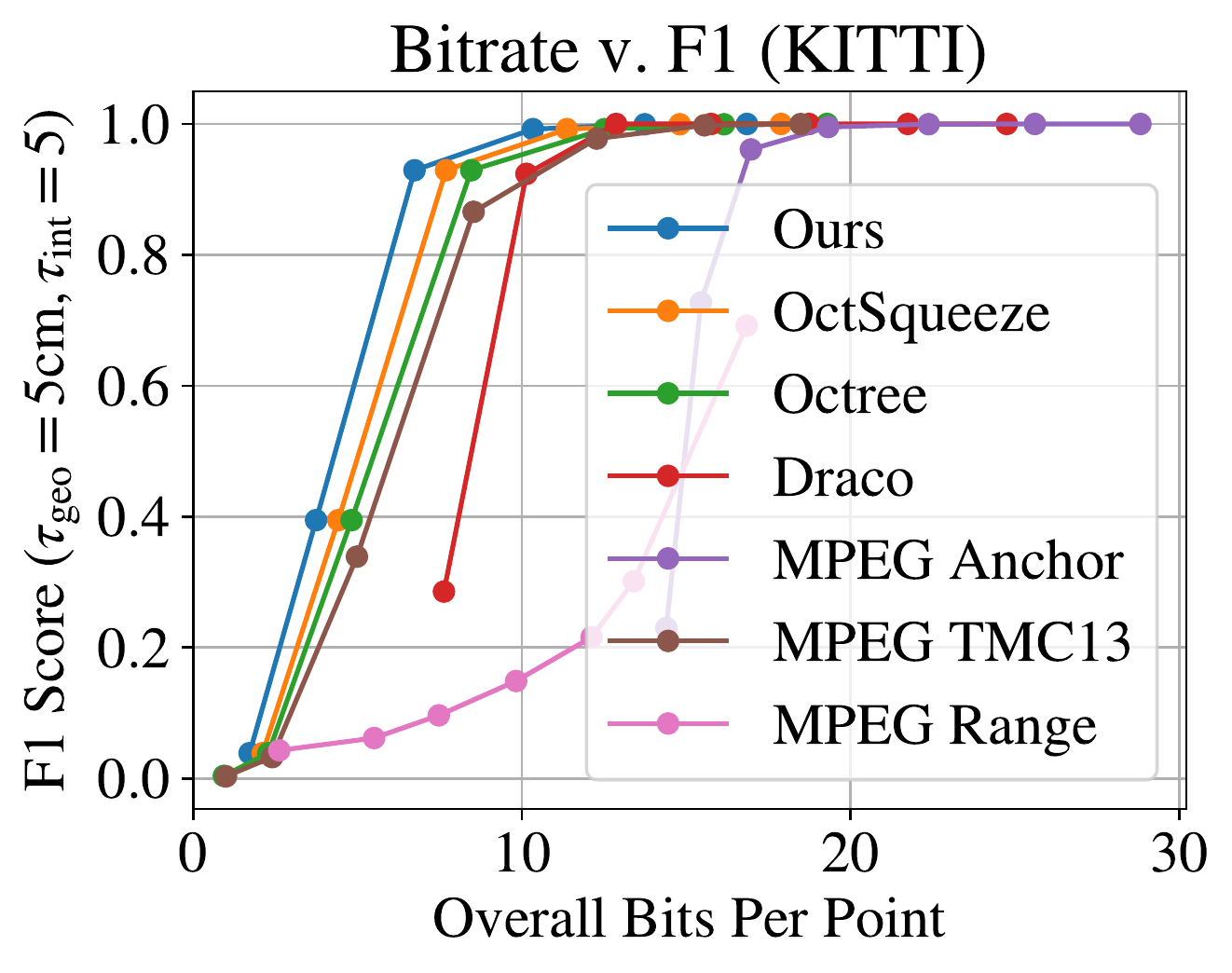}
\includegraphics[height=0.25\textwidth]{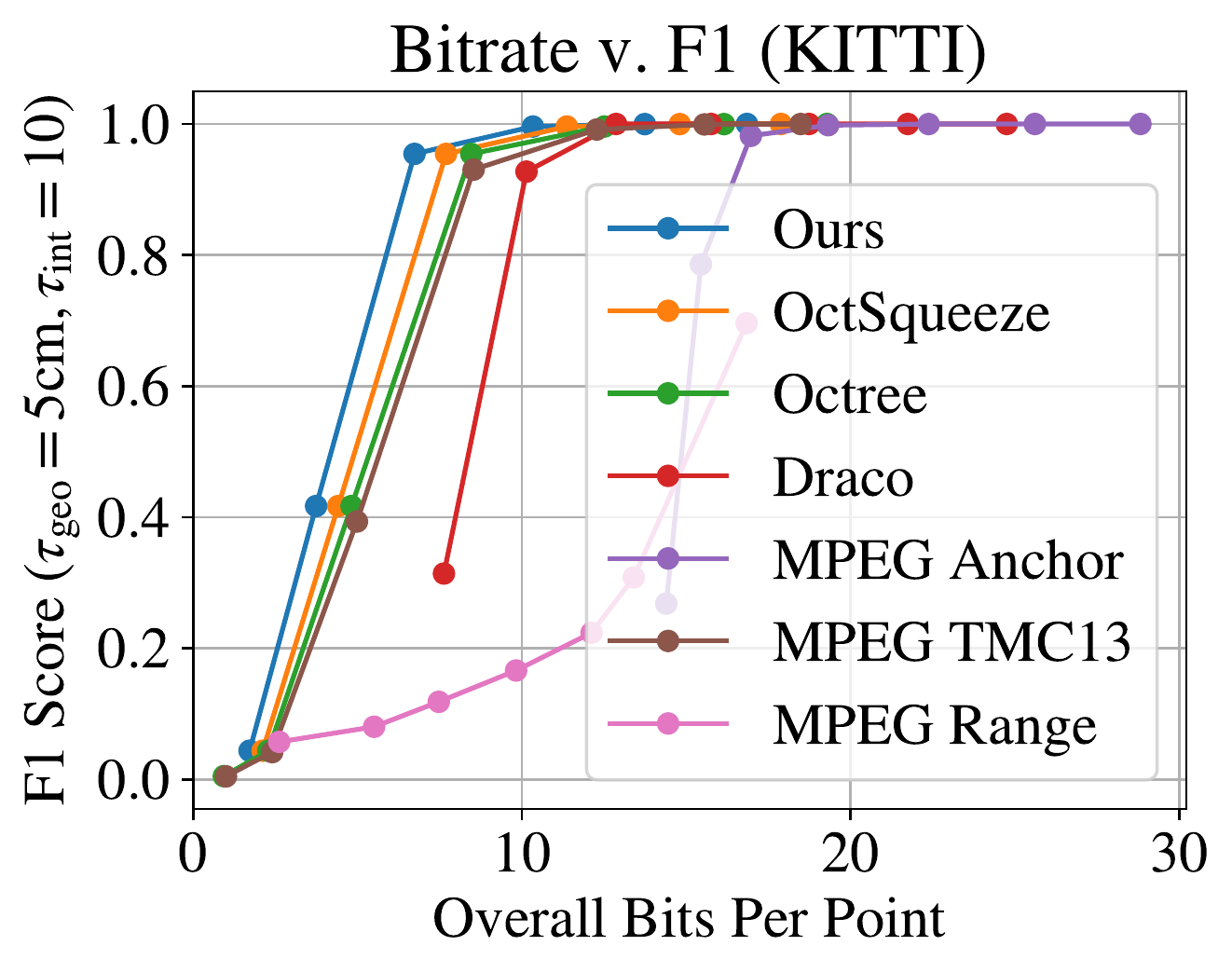} \\
\includegraphics[height=0.25\textwidth]{./figures/quantitative_results/kitti_supp_f1_010_000.pdf}
\includegraphics[height=0.25\textwidth]{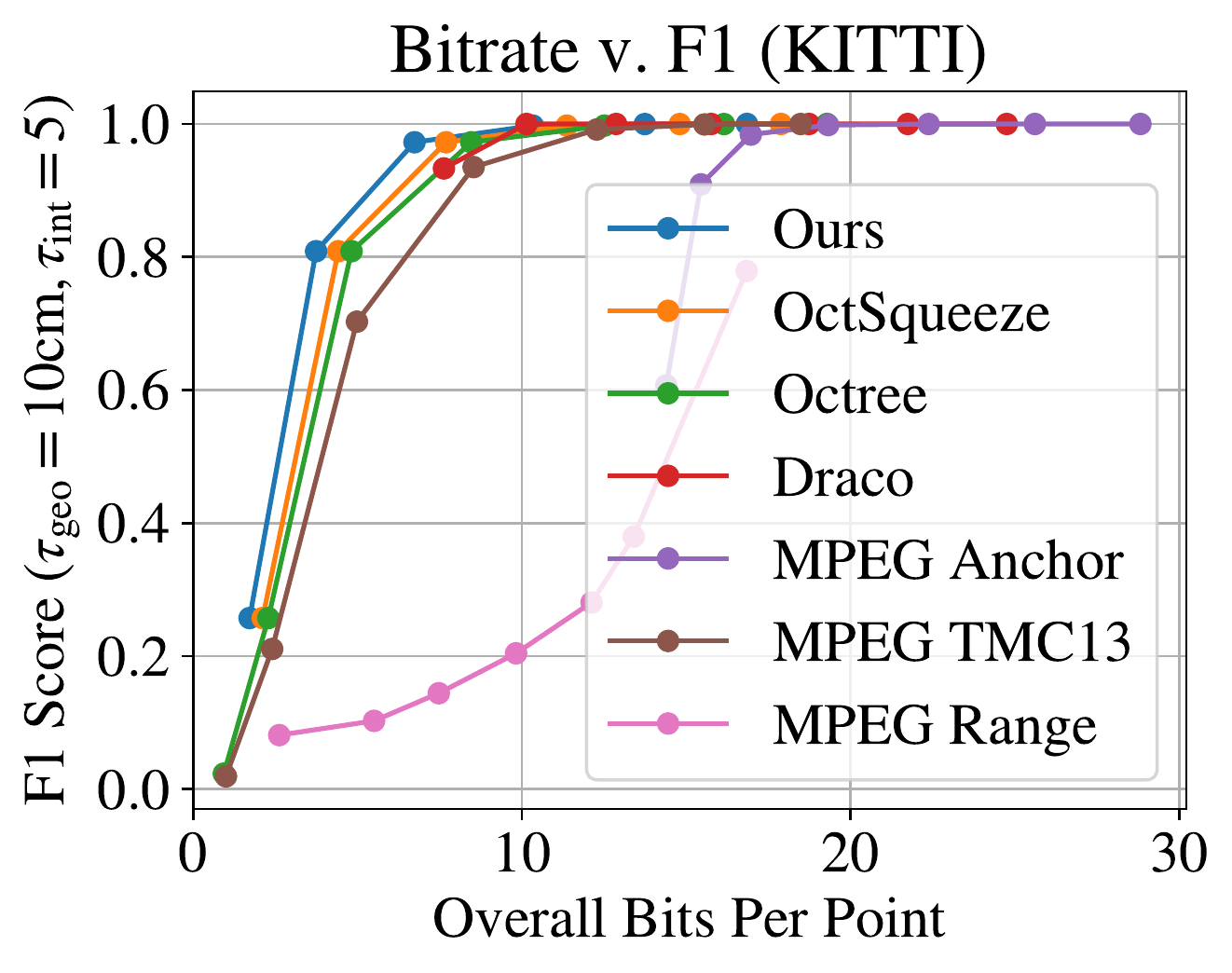}
\includegraphics[height=0.25\textwidth]{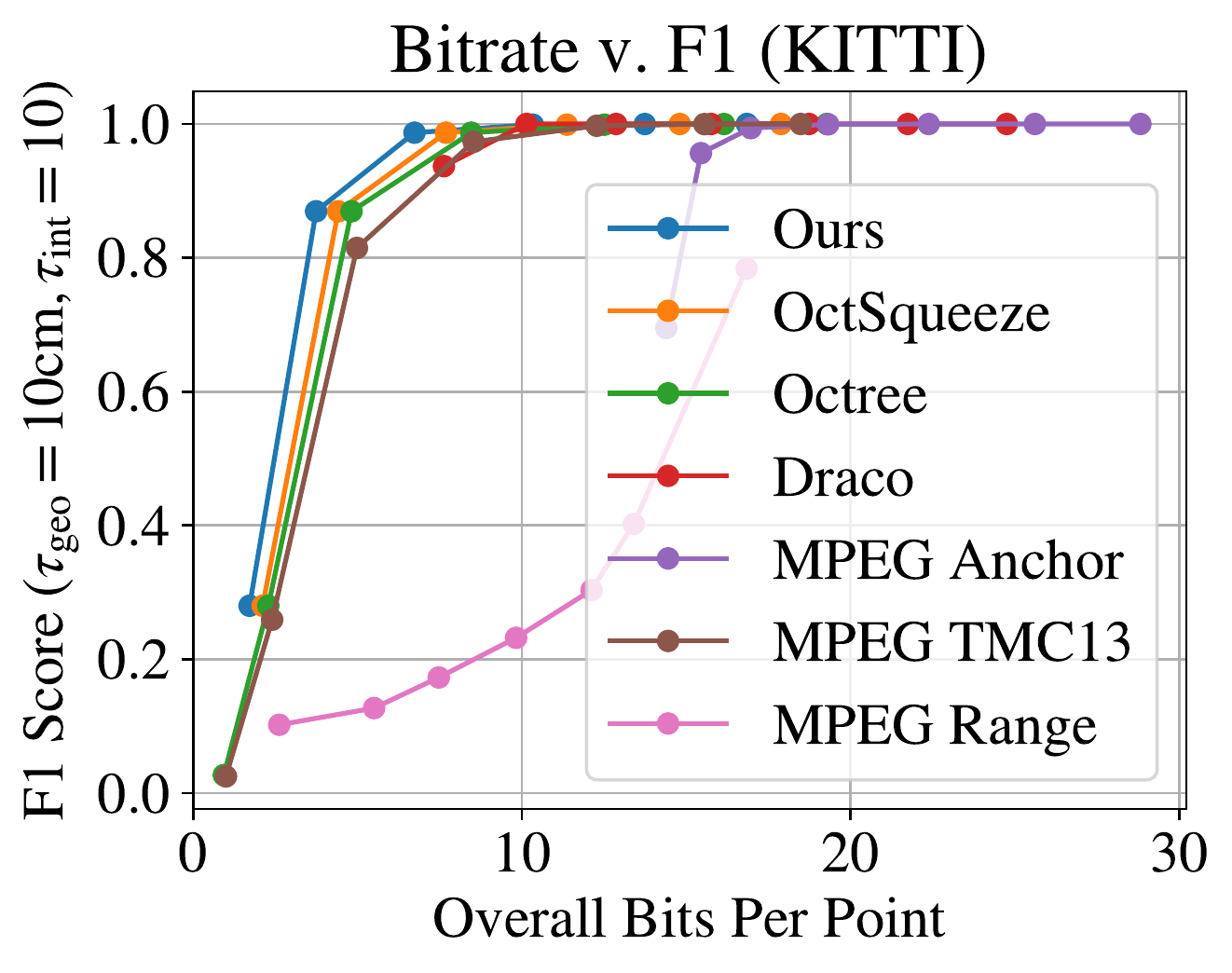} \\
\includegraphics[height=0.25\textwidth]{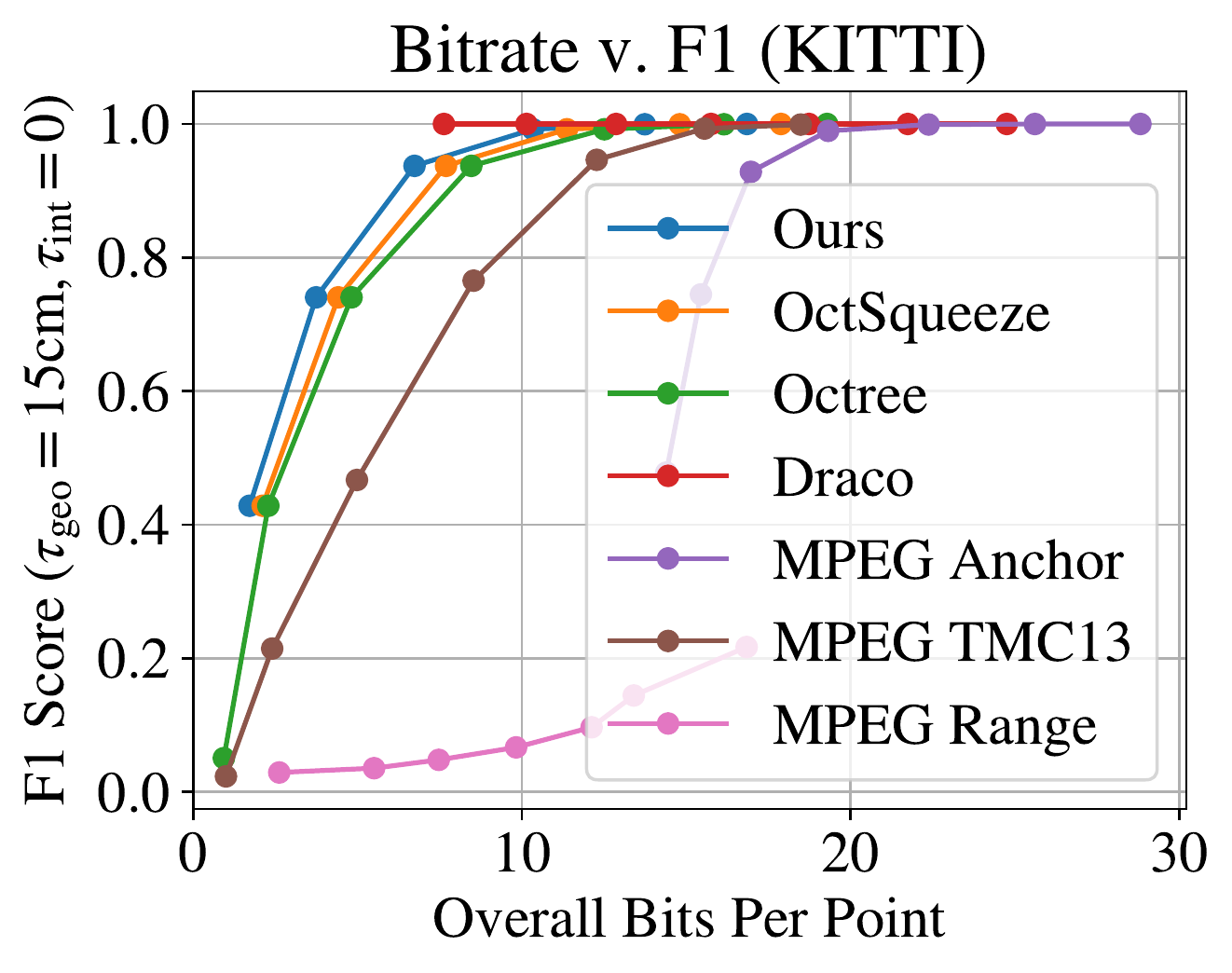}
\includegraphics[height=0.25\textwidth]{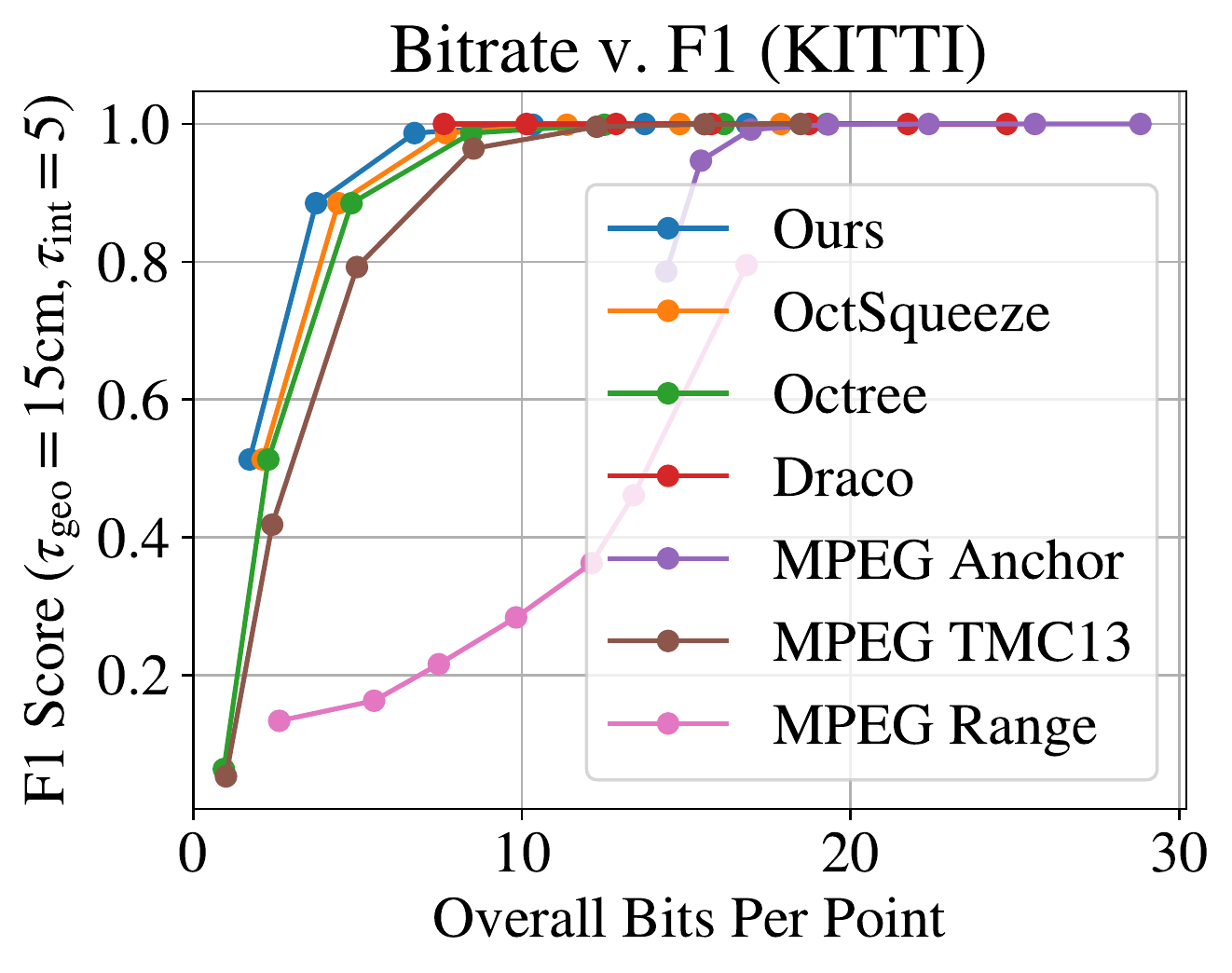}
\includegraphics[height=0.25\textwidth]{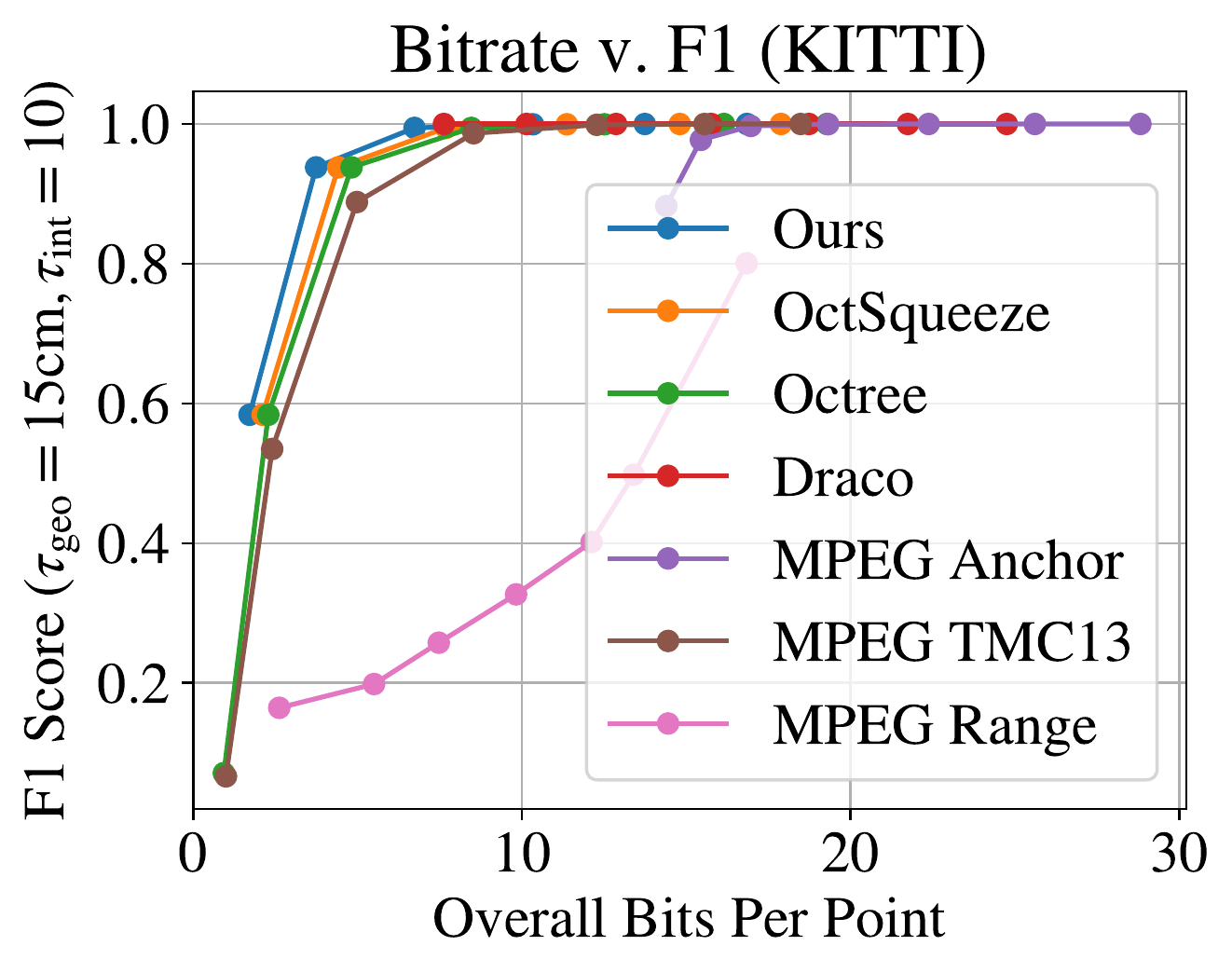} \\
\caption{
\small
Bitrate \vs $ \mathrm{F}_1 $ curves on \northamerica~(top three rows) and KITTI (bottom three rows).
We report $ \mathrm{F}_1 $ across various spatial and intensity thresholds:
$ \tau_\mathrm{geo} \in \{5\mathrm{cm}, 10\mathrm{cm}, 15\mathrm{cm}\} $ and
$ \tau_\mathrm{int} \in \{0, 5, 10\} $.
}
\label{figure:f1-hyperparameter-sweep}
\vspace{-10pt}
\end{figure}

\subsection{Reconstruction Metrics}
\label{section:reconstruction-metrics}
In Sec. 3.3 of the main text, we report reconstruction quality in terms of three
metrics: $ \mathrm{F}_1 $ score, point-to-point Chamfer Distance~\cite{huang2020},
and point-to-plane PSNR~\cite{tian2017}.
In the following, we explain each metric in detail.
Let $ \cP = \{(\bp_i, r_i)\}_{i = 1}^{N} $ be an input LiDAR point cloud, where
each $ \bp_i \in \mathbb{R}^3 $ denotes a point's spatial coordinates and
$ \bbr_i \in \{0, \ldots, 255\} $ its intensity.
Furthermore, let $ \hat{\cP} = \{(\hat{\bp}_j, \hat{\bbr}_j)\}_{j = 1}^{M} $ be
its reconstruction, where $ \hat{\bp}_j $ and $ \hat{\bbr}_j $ are similarly
defined.

Our first metric is an $ \mathrm{F}_1 $ score that measures reconstruction quality in
terms of both geometry and intensity:
\begin{equation}
    \mathrm{F}_1(\cP, \hat{\cP}) = \frac{2 \times \text{\# true positives}}{2 \times \text{\# true positives} + \text{\# false positives} + \text{\# false negatives}}
\end{equation}
where a reconstructed point $ (\hat{\bp}_j, \hat{\bbr}_j) \in \hat{\cP} $
is a \emph{true positive} if and only if there exists a point $ (\bp_i, \bbr_i) \in \cP $
such that $ \|\bp_i - \hat{\bp}_j\|_2 \leq \tau_\mathrm{geo} $ and
$ |\bbr_i - \hat{\bbr}_j| \leq \tau_\mathrm{int} $.
\emph{False positives} are the reconstructed points in $ \hat{\cP} $ that are
not true positives, and \emph{false negatives} are the original points in
$ \cP $ for which no reconstructed point is a true positive.
In our experiments, we use $ \tau_{\mathrm{geo}} = 10\mathrm{cm} $ and
$ \tau_{\mathrm{int}} = 0 $, and we report $ \mathrm{F}_1 $ as a function of
overall bitrates; \ie, the number of bits to store $ \bp $ and $ \bbr $.
We further report the $ \mathrm{F}_1 $ score for
$ \tau_\mathrm{geo} \in \{5\mathrm{cm}, 10\mathrm{cm}, 15\mathrm{cm}\} $ and
$ \tau_\mathrm{int} \in \{0, 5, 10\} $ in Fig.~\ref{figure:f1-hyperparameter-sweep}.

Following the MPEG standards, we also use two standard metrics that measure
reconstruction quality in terms of geometry only~\cite{schwarz2019}.
We report these metrics as a function of spatial bitrates; \ie, the number of
bits to store $ \bp $.
The first such metric measures the point-to-point error between the
original point cloud $ \cP $ and the reconstructed point cloud $ \hat{\cP} $;
this metric is often called the D1 error in the MPEG standards.
In our paper, we report this metric as a symmetric Chamfer distance:
\begin{align}
&\mathrm{CD}_\mathrm{sym}(\cP, \hat{\cP}) = \max \left\{\mathrm{CD}(\cP, \hat{\cP}), \mathrm{CD}(\hat{\cP}, \cP)\right\} \\
\text{where}~&\mathrm{CD}(\cP, \hat{\cP}) = \frac{1}{|\cP|} \sum_{\bp_i \in \cP} \min_{\hat{\bp}_j \in \hat{\cP}} \|\bp_i - \hat{\bp}_j \|_2
\end{align}

The second metric measures the point-to-place error between the original
point cloud $ \cP $ and the reconstructed point cloud $ \hat{\cP} $;
this metric is often called the D2 error in the MPEG standards.
In our paper, we report this metric in terms of its peak signal-to-noise ratio (PSNR):
\begin{align}
\mathrm{PSNR}(\cP, \hat{\cP}) = 10 \log_{10} \frac{3 r^2}{\max \{\mathrm{MSE}(\cP, \hat{\cP}), \mathrm{MSE}(\cP, \hat{\cP})\}}
\end{align}
where $ \mathrm{MSE}(\mathcal{P}, \hat{\mathcal{P}}) =
\frac{1}{|\mathcal{P}|} \sum_{i} ((\bp_i - \hat{\bp}_i) \cdot \hat{\bn}_i)^2 $
~is the mean squared point-to-plane distance,
$ \hat{\bn}_i $ is the normal vector on $ \hat{\bp}_i $,
~$ \hat{\bp}_i = \mathrm{argmin}_{\hat{\bp} \in \hat{\cP}} \|\bp_i - \hat{\bp}\|_2^2 $
~is $ \bp_i $'s nearest neighbor point in $ \hat{\cP} $,
and $ r $ is the peak constant value.
We estimate the normal $ \bm{n}_i $ at each point $ \bp_i \in \cP $
using the Open3D function \verb+estimate_normals+ with $ k = 12 $ nearest
neighbors~\cite{zhou2018}, and we compute the normal $ \hat{\bm{n}}_i $
corresponding to each point $ \hat{\bp}_i \in \hat{\cP} $ by taking the
normal of its nearest neighbor in the original point cloud $ \cP $.
Following the MPEG standard, for each dataset, we compute $ r $ as the maximum
nearest neighbor distance among all point clouds in the dataset:
\begin{align}
    r = \max_{\cP } \max_{\bp_i \in \cP} \min_{j \neq i } \|\bp_i - \bp_j\|_2
\end{align}

For {\northamerica}, we use $ r = 98.69 $ and for SemanticKITTI, we use $ r = 59.70 $.

\begin{figure}[!t]
\centering
\includegraphics[height=0.25\textwidth]{./figures/quantitative_results/na_supp_d1_cd.pdf}
\includegraphics[height=0.25\textwidth]{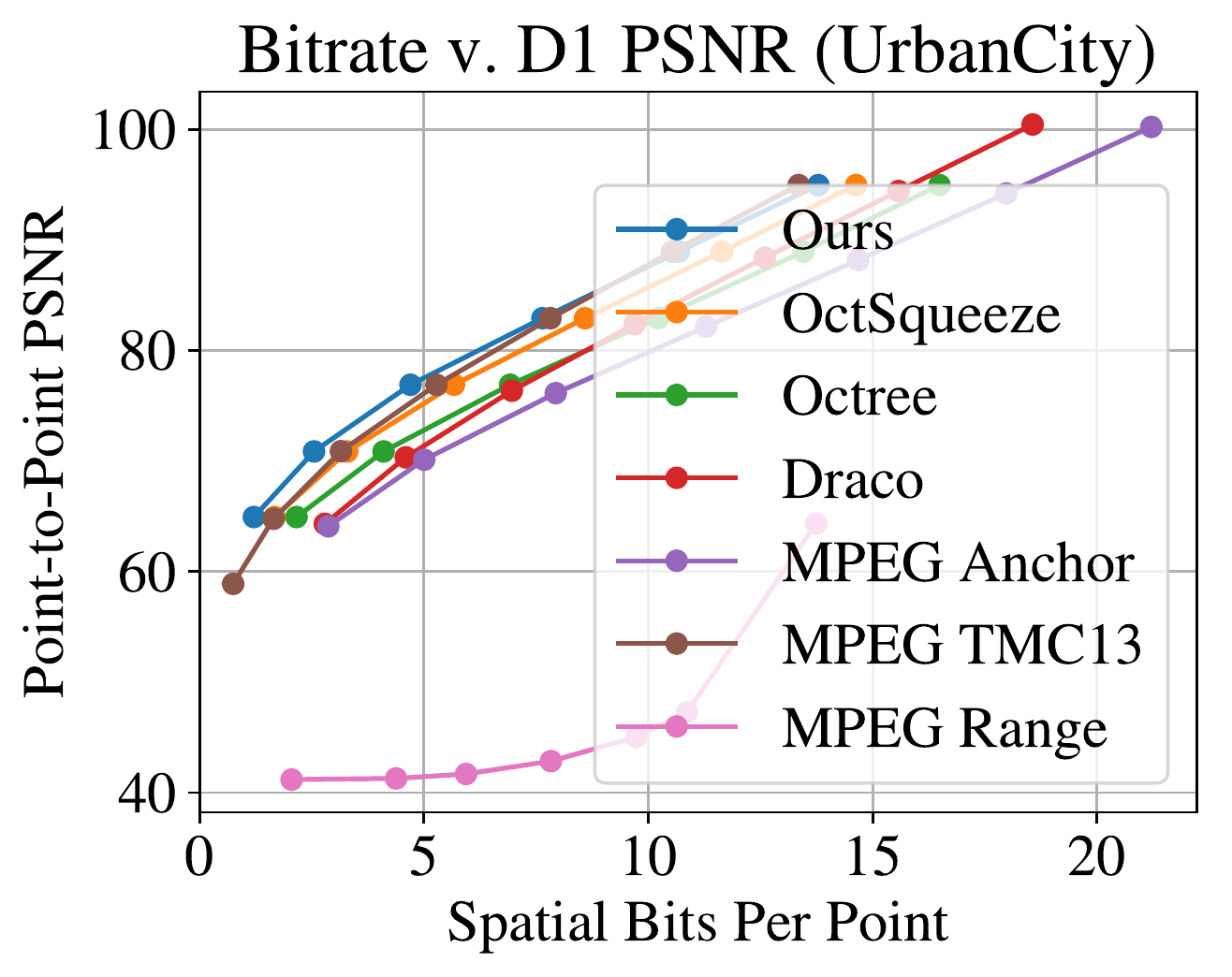} \\
\includegraphics[height=0.25\textwidth]{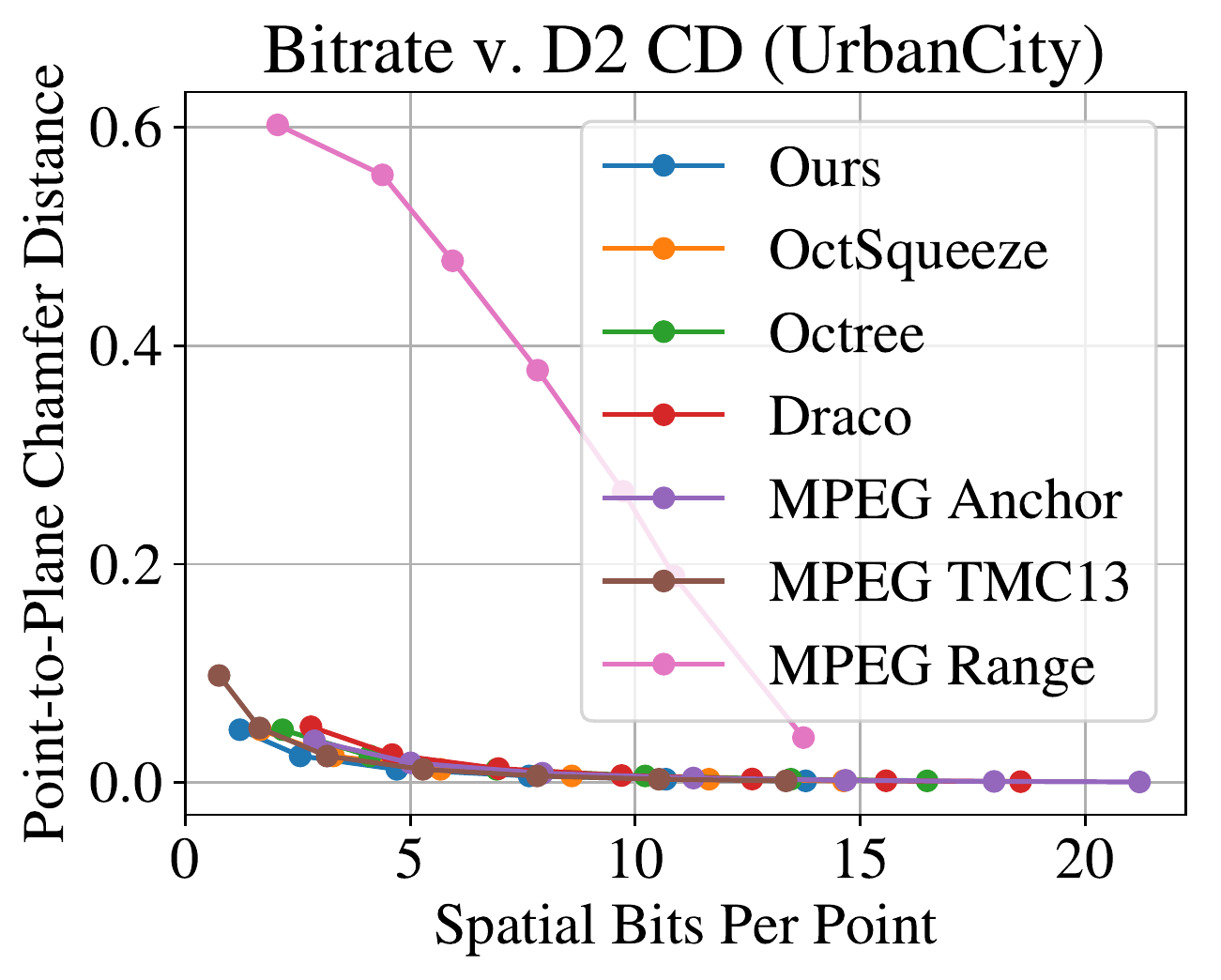}
\includegraphics[height=0.25\textwidth]{./figures/quantitative_results/na_supp_d2_psnr.pdf} \\
\includegraphics[height=0.25\textwidth]{./figures/quantitative_results/kitti_supp_d1_cd.pdf}
\includegraphics[height=0.25\textwidth]{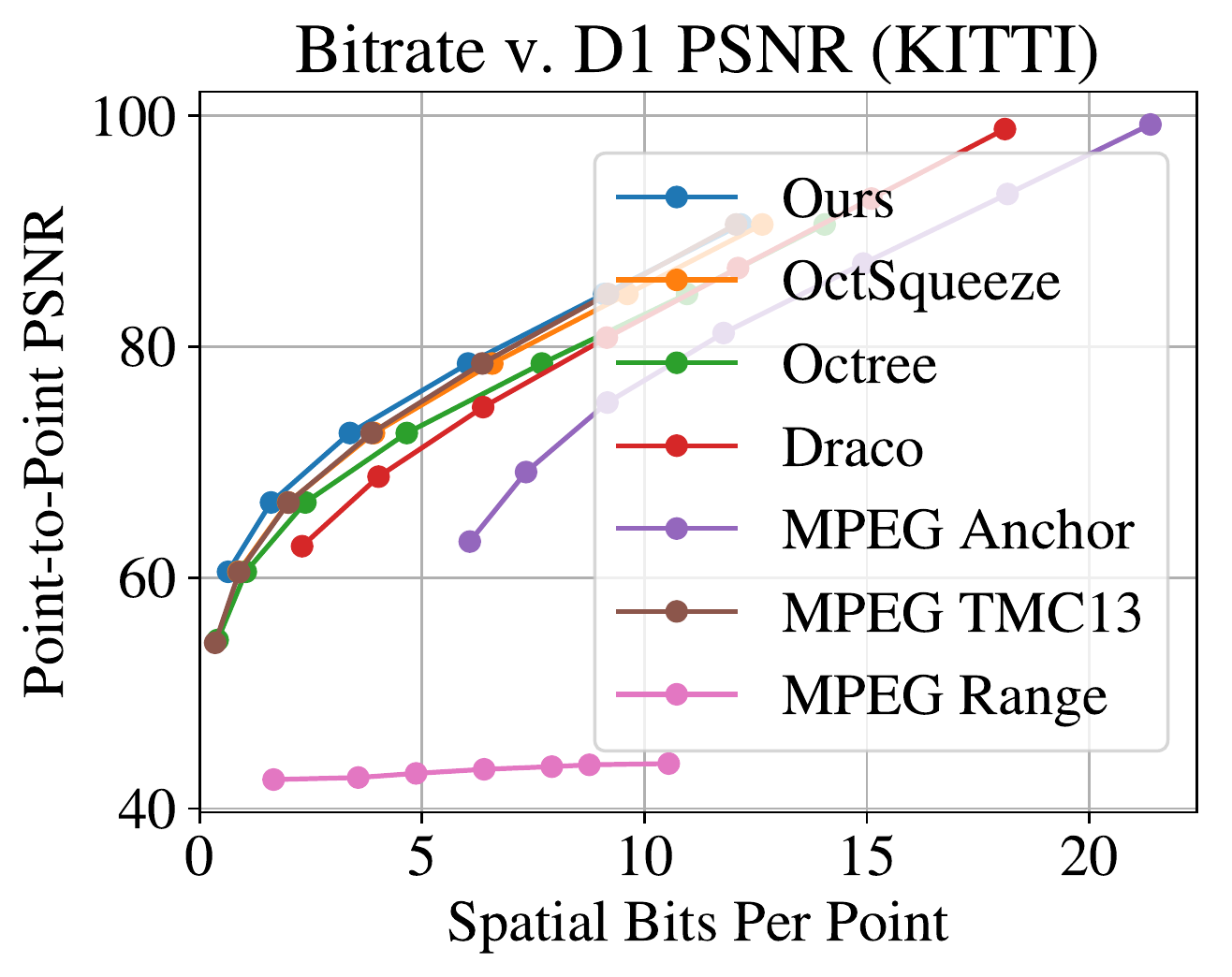} \\
\includegraphics[height=0.25\textwidth]{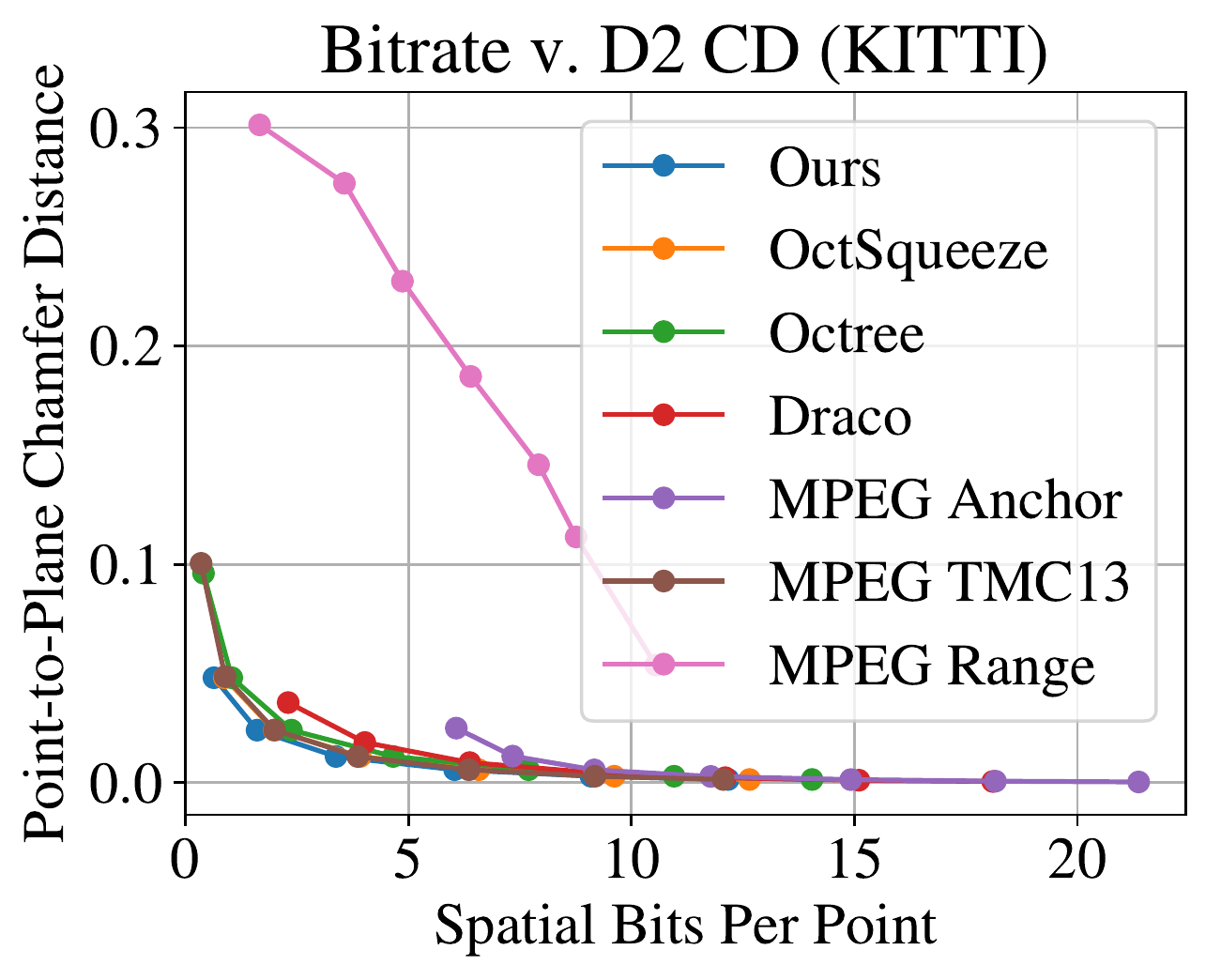}
\includegraphics[height=0.25\textwidth]{./figures/quantitative_results/kitti_supp_d2_psnr.pdf} \\
\caption{
\small
Bitrate \vs reconstruction curves on \northamerica~(top two rows) and KITTI (bottom two rows).
We report point-to-point (D1) and point-to-plane (D2) errors in terms
of Chamfer distances (left) and PSNR (right).
}
\label{figure:geometry-distortion-sweep}
\vspace{-10pt}
\end{figure}

For completeness, we also report the point-to-point error in terms of its
peak signal-to-noise ratio and the point-to-plane error as a symmetric
Chamfer distance in Fig.~\ref{figure:geometry-distortion-sweep}.

\subsection{Downstream Experiment Details}
\label{section:downstream-details}

In this section, we provide additional details for our downstream perception
experiments.

\subsubsection{Semantic Segmentation}
We use a modified version of the LiDAR semantic segmentation model described
in~\cite{huang2020}.

\paragraph{Input Representation:}
Our model takes as input $ T $ bird's eye view (``BEV'') occupancy grids of
the past $ T $ input LiDAR point clouds $ \{\cP^{(t - T + 1)}, \ldots, \cP^{(t)}\} $,
stacked along the height dimension (\ie, the $ z $-axis).
By treating the height dimension as multi-dimensional input features,
we have a compact input representation on which we can use
2D convolutions~\cite{zhang2018}.
Each voxel in the occupancy grids store the average intensity value of the
points occupying its volume, or 0 if it contains no points.
We use a region of interest of $ 160\mathrm{m} \times 160\mathrm{m} \times 5\mathrm{m} $
centered on the ego-vehicle, $ T = 5 $ past LiDAR point clouds,
and a voxel resolution of $ 0.15625\mathrm{cm} $,
yielding an input volume $ \boldsymbol{x} $ of size
$ (T \times Z) \times W \times H = 160 \times 1024 \times 1024$.

\paragraph{Architecture Details:}
Our model architecture consists of two components: (1) a backbone feature
extractor; and (2) a semantic segmentation head.
The backbone feature extractor $ \mathrm{CNN}_\mathrm{BEV} $ is a feature pyramid network based on the
backbone architecture of~\cite{liang_pnpnet}:
\begin{equation}
    \boldsymbol{f}_\mathrm{BEV} = \mathrm{CNN}_\mathrm{BEV}(\boldsymbol{x})
\end{equation}
where $ \boldsymbol{f}_\mathrm{BEV} \in \mathbb{R}^{C_\mathrm{BEV} \times W / 4 \times H / 4} $
and $ C_\mathrm{BEV} = 256 $.

The semantic segmentation head $ \mathrm{CNN}_\mathrm{sem} $ consists of four
2D convolution blocks with 128 hidden channels~\footnote{Each 2D convolution block consists of a
$ 3 \times 3 $ convolution, GroupNorm~\cite{wu2018}, and ReLU.},
followed by a $ 1 \times 1 $ convolution layer:
\begin{equation}
    \boldsymbol{f}_\mathrm{sem} = \mathrm{CNN}_\mathrm{sem}(\boldsymbol{f}_\mathrm{BEV})
\end{equation}
where $ \boldsymbol{f}_\mathrm{sem} \in \mathbb{R}^{(K \times Z) \times W / 4 \times H / 4} $
and $ K $ is the number of classes plus an additional \texttt{ignore} class.
To extract per-point predictions, we first reshape $ \boldsymbol{f}_\mathrm{sem} $
into a $ K \times Z \times W / 4 \times H / 4 $ logits tensor,
then use trilinear interpolation to extract per-point $ K $-dimensional logits,
and finally apply softmax.

\paragraph{Training Details:}
We use the cross-entropy loss to train our semantic segmentation model.
For SemanticKITTI, we follow~\cite{behley2019} and reweight the loss at each
point by the inverse of the frequency of its ground truth class; this helps to
counteract the effects of severe class imbalance.
Moreover, we use data augmentation by randomly scaling the point cloud by
$ s \sim \mathrm{Uniform}(0.95, 1.05) $,
rotating it by $ \theta \sim \mathrm{Uniform}(- \pi / 4, \pi / 4) $,
and reflecting it along the $ x $ and $ y $-axes.
We use the Adam optimizer~\cite{kingma2015} with a learning rate of
$ 4\mathrm{e}{-4} $ and a batch size of 12, and we train until convergence.

\subsubsection{Object Detection}
We use a modified version of the LiDAR object detection model described
in~\cite{huang2020}.
It largely follows the same architecture as our semantic segmentation model,
with a few modifications to adapt it for object detection.
We describe these modifications below.

\paragraph{Architecture Details:}
Our object detection model consists of two components: (1) a backbone feature
extractor; and (2) an object detection head.
The backbone feature extractor here shares an identical architecture to that
of the semantic segmentation model.
The object detection head consists of four 2D convolution blocks with 128 hidden
channels followed by a $ 1 \times 1 $ convolution layer to predict a bounding
box $ \boldsymbol{b}_{i, k} $ and detection score $ \alpha_{i, k} $ for every
BEV pixel $ i $ and class $ k $.
Each bounding box $ \boldsymbol{b}_{i, k} $ is parameterized by
$ (\Delta x, \Delta y, \log w, \log h, \sin \theta, \cos \theta) $, where
$ (\Delta x, \Delta y) $ are the position offsets to the object's center,
$ (w, h) $ are the width and height of its bounding box, and $ \theta $
is its heading angle.
To remove duplicate bounding boxes predictions, we use non-maximum suppression.

\paragraph{Training Details:}
We use a combination of classification and regression losses to train our
detection model.
In particular, for object classification, we use a binary cross-entropy loss
with online hard negative mining, where positive and negative BEV pixels are
determined based on their distance to an object center~\cite{yang2019}.
For bounding box regression, we use a smooth $ \ell_1 $ loss on
$ \Delta x, \Delta y, \log w, \log h, \sin \theta, \cos \theta $.
We use the Adam optimizer~\cite{kingma2015} with a learning rate of
$ 4\mathrm{e}{-4} $ and a batch size of 12, and we train until convergence.


\section{Additional Qualitative Results}
\label{section:qualitative-results}

\begin{figure}[!t]
\centering

\begin{overpic}[clip,trim=0cm 15cm 0cm 0cm,width=0.49\textwidth]{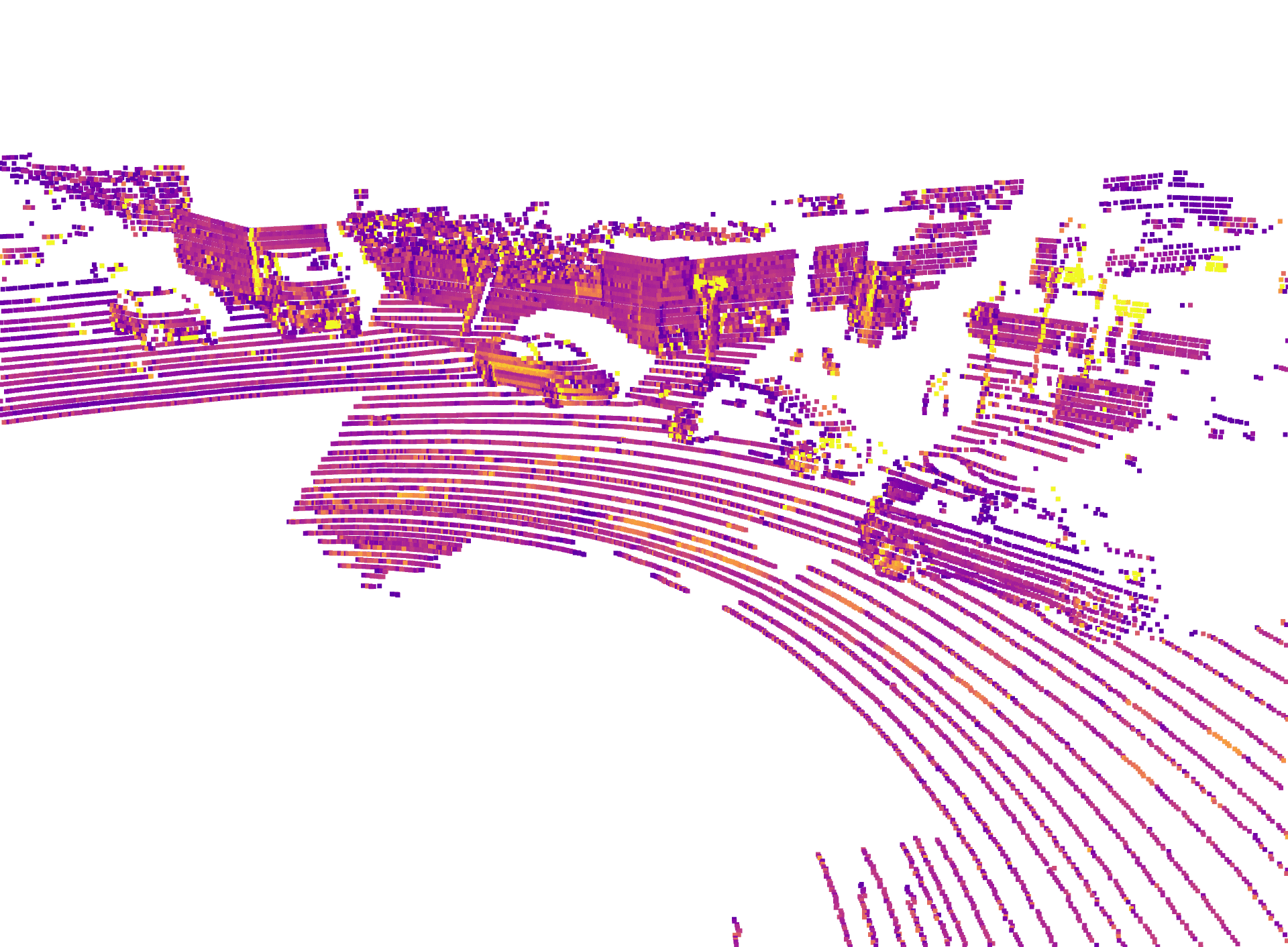}
    \put (0,48) {\colorbox{gray!30}{\scriptsize Oracle (\northamerica): Bitrate 104.0}}
\end{overpic}
\begin{overpic}[clip,trim=0cm 15cm 0cm 0cm,width=0.49\textwidth]{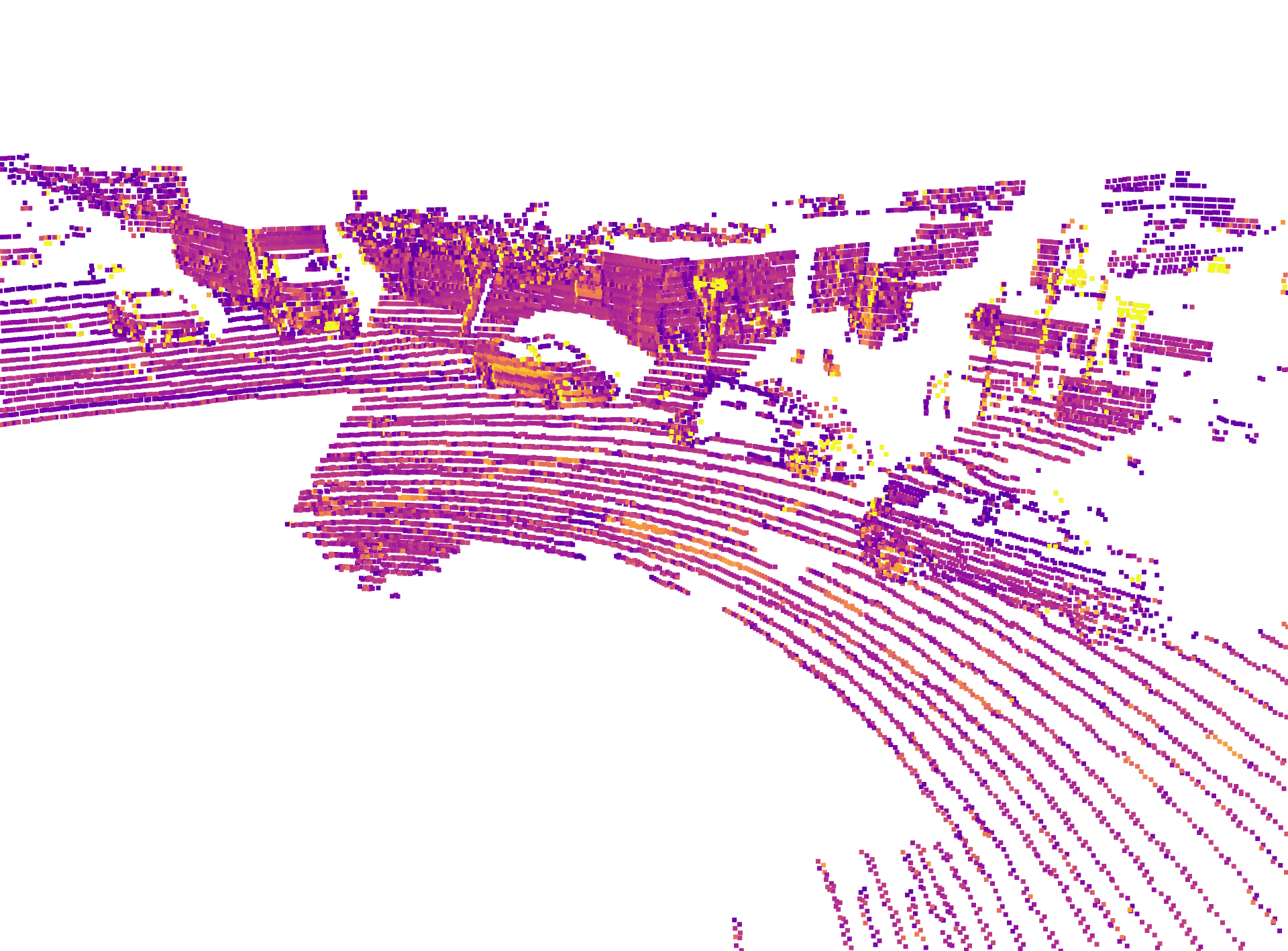}
    \put (0,48) {\colorbox{gray!30}{\scriptsize Ours: F1 92.9 Bitrate 10.0}}
\end{overpic}
\begin{overpic}[clip,trim=0cm 15cm 0cm 0cm,width=0.49\textwidth]{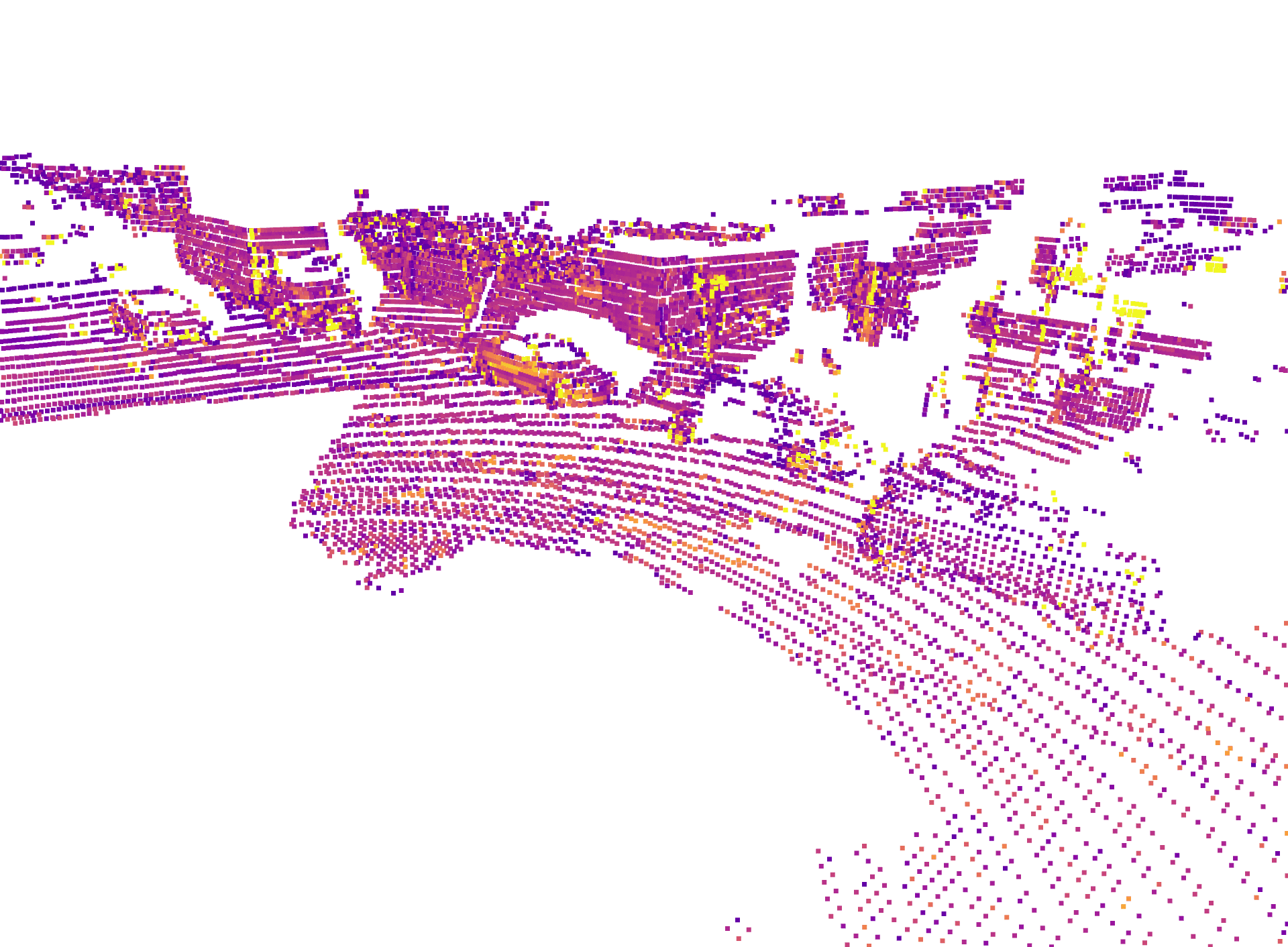}
    \put (0,48) {\colorbox{gray!30}{\scriptsize Draco: F1 85.1 Bitrate 10.9}}
\end{overpic}
\begin{overpic}[clip,trim=0cm 15cm 0cm 0cm,width=0.49\textwidth]{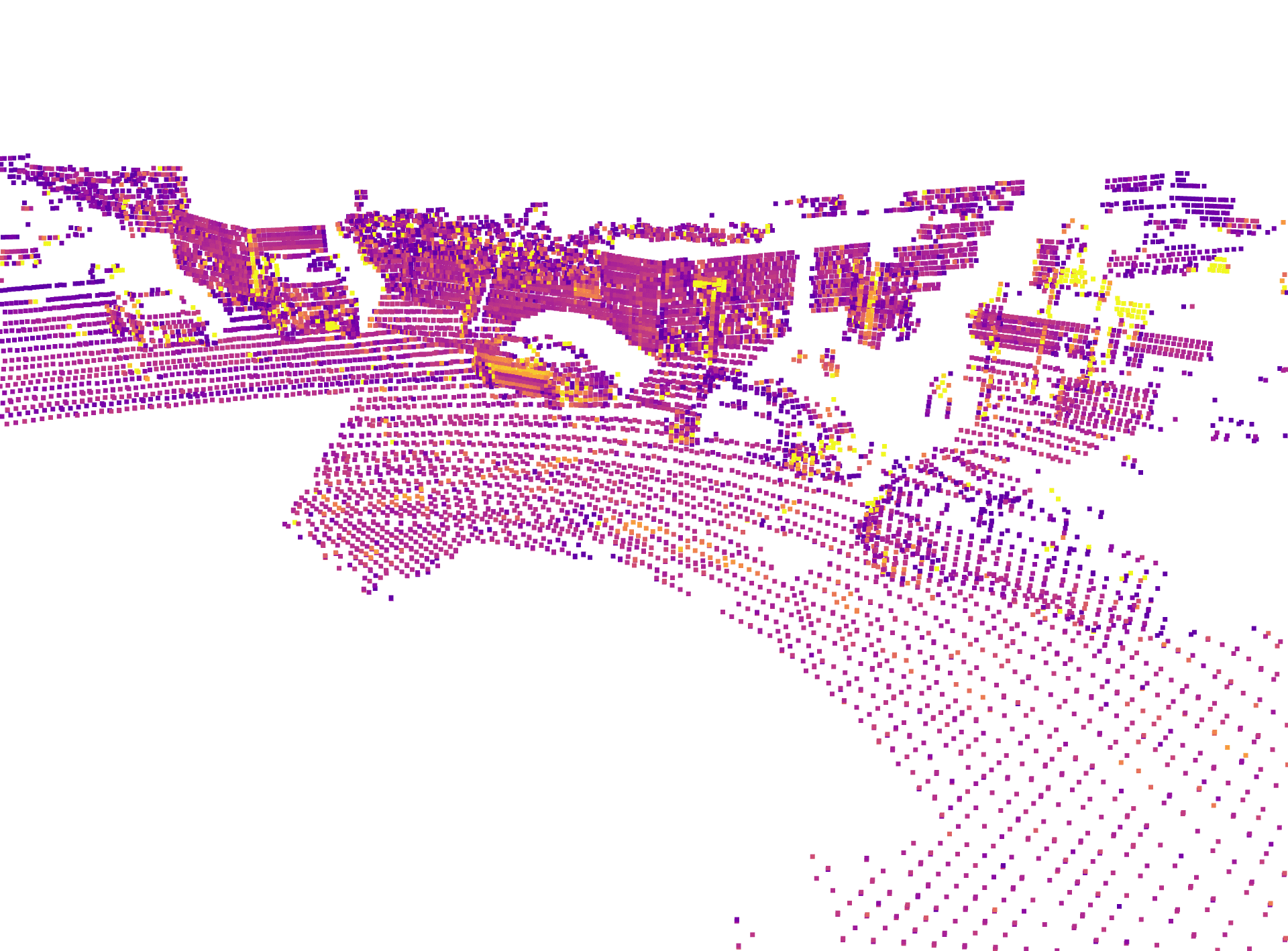}
    \put (0,48) {\colorbox{gray!30}{\scriptsize MPEG: F1 53.4 Bitrate 10.4}}
\end{overpic}

\caption{
\small Qualitative results on \northamerica.
Points are colored by intensity.
}
\label{figure:northamerica-additional-qualitative-results}
\end{figure}

\begin{figure}[!t]
\centering

\begin{overpic}[clip,trim=0cm 15cm 0cm 0cm,width=0.49\textwidth]{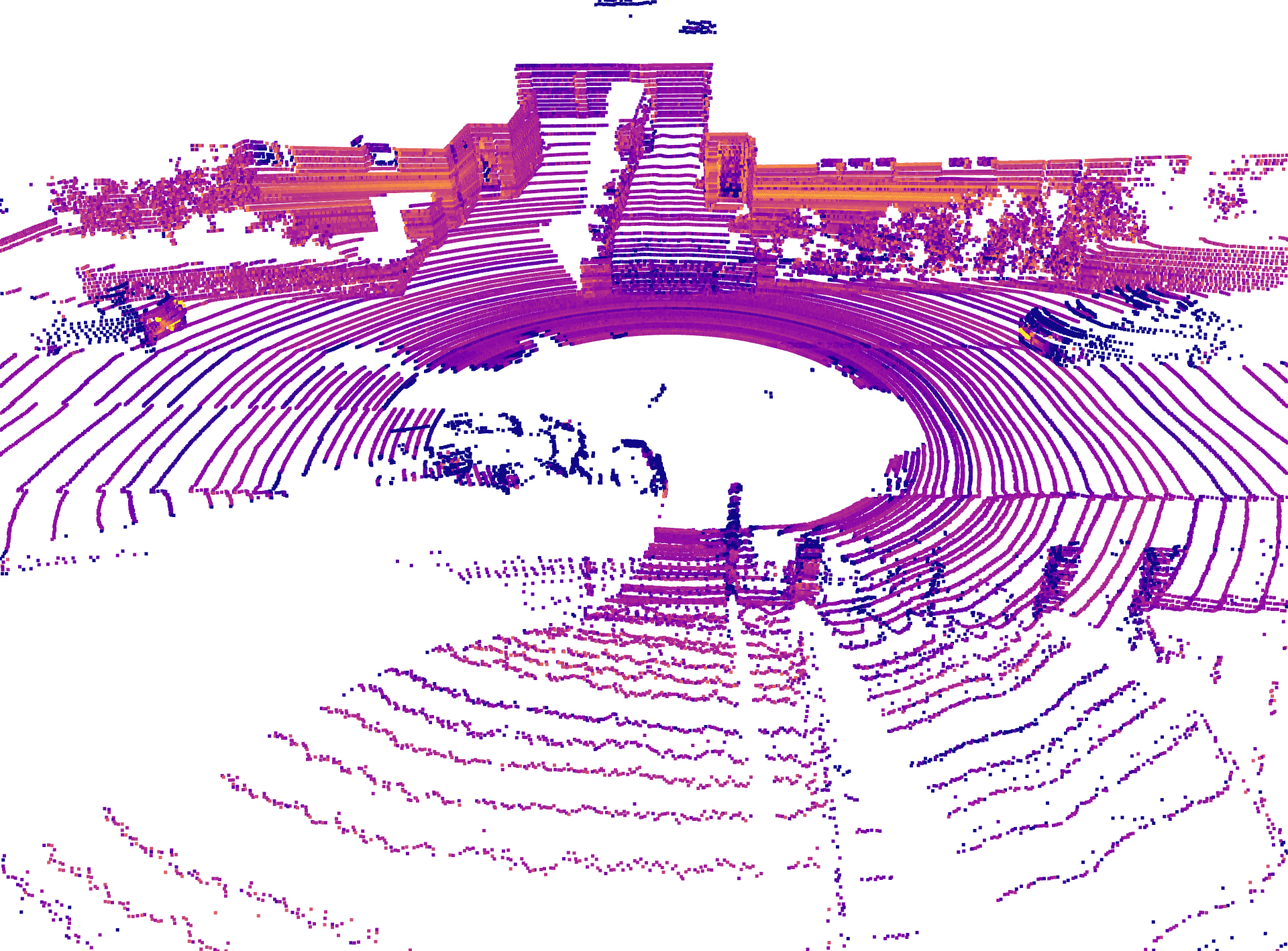}
    \put (0,48) {\colorbox{gray!30}{\scriptsize Oracle (KITTI): Bitrate 104.0}}
\end{overpic}
\begin{overpic}[clip,trim=0cm 15cm 0cm 0cm,width=0.49\textwidth]{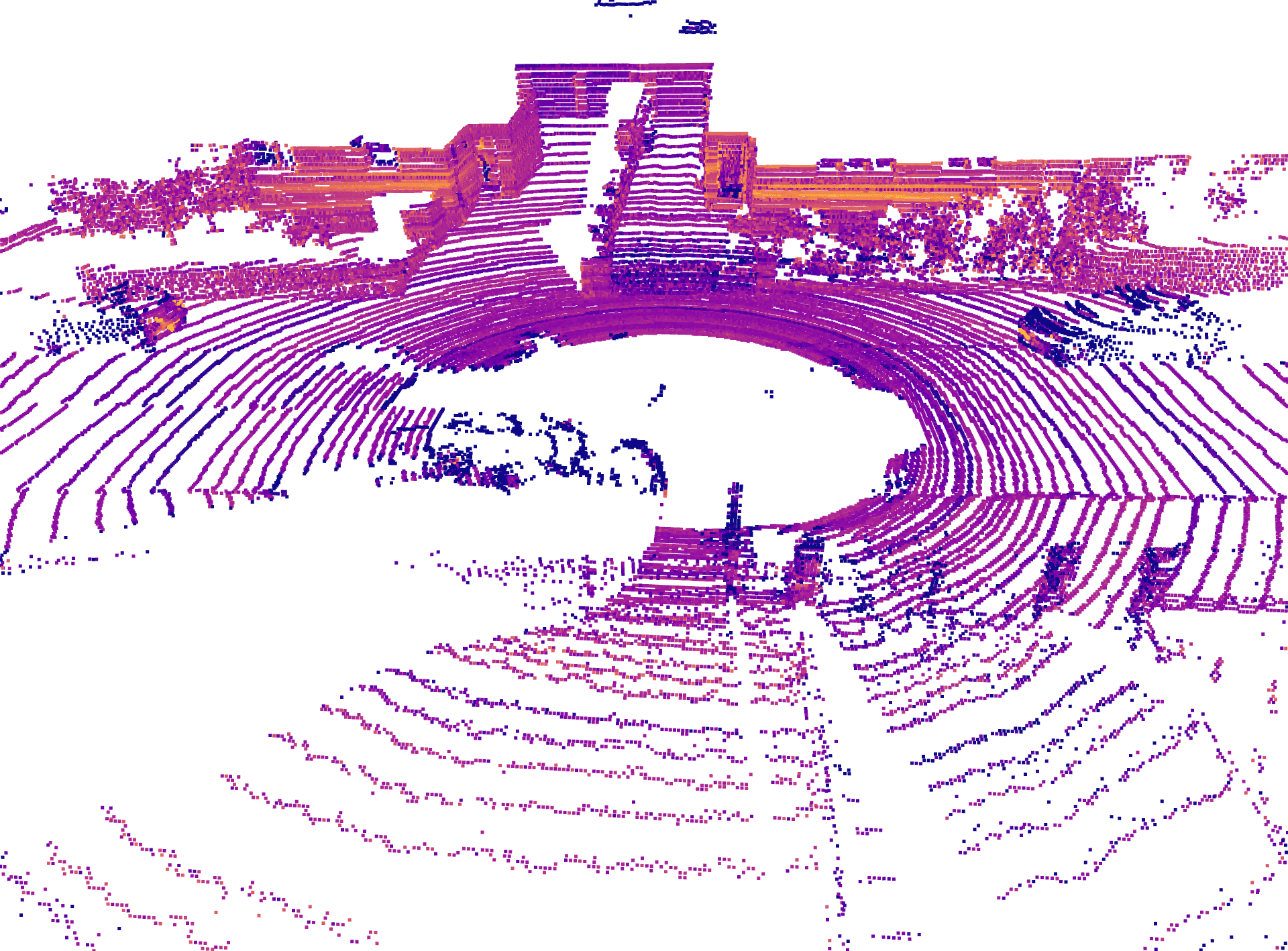}
    \put (0,48) {\colorbox{gray!30}{\scriptsize Ours: F1 90.8 Bitrate 5.6}}
\end{overpic}
\begin{overpic}[clip,trim=0cm 15cm 0cm 0cm,width=0.49\textwidth]{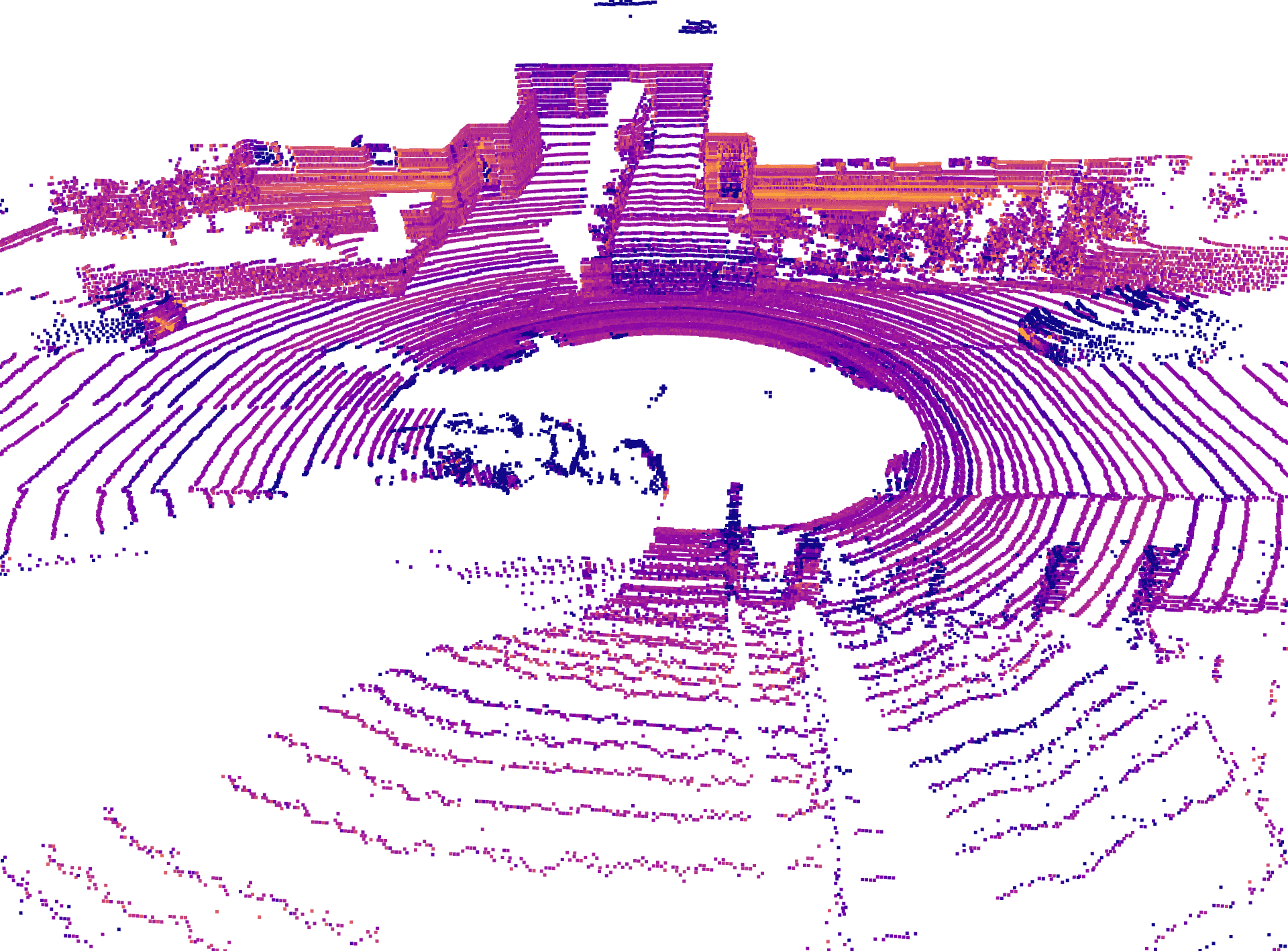}
    \put (0,48) {\colorbox{gray!30}{\scriptsize Draco: F1 89.2 Bitrate 5.8}}
\end{overpic}
\begin{overpic}[clip,trim=0cm 15cm 0cm 0cm,width=0.49\textwidth]{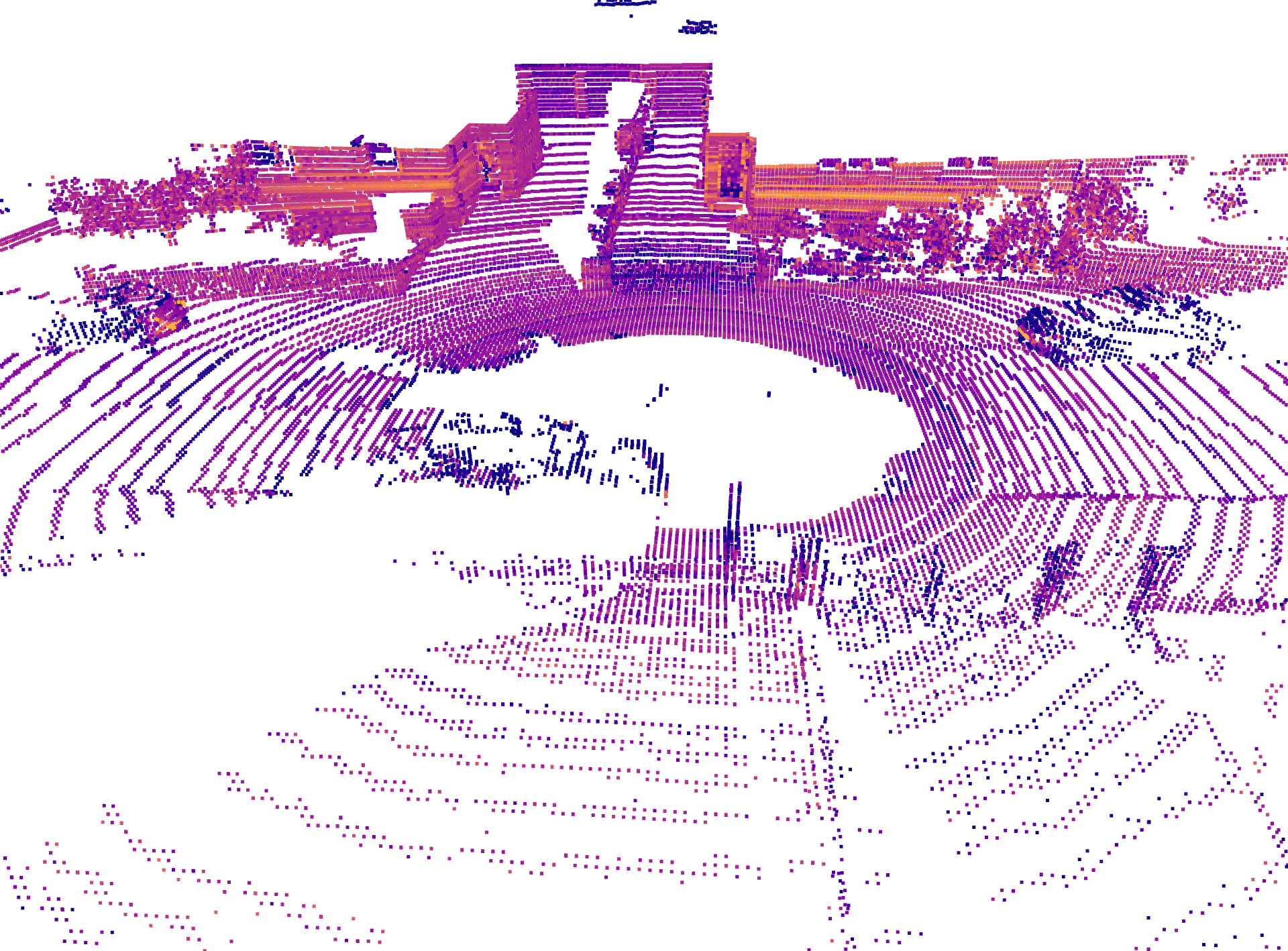}
    \put (0,48) {\colorbox{gray!30}{\scriptsize MPEG: F1 69.2 Bitrate 11.0}}
\end{overpic}

\caption{
\small Qualitative results on SemanticKITTI.
Points are colored by intensity.
}
\label{figure:kitti-additional-qualitative-results}
\end{figure}

\begin{figure}[!t]
\centering

\begin{overpic}[clip,trim=0cm 10cm 0cm 5cm,width=0.49\textwidth]{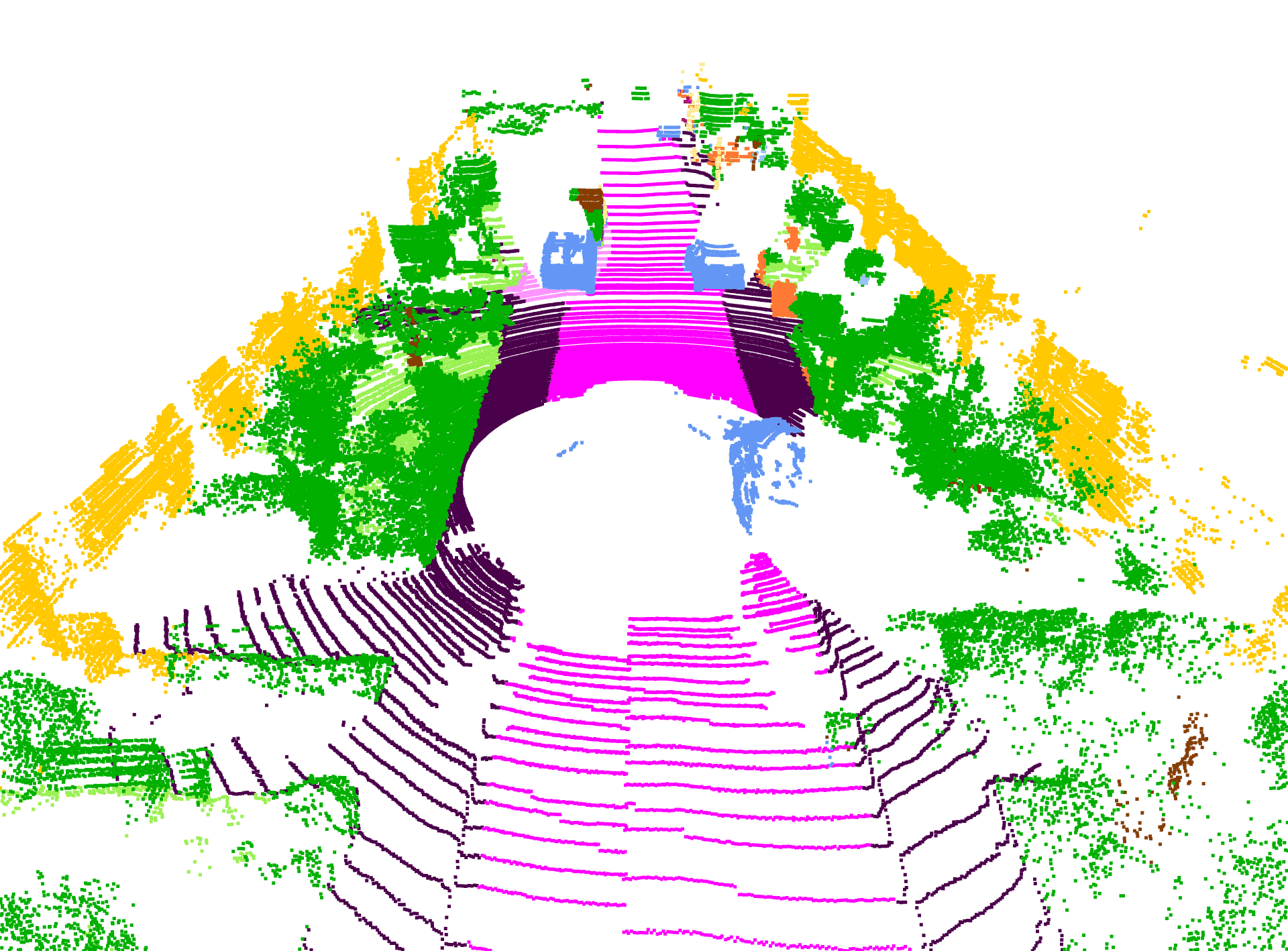}
    \put (0,48) {\colorbox{gray!30}{\scriptsize Oracle (KITTI): IOU 31.3 Bitrate 104.0}}
\end{overpic}
\begin{overpic}[clip,trim=0cm 10cm 0cm 5cm,width=0.49\textwidth]{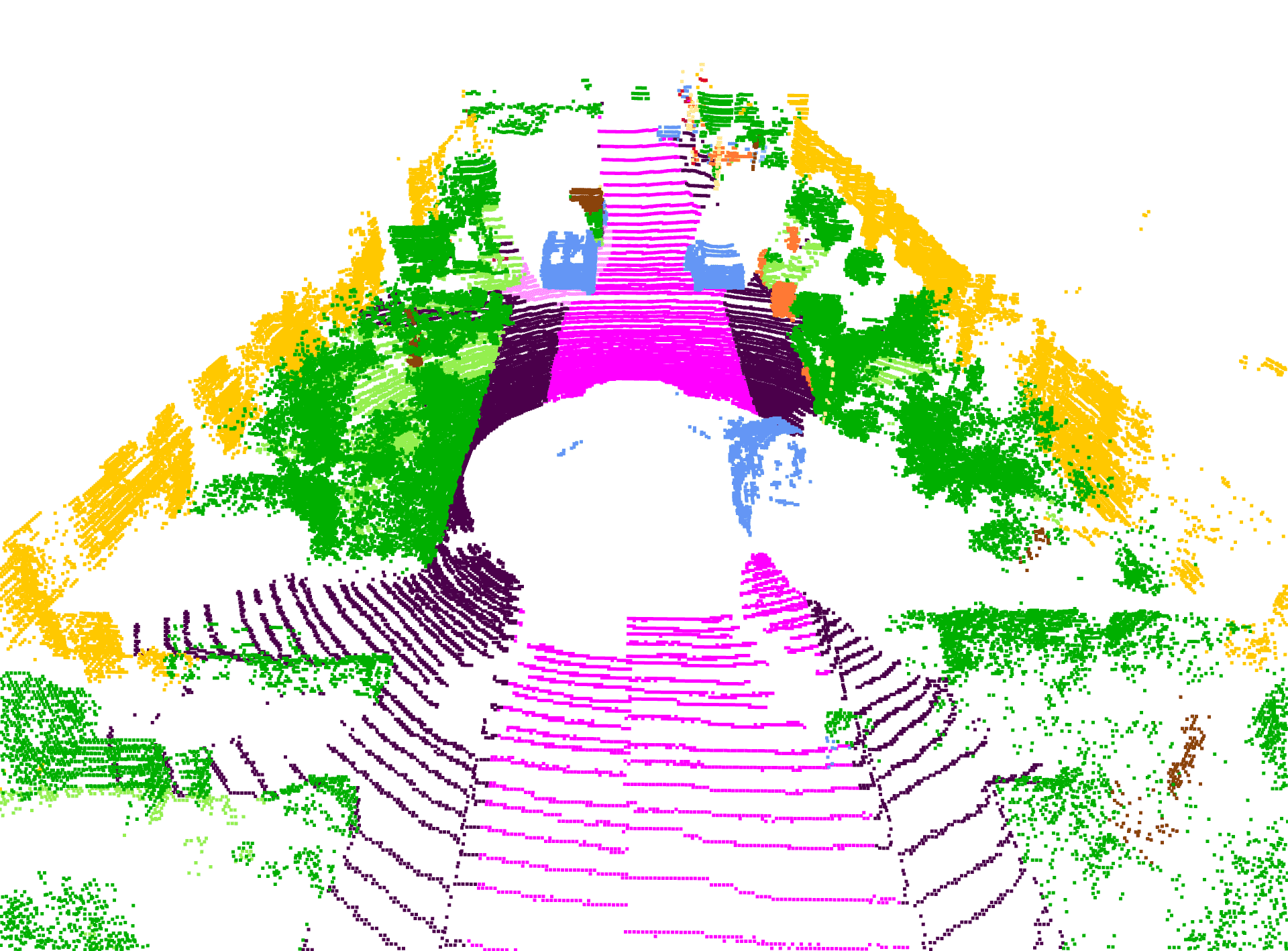}
    \put (0,48) {\colorbox{gray!30}{\scriptsize Ours: IOU 29.5 Bitrate 6.7}}
\end{overpic}
\begin{overpic}[clip,trim=0cm 10cm 0cm 5cm,width=0.49\textwidth]{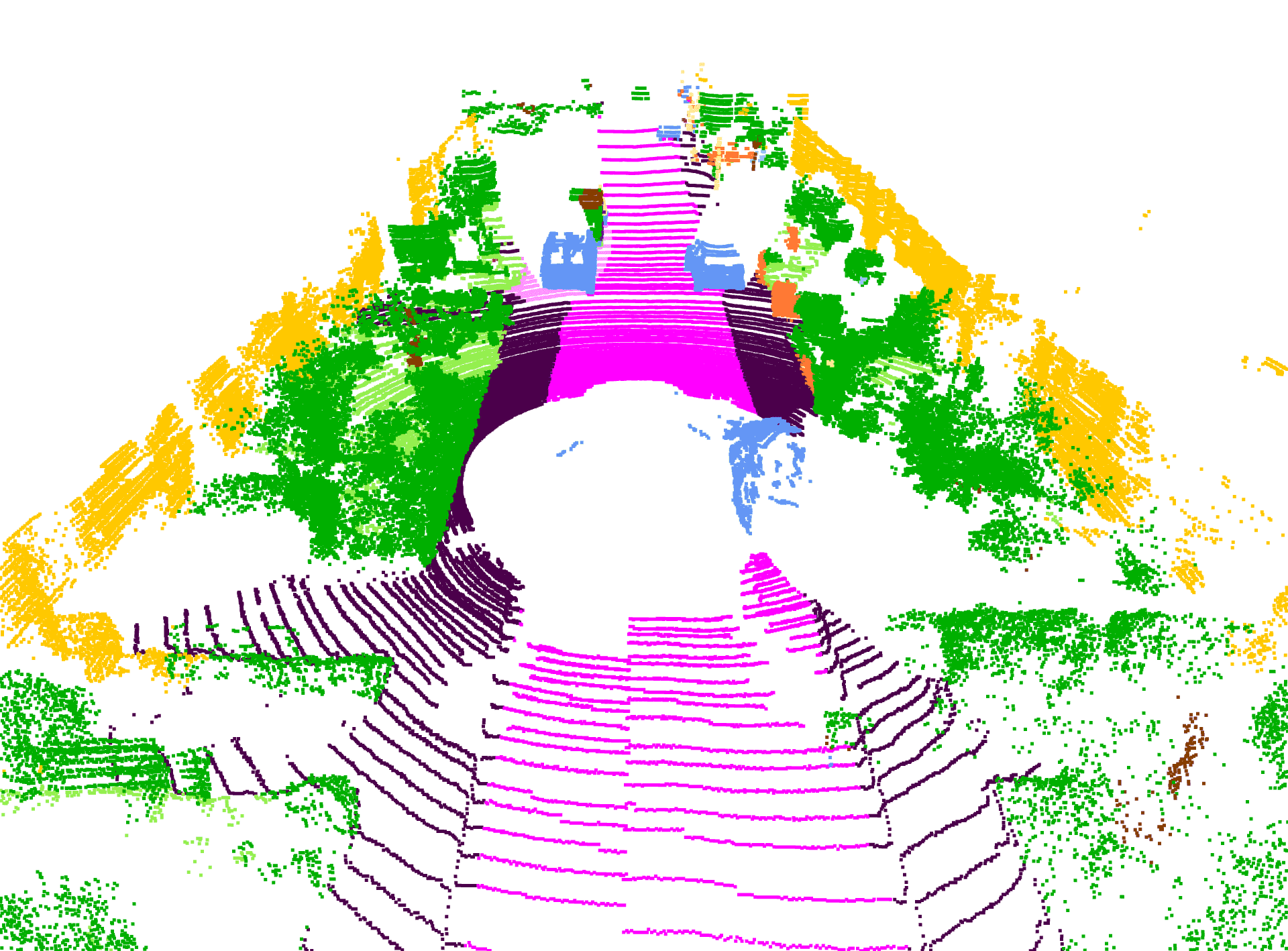}
    \put (0,48) {\colorbox{gray!30}{\scriptsize Draco: IOU 29.0 Bitrate 8.4}}
\end{overpic}
\begin{overpic}[clip,trim=0cm 10cm 0cm 5cm,width=0.49\textwidth]{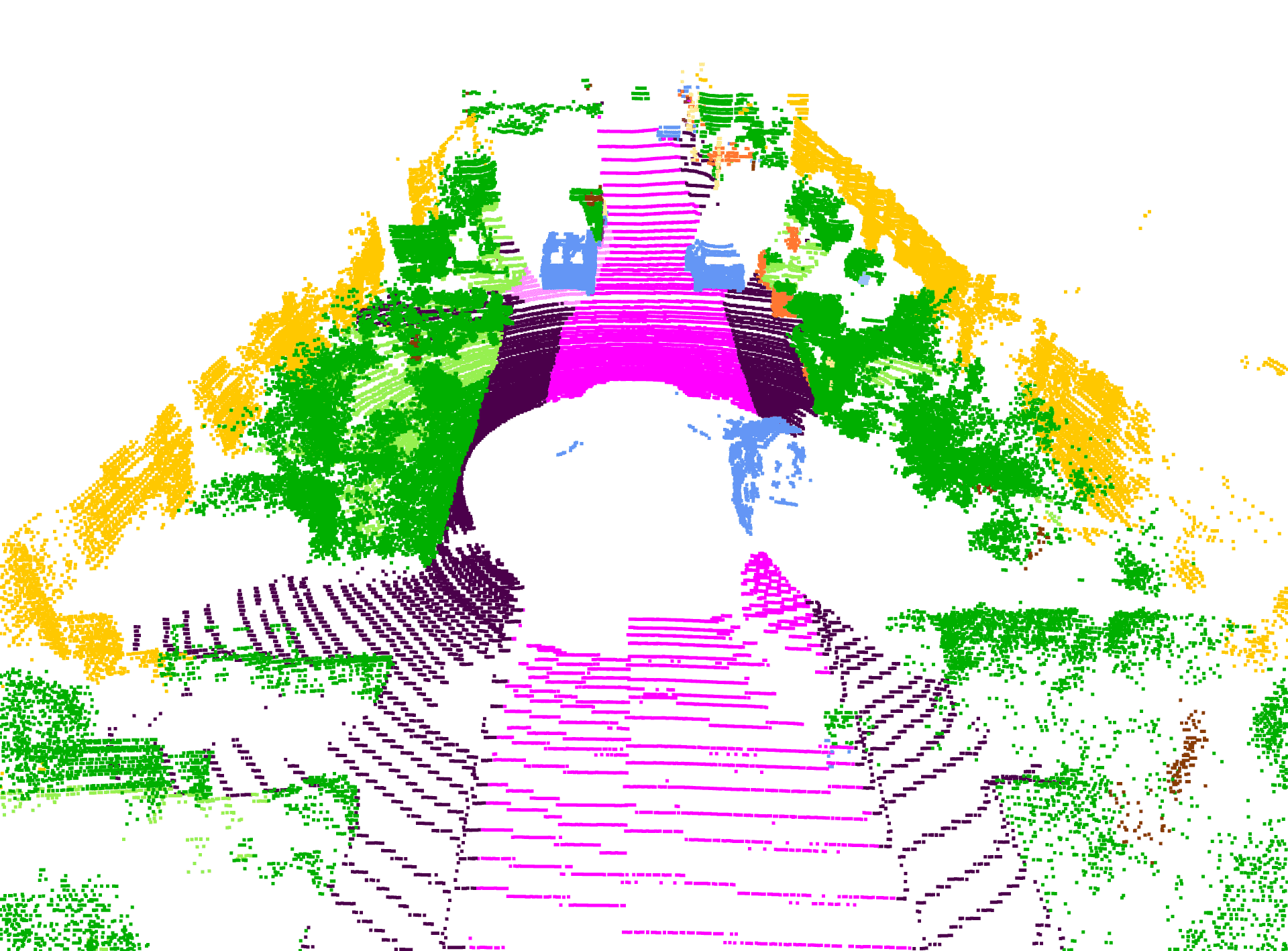}
    \put (0,48) {\colorbox{gray!30}{\scriptsize MPEG: IOU 26.3 Bitrate 13.0}}
\end{overpic}

\caption{
\small Semantic segmentation results on SemanticKITTI.
IOU is averaged over all classes.
}
\label{figure:kitti-additional-downstream-qualitative-results}
\end{figure}

\begin{figure}[!t]
\centering

\begin{overpic}[clip,trim=0cm 10cm 0cm 5cm,width=0.49\textwidth]{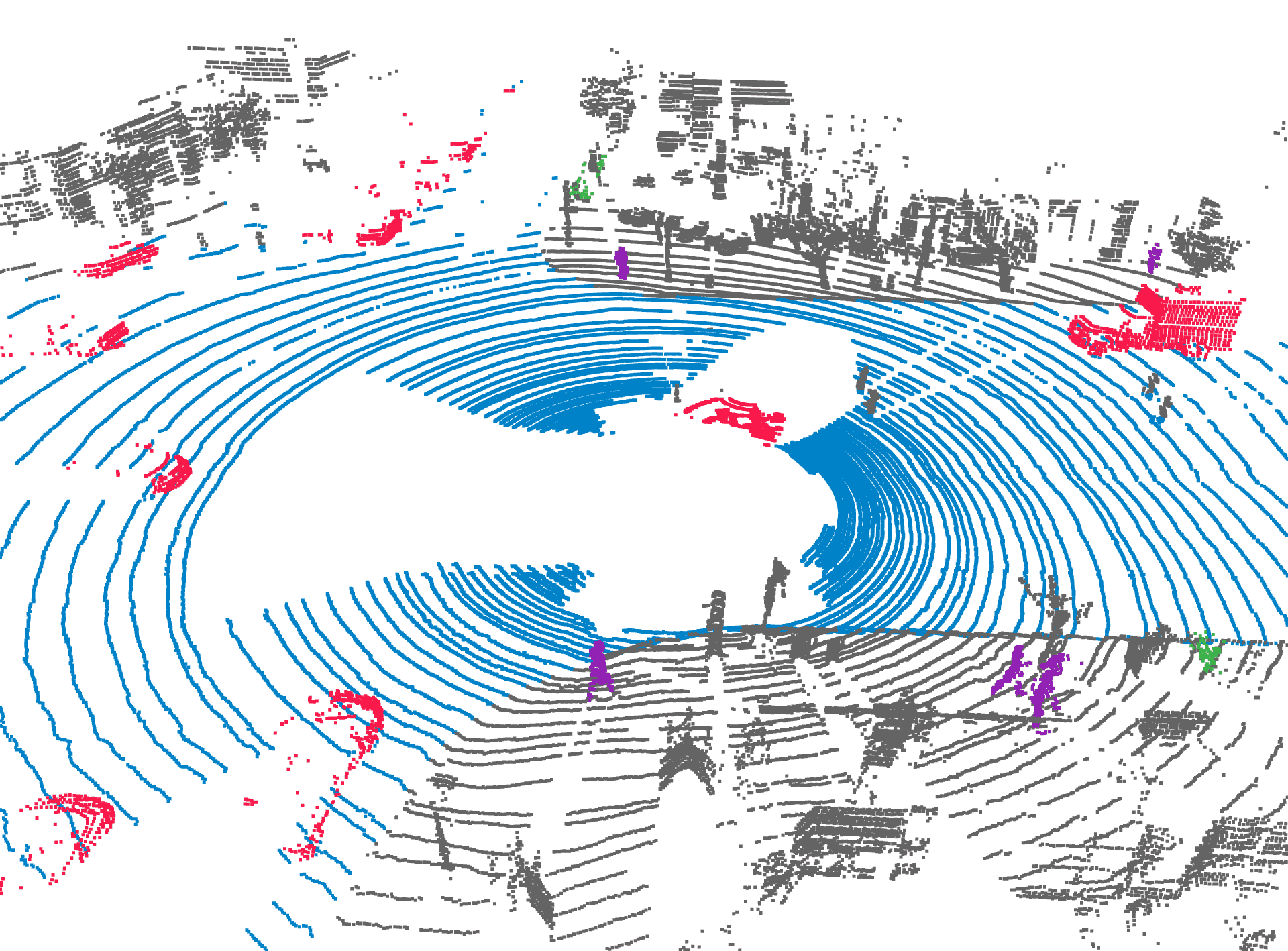}
    \put (0,48) {\colorbox{gray!30}{\scriptsize Oracle (KITTI): IOU 97.2 Bitrate 104.0}}
\end{overpic}
\begin{overpic}[clip,trim=0cm 10cm 0cm 5cm,width=0.49\textwidth]{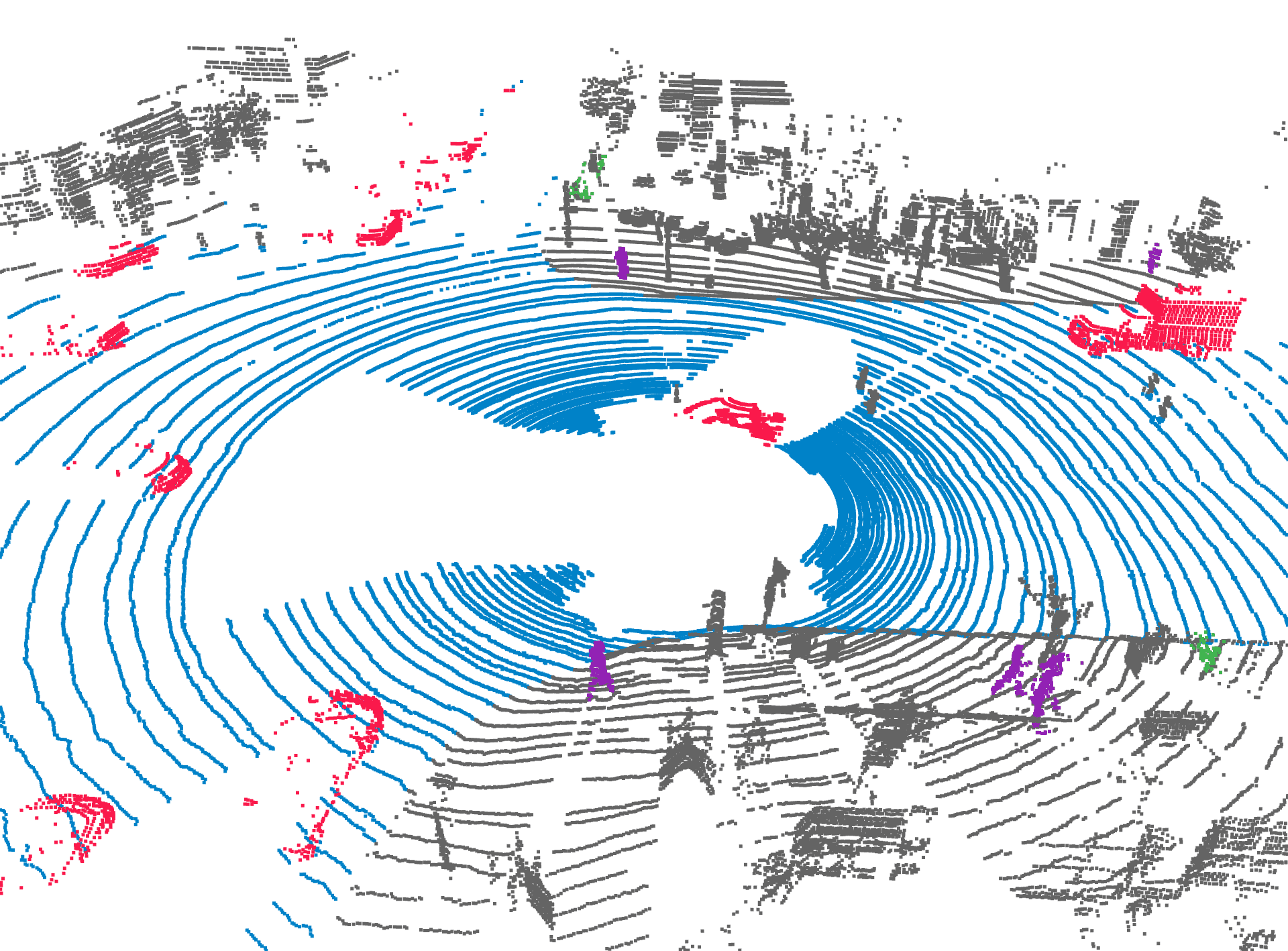}
    \put (0,48) {\colorbox{gray!30}{\scriptsize Ours: IOU 94.3 Bitrate 19.8}}
\end{overpic}
\begin{overpic}[clip,trim=0cm 10cm 0cm 5cm,width=0.49\textwidth]{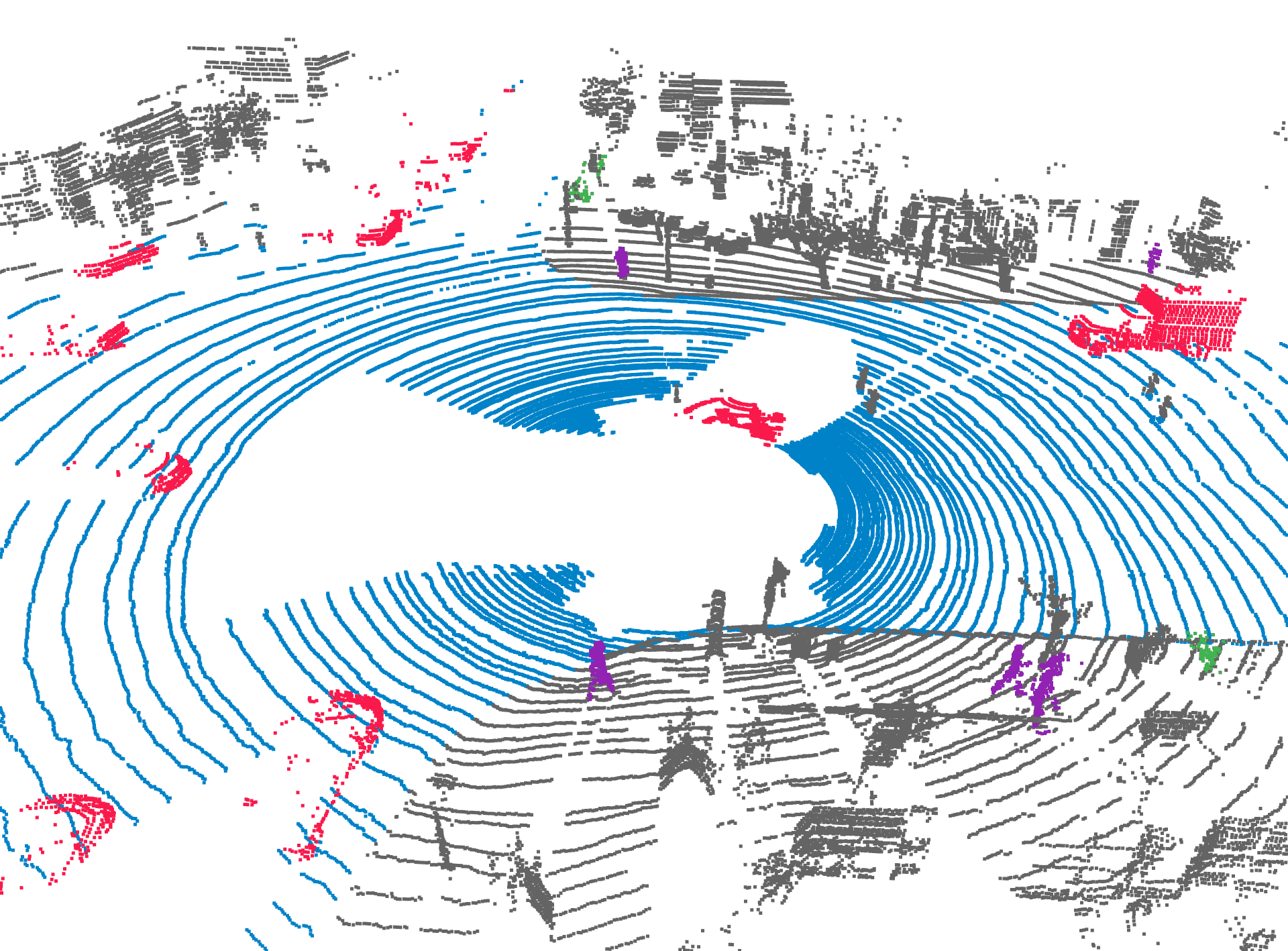}
    \put (0,48) {\colorbox{gray!30}{\scriptsize Draco: IOU 88.3 Bitrate 20.1}}
\end{overpic}
\begin{overpic}[clip,trim=0cm 10cm 0cm 5cm,width=0.49\textwidth]{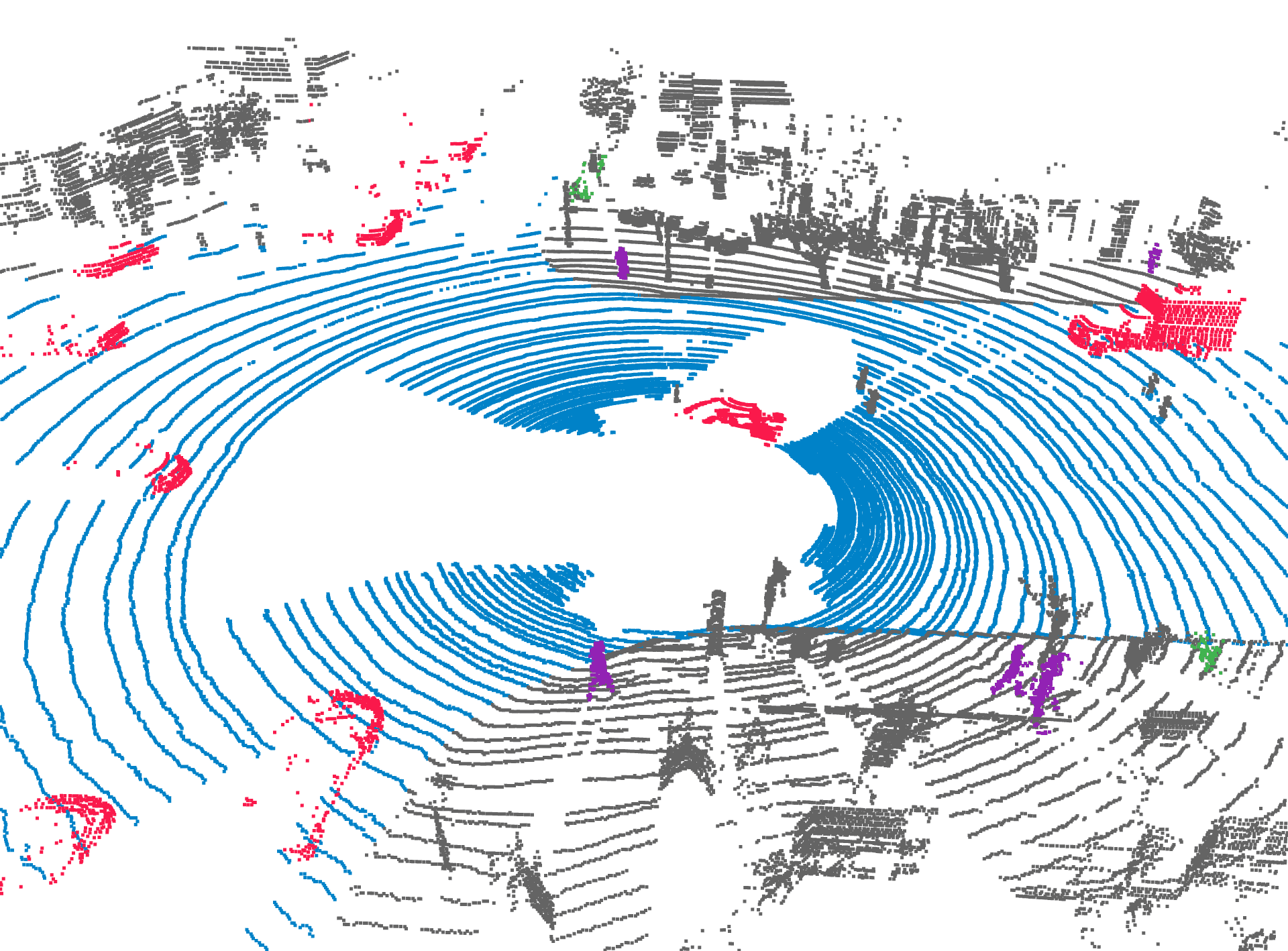}
    \put (0,48) {\colorbox{gray!30}{\scriptsize MPEG: IOU 85.2 Bitrate 20.5}}
\end{overpic}

\caption{
\small Semantic segmentation results on \northamerica.
IOU is averaged over all classes.
}

\label{figure:northamerica-segmentation-additional-downstream-qualitative-results}
\end{figure}

\begin{figure}[!t]
\centering

\begin{overpic}[clip,trim=0cm 10cm 0cm 5cm,width=0.49\textwidth]{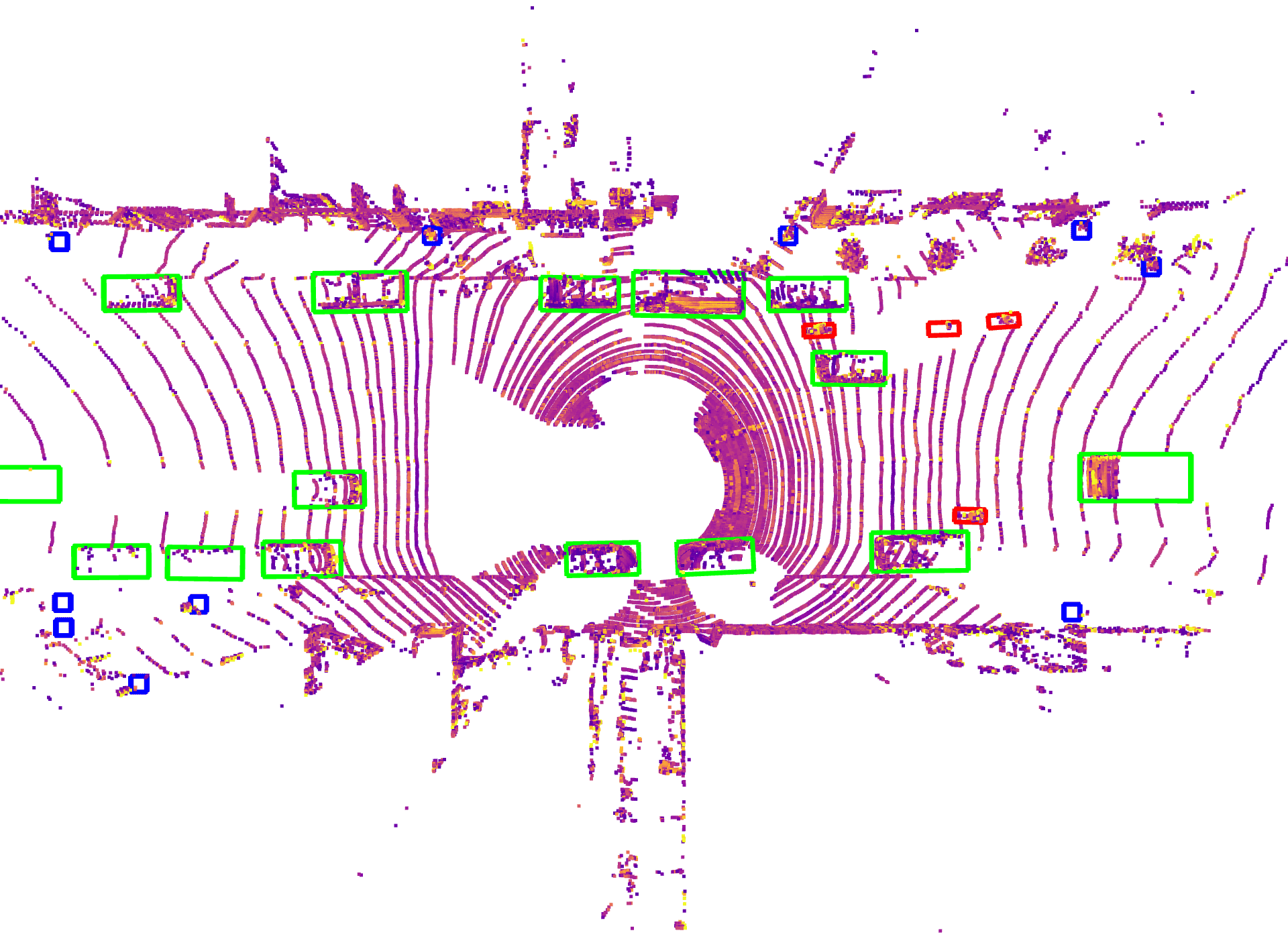}
    \put (0,48) {\colorbox{gray!30}{\scriptsize Oracle (KITTI): AP 90.6 Bitrate 104.0}}
\end{overpic}
\begin{overpic}[clip,trim=0cm 10cm 0cm 5cm,width=0.49\textwidth]{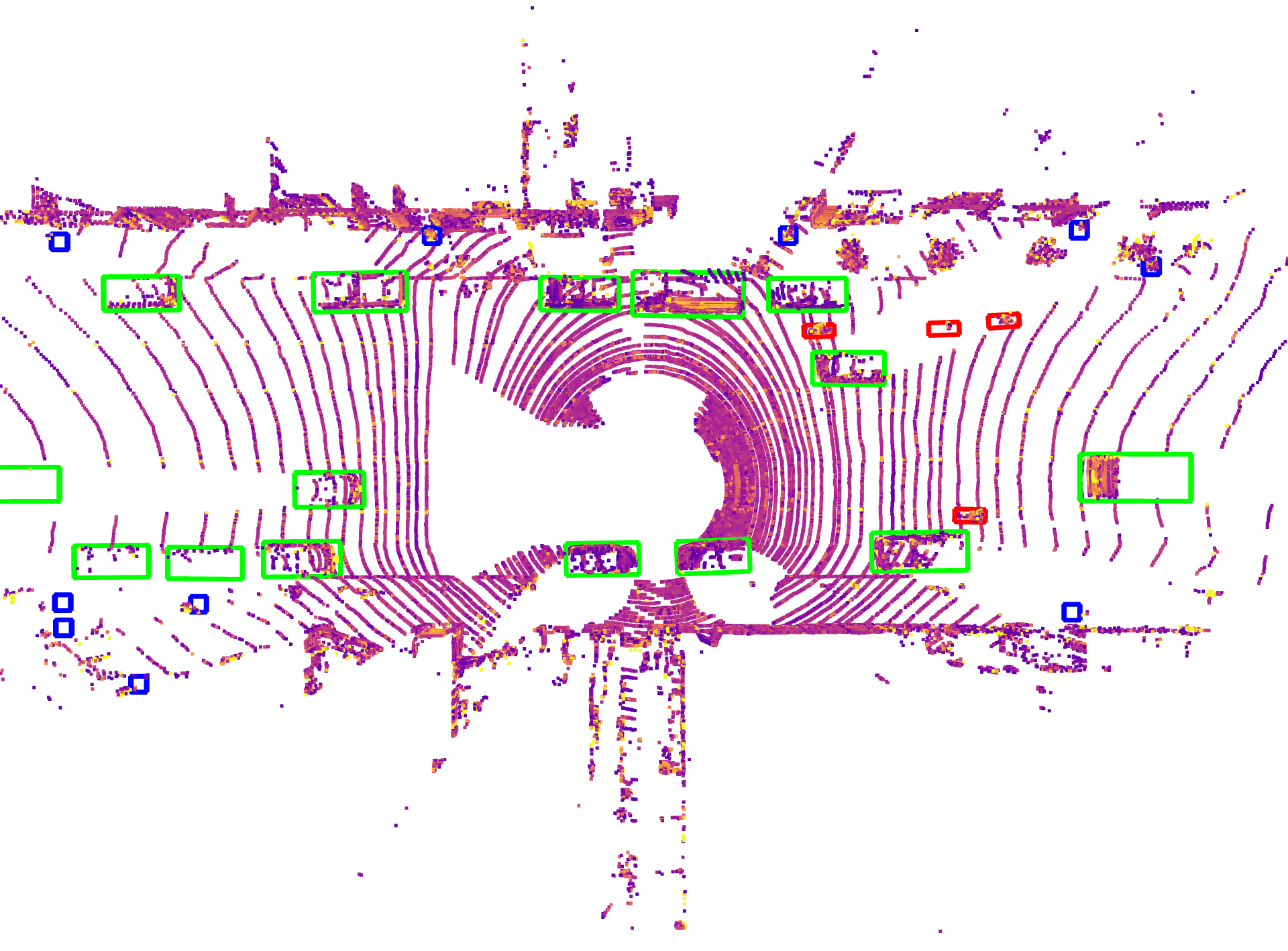}
    \put (0,48) {\colorbox{gray!30}{\scriptsize Ours: AP 90.4 Bitrate 14.5}}
\end{overpic}
\begin{overpic}[clip,trim=0cm 10cm 0cm 5cm,width=0.49\textwidth]{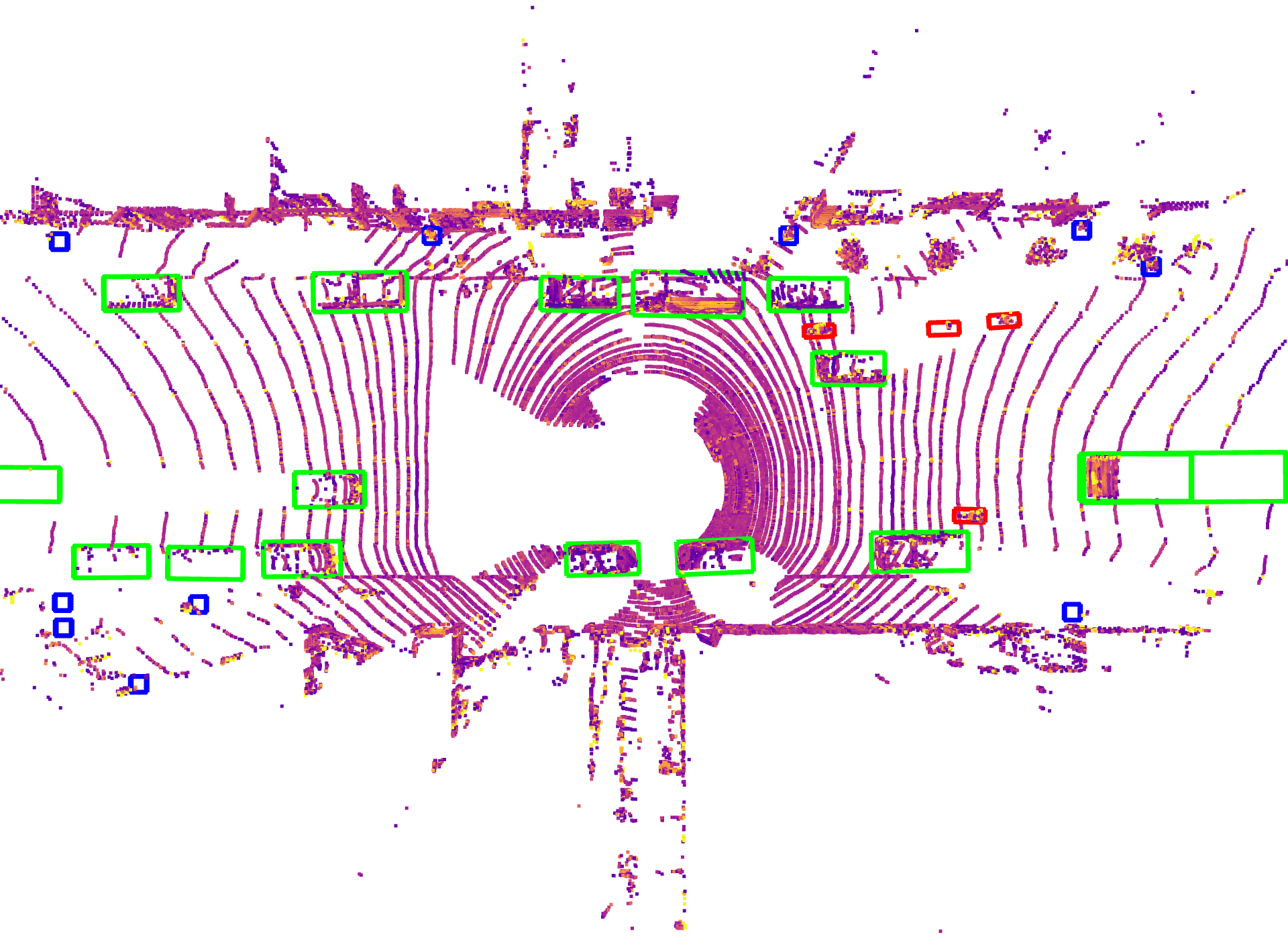}
    \put (0,48) {\colorbox{gray!30}{\scriptsize Draco: AP 89.4 Bitrate 16.6}}
\end{overpic}
\begin{overpic}[clip,trim=0cm 10cm 0cm 5cm,width=0.49\textwidth]{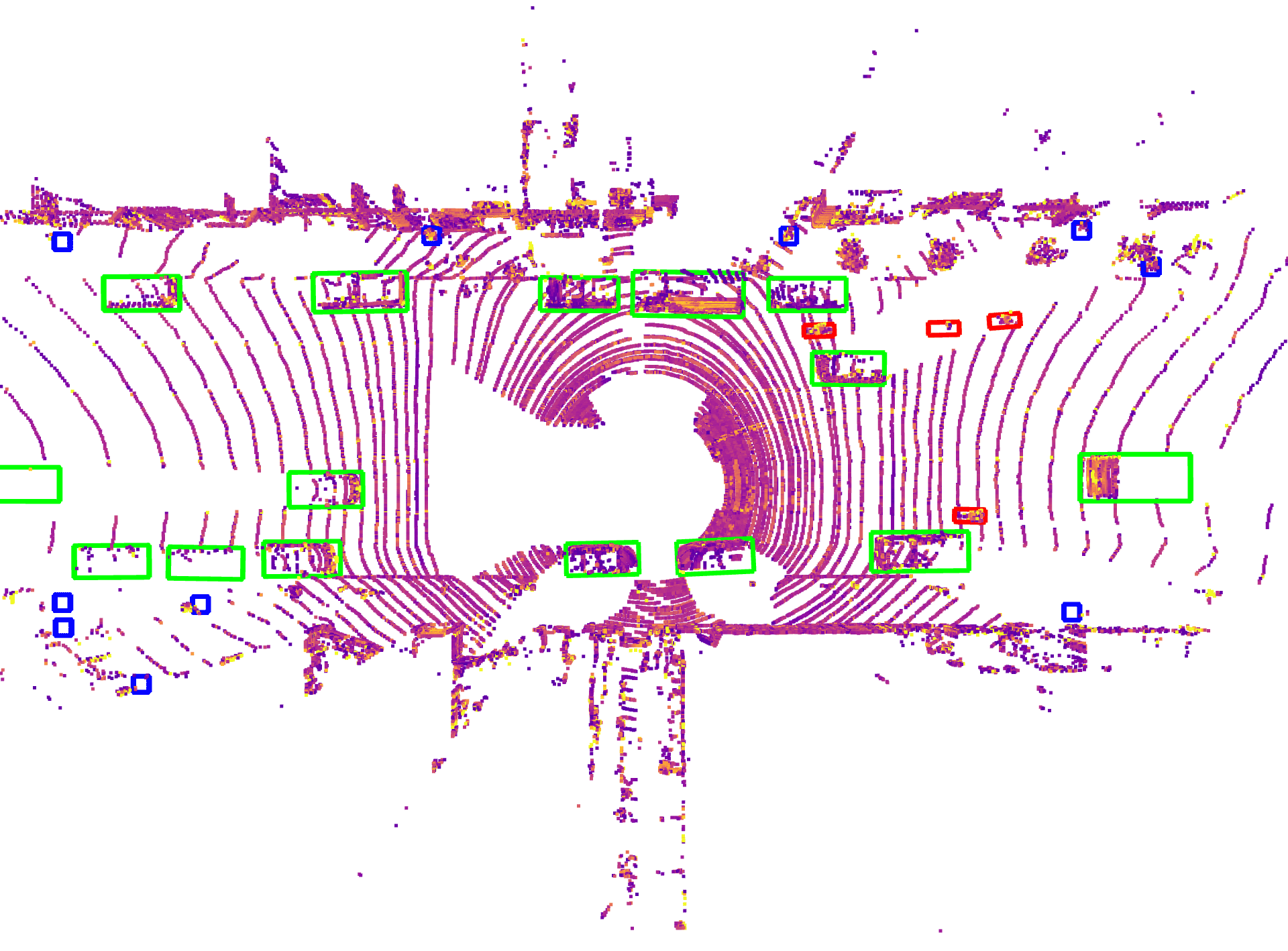}
    \put (0,48) {\colorbox{gray!30}{\scriptsize MPEG: AP 91.2 Bitrate 17.6}}
\end{overpic}

\caption{
\small Object detection results on \northamerica.
AP is averaged over the vehicle, pedestrian, and motorbike classes.
}
\label{figure:northamerica-detection-additional-downstream-qualitative-results}
\end{figure}

In Fig.~\ref{figure:northamerica-additional-qualitative-results} and
\ref{figure:kitti-additional-qualitative-results}, we compare the reconstruction
quality of our method versus Draco~\cite{draco} and MPEG anchor~\cite{mekuria2016}.
Then, in Figs.~\ref{figure:kitti-additional-downstream-qualitative-results},
\ref{figure:northamerica-segmentation-additional-downstream-qualitative-results},
and \ref{figure:northamerica-detection-additional-downstream-qualitative-results},
we visualize results from semantic segmentation and object detection on SemanticKITTI
and \northamerica.
As shown in these figures, our compression algorithm yields the best
reconstruction quality at comparable or lower bitrates than the competing methods.


\section{Change Log}

\paragraph{ArXiv v2:}
We updated our reconstruction metrics to use the standard MPEG
definitions~\cite{schwarz2019}.
Furthermore, we added bitrate \vs $ \mathrm{F}_1 $ curves
for a number of spatial and intensity thresholds.
We also updated our Draco baseline to use its built-in attributes coder.


\end{document}